\documentclass[letterpaper,twocolumn,10pt]{article}
\usepackage{styles/usenix2019_v3}

\usepackage{nth}
\usepackage{tikz}
\usepackage{amsfonts}
\usepackage{amsmath}
\usepackage{amsthm}
\usepackage{todonotes}
\usepackage[inline]{enumitem}
\usepackage{soul}
\usepackage{graphicx}
\usepackage{subcaption}
\usepackage[hang,flushmargin,bottom]{footmisc}
\usepackage{verbatim}
\usepackage{fancyvrb}
\usepackage{adjustbox}
\usepackage{booktabs}
\usepackage[ruled,linesnumbered]{algorithm2e}
\usepackage{scrextend}
\usepackage{multirow}
\usepackage{array}
\usepackage{amssymb}
\usepackage{pifont}
\usepackage{threeparttable}
\usepackage{footnote}

\usepackage{authblk}
\usepackage{combelow}

\newcommand{\PreserveBackslash}[1]{\let\temp=\\#1\let\\=\temp}
\newcolumntype{C}[1]{>{\PreserveBackslash\centering}p{#1}}
\newcolumntype{R}[1]{>{\PreserveBackslash\raggedleft}p{#1}}
\newcolumntype{L}[1]{>{\PreserveBackslash\raggedright}p{#1}}

\newcommand{\cmark}{{\color{black}\ding{51}}}
\newcommand{\xmark}{{\color{black}\ding{55}}}

\newcommand{\sfvrule}{\unskip\ \vrule\ }

\newcommand{\topic}[1]{\vspace{0.2em} \noindent \textbf{#1}}
\newcommand{\subbullet}[1]{\noindent $\bullet$ \emph{#1}}

\SetCommentSty{commentfont}

\theoremstyle{definition}

\newcommand{\tcircle}[1]{%
	\begin{tikzpicture}
	\draw (0,0) circle (0.8ex);\fill (1ex,0) arc (0:#1:1ex) -- (0,0) -- cycle;
	\end{tikzpicture}%
}

\newcommand{\rotate}[1]{\rotatebox[origin=c]{90}{#1}}

\definecolor{moss}{RGB}{0,144,81}
\definecolor{brown}{RGB}{132,60,12}

\usepackage{afterpage}

\renewcommand{\paragraph}{\vspace{3pt}\noindent\textbf}
\usepackage{titlesec}
\titlespacing*{\section}{0pt}{3.0ex plus 0.0ex minus 0.0ex}{1.4ex plus 0.0ex}
\titlespacing*{\subsection}{0pt}{2.4ex plus 0.0ex minus 0.0ex}{1.2ex plus 0.0ex}
\titlespacing*{\subsubsection}{0pt}{2.0ex plus 0.0ex minus 0.0ex}{1.0ex plus 0.0ex}

\begin{document}

\def\usenixcamready{1}


\title{On the Effectiveness of Mitigating Data Poisoning Attacks with Gradient Shaping}

\if\usenixcamready1
	\author{
		{\rm Sanghyun Hong, Varun Chandrasekaran\textsuperscript{\textdagger}, Yi\u{g}itcan Kaya, Tudor Dumitra\cb{s}, Nicolas Papernot\textsuperscript{\textasteriskcentered}}\\
		\textit{University of Maryland, College Park}\\
		\textsuperscript{\textdagger}\textit{University of Wisconsin-Madison}\\
		\textsuperscript{\textasteriskcentered}\textit{University of Toronto and Vector Institute}
	}
\else
	\author{}
\fi

\maketitle


\newcommand{\norm}[1]{\left\|#1\right\|}
\def\minop{\mathop{\rm min}\limits}
\def\maxop{\mathop{\rm max}\limits}
\newcommand{\sign}{\ensuremath{\text{sign}}}

\newcommand{\myparagraph}[1]{\vspace{0.1in}\noindent\textbf{#1}}
\newcommand{\myparnovspace}[1]{\vspace{0.0in}\noindent\textbf{#1}}
\newcommand{\myvspacenopar}{\vspace{0.1in}\noindent }

\newtheorem{prop}{Proposition}
\newtheorem{fact}{Fact}
\newtheorem{lem}{Lemma}
\newtheorem{rem}{Remark}
\newcommand{\One}{\mathds{1}}

\newtheorem{theorem}{Theorem}
\newtheorem{proposition}{Proposition}
\newtheorem{lemma}{Lemma}

\newtheorem{example}{Example}
\newtheorem{remark}{Remark}

\newcommand{\ie}{\textit{i.e.,}\@\xspace}
\newcommand{\eg}{\textit{e.g.,}\@\xspace}
\newcommand{\etal}{\textit{et al.}\@\xspace}
\newcommand{\Z}{\ensuremath{\mathbb{Z}}}
\newcommand{\E}{\ensuremath{\mathbb{E}\nicolas}}
\newcommand{\D}{\ensuremath{\mathcal{D}}}
\newcommand{\Q}{\ensuremath{\mathbb{Q}}}
\newcommand{\Y}{\ensuremath{\mathbf{Y}}}
\newcommand{\X}{\ensuremath{\mathbf{X}}}
\newcommand{\F}{\ensuremath{\mathcal{F}}}
\renewcommand{\L}{\ensuremath{\mathcal{L}}}
\renewcommand*{\O}{\ensuremath{\mathcal{O}}}
\newcommand{\A}{\ensuremath{\mathcal{A}}}
\newcommand{\Ser}{\ensuremath{S}}
\newcommand{\uErr}{\ensuremath{\textup{Err}_u}}
\newcommand{\gErr}{\ensuremath{\textup{Err}_2}}
\newcommand{\Err}{\ensuremath{\textup{Err}}}
\newcommand{\negl}{\ensuremath{\textup{negl}}}
\newcommand{\tr}{\top}
\newcommand{\at}{\makeatletter @\makeatother}
\newcommand{\qquote}[1]{``#1''}

\newcommand{\Varun}[1]{\textcolor{orange}{[\textbf{Varun}: #1]}}
\newcommand{\Tudor}[1]{\textcolor{blue}{[\textbf{Tudor}: #1]}}
\newcommand{\Sanghyun}[1]{\noindent \textcolor{red}{[\textbf{Sanghyun}: #1]}}
\newcommand{\Nicolas}[1]{\textcolor{cyan}{[\textbf{Nicolas}: #1]}}
\newcommand{\John}[1]{\textcolor{purple}{[\textbf{John}: #1]}}

\newcommand{\lt}{\ensuremath <}
\newcommand{\gt}{\ensuremath >}
\newcommand*\bcircle[1]{\tikz[baseline=(char.base)]{
            \node[shape=circle,fill,inner sep=0.2pt] (char) {\textcolor{white}{#1}};}}

\newcommand{\btcircle}[2][black,fill=black]{\tikz[baseline=-0.7ex]\draw[#1,radius=#2] (0,0) circle ;}%
\newcommand{\wtcircle}[2][black,fill=white]{\tikz[baseline=-0.7ex]\draw[#1,radius=#2] (0,0) circle ;}%
\newcommand{\nomark}{\ding{53}}%

\newcommand\blfootnote[1]{%
  \begingroup
  \renewcommand\thefootnote{}\footnote{#1}%
  \addtocounter{footnote}{-1}%
  \endgroup
}


\begin{abstract}
\noindent
Machine learning algorithms are vulnerable to data poisoning attacks.
Prior taxonomies that focus on specific scenarios, \eg indiscriminate or targeted, have enabled defenses for the corresponding subset of known attacks. Yet, this  introduces an inevitable arms race between  adversaries and defenders.
In this work, we study the feasibility of an attack-agnostic defense relying on artifacts that are common to all poisoning attacks.
Specifically, we focus on a common element between all attacks: they modify gradients computed to train the model. We identify two main artifacts of gradients computed in the presence of poison: (1) their $\ell_2$ norms have significantly higher magnitudes than those of clean gradients, and (2) their orientation differs from clean gradients.
Based on these observations, we propose the prerequisite for a generic poisoning defense: it must bound  gradient magnitudes and minimize differences in orientation. We call this \emph{gradient shaping}.
As an exemplar tool to evaluate the feasibility of gradient shaping, we use differentially private stochastic gradient descent (DP-SGD), which clips and perturbs individual gradients during training to obtain privacy guarantees.
We find that DP-SGD, even in configurations that do not result in meaningful privacy guarantees, increases the model's robustness to indiscriminate attacks. It also mitigates worst-case targeted attacks and increases the adversary's cost in multi-poison scenarios.
The only attack we find DP-SGD to be ineffective against is a strong, yet unrealistic, indiscriminate attack.
Our results suggest that, while we currently lack a generic poisoning defense, gradient shaping is a promising direction for future research. 
\end{abstract}

\if\usenixcamready1
	\blfootnote{This work was carried out when Varun Chandrasekaran was a visiting researcher at the University of Toronto and the Vector Institute.}
\else\fi


\section{Introduction}
\label{sec:intro}
A common paradigm for building a new machine learning (ML) system is to collect the required training data from several sources.
Ideally, the data should come from trustworthy sources. However, the scale of modern ML tasks and challenges in establishing trust often force practitioners to resort to unvetted sources.
This exposes ML systems to potentially dangerous training data and enables \emph{data poisoning attacks}.
In poisoning attacks, the attacker inserts poison in the victim's training set to induce it into learning a model whose behavior is advantageous to the attacker.
Data poisoning has been demonstrated in malware classification~\cite{perdisci2006misleading, xiao2015feature}, spam filtering~\cite{Nelson:LEET08}, and DoS detection~\cite{Antidote:IMC09}.
%

Poisoning has sparked an arms race between proposed defenses and  attacks that defeat them.
For example,  a data-sanitization defense called RONI relies on the assumption that a poison sample in the training set necessarily  hurts the model's accuracy~\cite{Barreno:ML10}.
While RONI could defend against some \emph{indiscriminate} attacks, it was shown to be ineffective against novel and adaptive \emph{targeted} attack~\cite{Suciu:USENIX18}.
%
In general, prior attacks and defenses emphasized the diversity of information available to the adversary. In this paradigm, safeguarding ML models against poisoning requires a strong understanding of the threat surface exposed by learning algorithms.

In this work, we challenge prior taxonomies of poisoning in ML and make a first step towards a unified view of the threat surface. 
Specifically, we ask: \emph{What are some essential characteristics shared across various poisoning attacks?}
To answer this question, we focus on the cornerstone of how most ML models are trained -- using gradients.
During training, the gradients computed on data 
dictate how a model's parameters should be updated; this determines  properties of the resulting model.
Stochasticity found in many popular optimizers results in poisoners having limited control over gradients computed during training.
%
This gives defenders an edge to reason about differences between clean and poisoned gradients.

Poisoning attacks rely on two broad approaches to manipulate a model's behavior: feature collision and feature insertion.
For example, to degrade performance, indiscriminate poisoning attacks~\cite{Barreno:ML10, Steinhardt:NIPS17, Jagielski:Oakland18, Diakonikolas:ICML19} rely on feature collision to overwrite clean features.
Targeted poisoning attacks~\cite{Suciu:USENIX18, Ali:NIPS18, Zhu:ICML19}, on the other hand, rely mainly on feature insertion from the target samples into the model to cause misclassification on a few test-time targets without hurting overall model performance.
Prior work treats these as distinct attacks, because they make different assumptions about the adversary's capabilities, such as the amount of poison that can be inserted in a dataset.

As we consider how poisoning affects gradients, we find that both scenarios craft poisons that share two properties. 
First, the gradients computed from the poison and the clean samples have observable magnitude and orientation differences.
Second, these differences grow as we introduce {\em stronger} poison samples.
We use these properties to unify the threat surface of both indiscriminate and targeted attacks.

These properties also suggest design guidelines for a generic defense strategy effective against more forms of poisoning.
First, gradient-level differences---magnitude and orientation---between the poison and the clean samples result in model parameter updates in favor of the adversary.
In consequence, an ideal defense should minimize such differences to ensure that the poison cannot dominate a model's behavior.
Further, as attacks can still be effective even when poison samples closely resemble clean ones, a defense also should not rely on data sanitization, \ie identifying and removing malicious samples.
Most prior defenses~\cite{Barreno:ML10, Steinhardt:NIPS17, Diakonikolas:ICML19} rely on a form of sanitization, which means they have to make attack-specific assumptions.
Based on these desiderata, we propose \emph{gradient shaping} as a step toward defenses that generalize.

In gradient shaping, a defense aims to mitigate poisoning at the gradient-level during training and remains agnostic to training samples.
As a concrete implementation of gradient shaping, we experiment with an off-the-shelf tool: differentially private stochastic gradient descent (DP-SGD)~\cite{Abadi:CCS16}.
DP-SGD is originally a training algorithm that provides differential privacy guarantees with respect to training data.
Because it clips the norm of individual gradients and adds noise to them; we find DP-SGD 
is a suitable candidate for a gradient shaping mechanism.
Thus, we study the feasibility of gradient shaping with DP-SGD against a wide range of attacks.

We evaluate DP-SGD against two indiscriminate poisoning attacks~\cite{Barreno:ML10, Steinhardt:NIPS17} and a strong clean-label targeted poisoning attack~\cite{Ali:NIPS18}.
Our results on three ML models, linear regression, multi-layer perceptrons, and convolutional neural networks---trained on three popular ML tasks, Purchase-100, FashionMNIST, and CIFAR-10---reveal that DP-SGD can be effective against multiple poisoning attacks, even when DP-SGD only provides trivial privacy guarantees.
For example, against an indiscriminate attack (random label-flipping), it 
reduces the performance degradation by half, and against a one-shot targeted attack, it prevents targets from being misclassified.
Furthermore, against a multi-poison targeted attack, it also forces the adversary to blend more poisons and, therefore, increases the attack's cost.
However, even though we still observe gradient-level differences, DP-SGD is relatively ineffective against a strong, albeit unrealistic, indiscriminate attack~\cite{Steinhardt:NIPS17}.
We believe this exposes the limitations of DP-SGD in performing gradient shaping; therefore, designing even more suitable mechanisms is an important direction for future research.

\topic{Contributions.} In summary, we make four contributions:
\begin{itemize}[noitemsep, topsep=0pt]
    \item We expose common gradient-level properties across various forms of poisoning, in contrast to previous taxonomies. In particular, we identify that poisoned gradients have higher magnitudes and are oriented differently when compared to clean gradients. 
    
    \item We take a step towards unifying the poisoning threat surface based on our gradient-level analysis.
    
    \item Based on our unified view, we discuss the desiderata for a generic, attack-agnostic, defense against poisoning and propose \emph{gradient shaping} as a defense approach that fulfills these requirements.
    
	\item We utilize DP-SGD as an off-the-shelf gradient shaping tool. We evaluate the effectiveness of DP-SGD against various poison attacks with a systematic study on three ML models and three ML tasks.
\end{itemize}

\section{Preliminaries on ML and Poisoning}
\label{sec:prelim}

In probably approximately correct (PAC) learning~\cite{Valiant:ACM84},
there is an underlying data distribution $\mathcal{Z}\!=\!\mathcal{X} \times \mathcal{Y}$ where $\mathcal{X}$ is the domain of inputs and $\mathcal{Y}$ the outputs.
For example, in spam email classification, inputs could be emails and outputs are labels indicating whether these emails are spam or ham. 
The objective of a learning algorithm $\mathcal{Q}$ is to learn a parameterized model $f_\theta \in \mathcal{H}$, where $\mathcal{H}$ is the space of hypotheses.
For instance, a restricted $\mathcal{H}$ could be the space of all parameters for a particular neural network whose weights and biases $\theta$ need to be learned to obtain a model $f_\theta$. 
The model itself is a mapping between inputs and labels \ie $f_\theta :\mathcal{X} \rightarrow \mathcal{Y}$.

To train such a model, $\mathcal{Q}$ has access to a dataset $\mathcal{D}$ that is drawn from the underlying data distribution $\mathcal{Z}$.
In the supervised learning setting~\cite{Russell:Book09}, $\mathcal{D}$ is partitioned into disjoint subsets: the training $\mathcal{D}_{tr}$ and test $\mathcal{D}_{ts}$ datasets.
We also assume the existence of a non-negative, real-valued loss function $\mathcal{L}(f_{\theta}(x),y)$, which quantifies how correct (or incorrect) the prediction of a model is given an input $x$ and label $y$.
%

\topic{Empirical Risk Minimization (ERM).}
Given such a loss function $\mathcal{L}$, the output of a supervised learning algorithm is a model $f_\theta$ which minimizes the risk $\mathbb{E}_{(x,y) \sim \mathcal{Z}}[\mathcal{L}(f_{\theta}(x),y)]$ where $f_{\theta}(x)$ is the predicted label and $y$ is the true label. 
Since the underlying data distribution $\mathcal{Z}\!=\!\mathcal{X} \times \mathcal{Y}$ is unknown, supervised learning algorithms use the training set $\mathcal{D}_{tr}$ of size $m$ to learn a hypothesis that minimizes the {\em empirical risk}~\cite{Vapnik:NIPS91}.
The empirical risk is defined as $\frac{1}{m} \sum_{i=1}^{m} \mathcal{L}(f_{\theta}(x_i), y_i)$ for $(x_i, y_i)\!\sim\!\mathcal{D}_{tr}$.
Commonly used loss functions include mean square error (MSE) and cross-entropy loss~\cite{Lecun:Book15}. 
%
The model is then tested based on how well it generalizes to the unseen samples in $\mathcal{D}_{ts}$.

\topic{Gradient Descent.}
%
Gradient descent has established itself as the de-facto approach, in particular when it comes to training neural networks.
Known as the backpropagation algorithm~\cite{rumelhart1988learning} in the context of neural networks, gradient descent updates model parameters with a multiplicative of the derivative of the empirical risk with respect to model parameters $\theta$ at each iteration $t$:
\vspace{-0.4em}			
\begin{equation*}
\vspace{-0.4em}			
\theta_{t+1} \leftarrow \theta_t - \eta \cdot \nabla\frac{1}{m} \sum_{i=1}^{m} \mathcal{L}(f_{\theta}(x_i), y_i)
\end{equation*}
where $\eta$ is the learning rate and controls the magnitude of changes made at each iteration.
When it is not feasible to compute the empirical risk over the entire training set, one samples a single example from the training set and uses it to compute empirical risk and update the model.
This variant is known as stochastic gradient descent (SGD).
To obtain an unbiased estimate of the true gradient that would be computed on the entire set~\cite{bottou2008tradeoffs}, one may alternatively sample a mini-batch, \ie a small number of examples, $(\mathbf{x}_t, \mathbf{y}_t)$  from the dataset at each iteration---rather than a single example or the whole set:
%
%
This learning procedure is called mini-batch SGD. 


\subsection{Data Poisoning}
\label{subsec:data-poisoning}

Data poisoning is a training-time attack which manipulates a ML model's behavior in favor of the attacker, by injecting maliciously crafted samples, \ie \emph{poisons}, into the training set.
If a ML model is trained on the contaminated training set, an attacker can prevent the trained model from generalizing well to the test data or cause misclassifications of specific test-time samples, \ie \emph{targets}, without degrading the generalization performance of the trained model.
The former attacks are known as \emph{indiscriminate poisoning} whereas the latter are referred to as \emph{targeted poisoning}.
These attacks are highly effective 
when an adversary cannot control the test-time samples, or when an attacker is not able to alter the training procedures.

\subsubsection{Poisoning Mechanisms}
\label{subsubsec:mechanisms}

Adversaries exploit two mechanisms to trigger the intended misclassifications: \emph{feature collisions} and \emph{feature insertions}.

\subbullet{Feature Collision:} An attacker can blend poisons such that a victim model learns the opposite of what it would from clean data.
For example, the attacker can create poisons by just flipping the label of a clean samples because both the clean and poison samples have the same input space information.
If the victim trains a model on the poisoned training data, the information learned from the clean and poison samples contrast each other. 
%
%
The indiscriminate attackers cause the collisions to occur with the features useful for the classification of most test-time samples, while the feature collisions happen locally by the targeted attackers so that they cause misclassifications on a small subset of testing samples.

\subbullet{Feature Insertion:} In addition, the attacker can make a victim model learn new latent representations useful for misclassification. 
For instance, in image classification, the attacker can identify the features (mostly in the input space) that commonly appear in the targets, but are not critical to classify the target samples.
The adversary can proceed to craft the poisons to include those features, but with the label that the attacker wants for the targets to be misclassified as.
If a model trained with the poisons encounters samples, and if it does not find the features added by the adversary, the model classifies them correctly, while the targets are classified into the attacker's label.
Thus, the mechanism is useful for targeted poisoning.

%

\subsubsection{Threat Model}
\label{subsubsec:threat-models}

In this subsection, we specify the threat model we operate in.

\topic{Capability:}
Our work assumes an attacker who cannot modify the victim model $f$ and its parameters $\theta$ directly, or control/modify the training procedure $\mathcal{Q}$ to cause the misclassification.
We consider an attacker who can craft poisons offline and blend them into the training data $\mathcal{D}_{tr}$.
%
%
More specifically, we consider two scenarios:
(1) the attacker blends poisons at the beginning of the training phase, or 
(2) adds poisons when the learner attempts to update the trained model.
Thus, what the attacker can control is (1) the number of poisons that will be added to the training data, and (2) the time (or the training stage) when the attacker decides to blend poisons.

\topic{Knowledge:}
Prior work~\cite{Suciu:USENIX18} characterized the poisoning attacker's knowledge using four dimensions: (1) the training data $\mathcal{D}_{tr}$, (2) the subset of features the victim uses $\mathcal{X}$, and (3) the learning algorithm $\mathcal{Q}$, and (4) the model parameters $\theta$.
The \emph{black-box} setting considers the attacker has no knowledge about the four dimensions whereas in the \emph{white-box} setting, where the attacker knows all the four components, at least partially.
As we discuss defensive mechanisms against data poisoning, we consider strong white-box attackers.

\section{Poisoning Mechanisms and Gradients}
\label{sec:grad}

In this section, we characterize the attack surface exploited by data poisoning attacks through a systematic analysis of gradients computed by training algorithms.
Here, we focus on the impact of the poisoning mechanisms that we specified in \S\ref{subsubsec:mechanisms} on the gradients computed during the training of a model.
Since the attacker cannot modify the victim model, perturb its parameters, and control the training procedures (specified in \S\ref{subsubsec:threat-models}), the attacker cannot directly manipulate the gradients computed on the poisons by the victim's training algorithm but can influence them through poisons.
In contrast to previous taxonomies structured around the knowledge and capabilities of adversaries~\cite{Suciu:USENIX18}, our characterization provide a unified perspective on data poisoning attacks in \S\ref{sec:taxonomy}.

\subsection{Experimental Setup and Methodology}
\label{subsec:analysis-gradients-methods}


\begin{figure*}[t]
	\centering
	\begin{subfigure}[t]{0.32\textwidth}
		\centering
		\caption*{\textbf{Feature Collision (LR, Scratch)}}
		\vspace{-0.6em}
		\includegraphics[width=\textwidth]{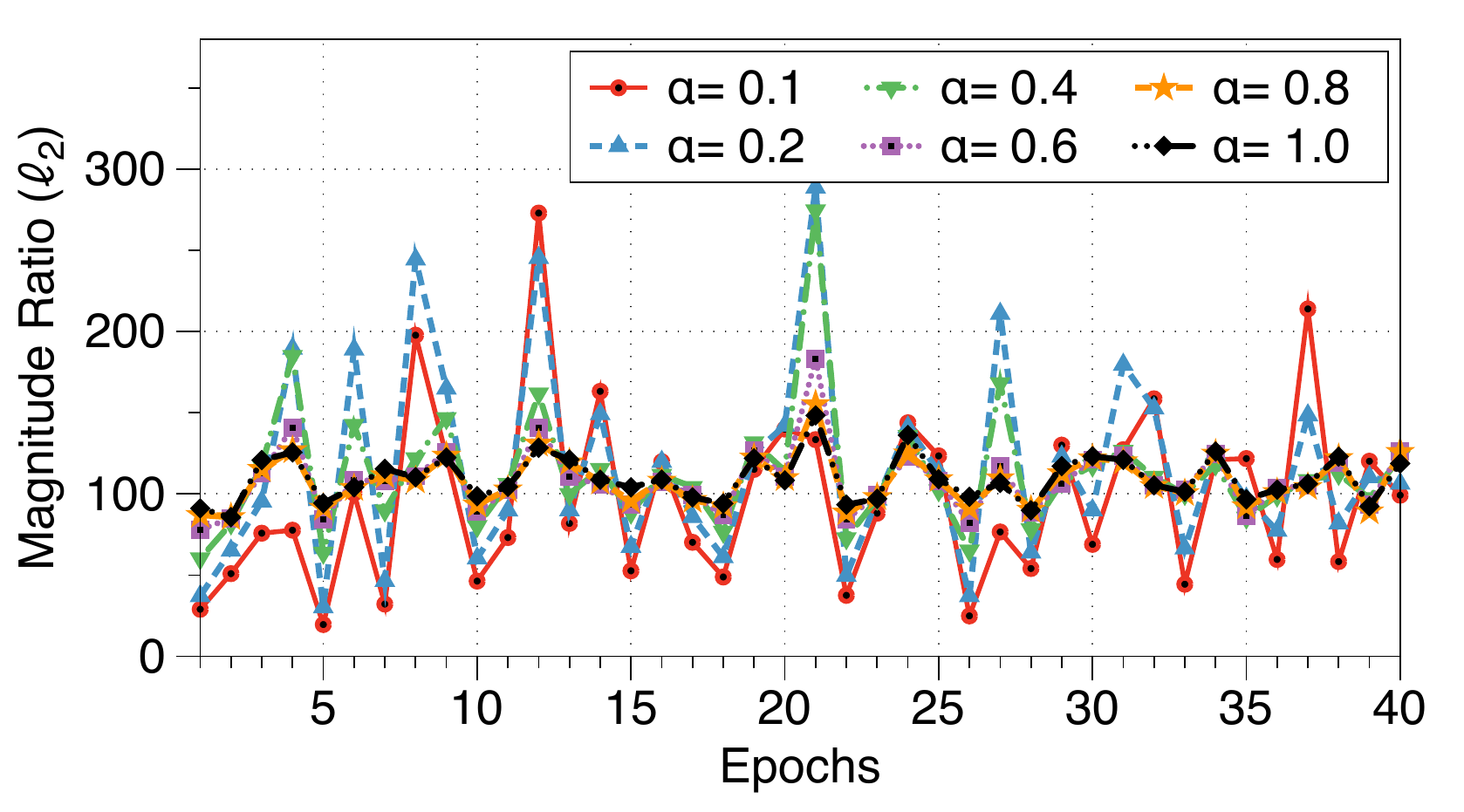}
	\end{subfigure}
	\begin{subfigure}[t]{0.32\textwidth}   
		\centering
		\caption*{\textbf{Feature Collision (MLP, Re-train)}}
		\vspace{-0.6em}
		\includegraphics[width=\textwidth]{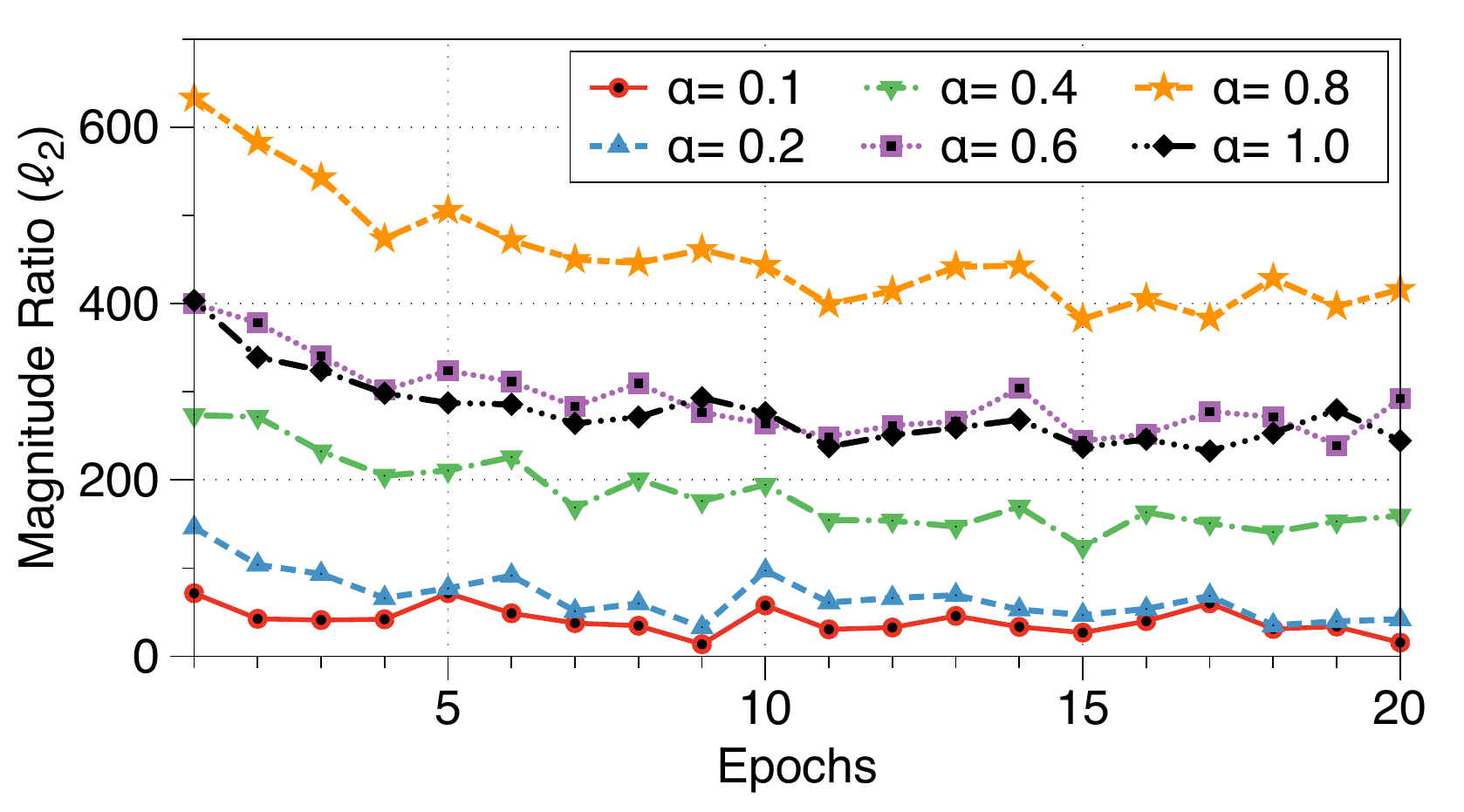}
	\end{subfigure}
	\begin{subfigure}[t]{0.32\textwidth}
		\centering
		\caption*{\textbf{Feature Insertion (MLP, Re-train)}}
		\vspace{-0.6em}
		\includegraphics[width=\textwidth]{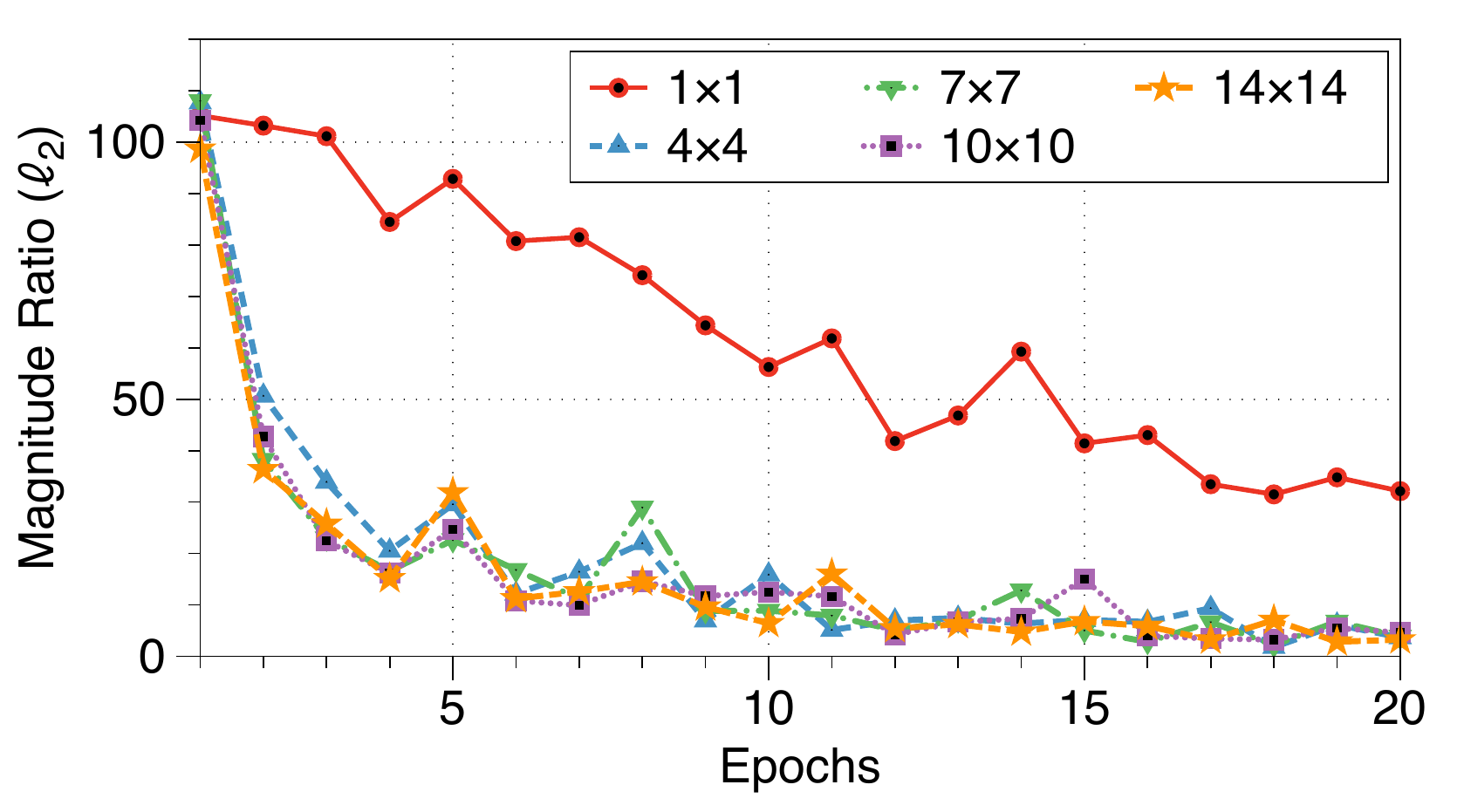}
	\end{subfigure}
	\begin{subfigure}[t]{0.32\textwidth}
		\centering
		\includegraphics[width=\textwidth]{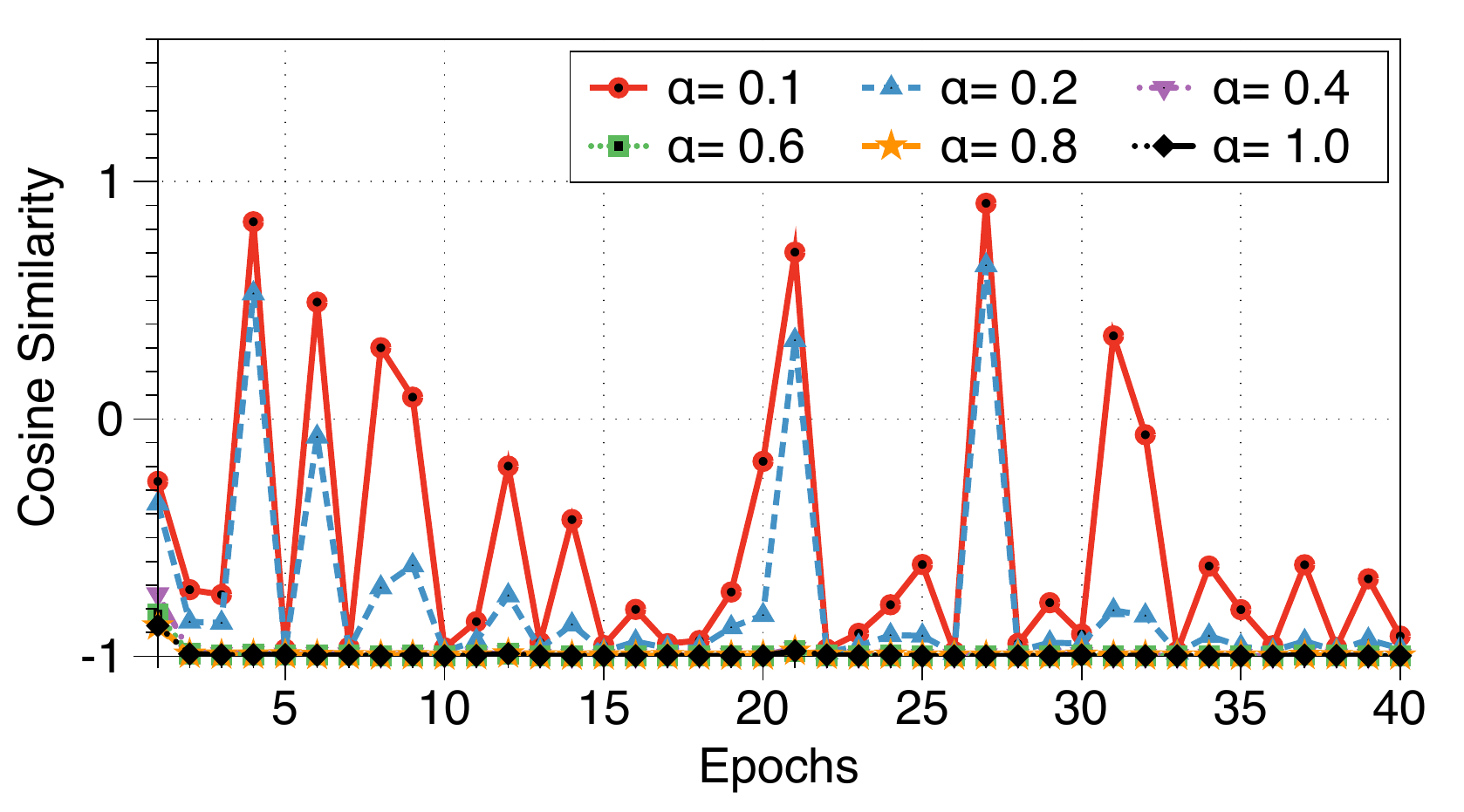}
	\end{subfigure}
	\begin{subfigure}[t]{0.32\textwidth}   
		\centering 
		\includegraphics[width=\textwidth]{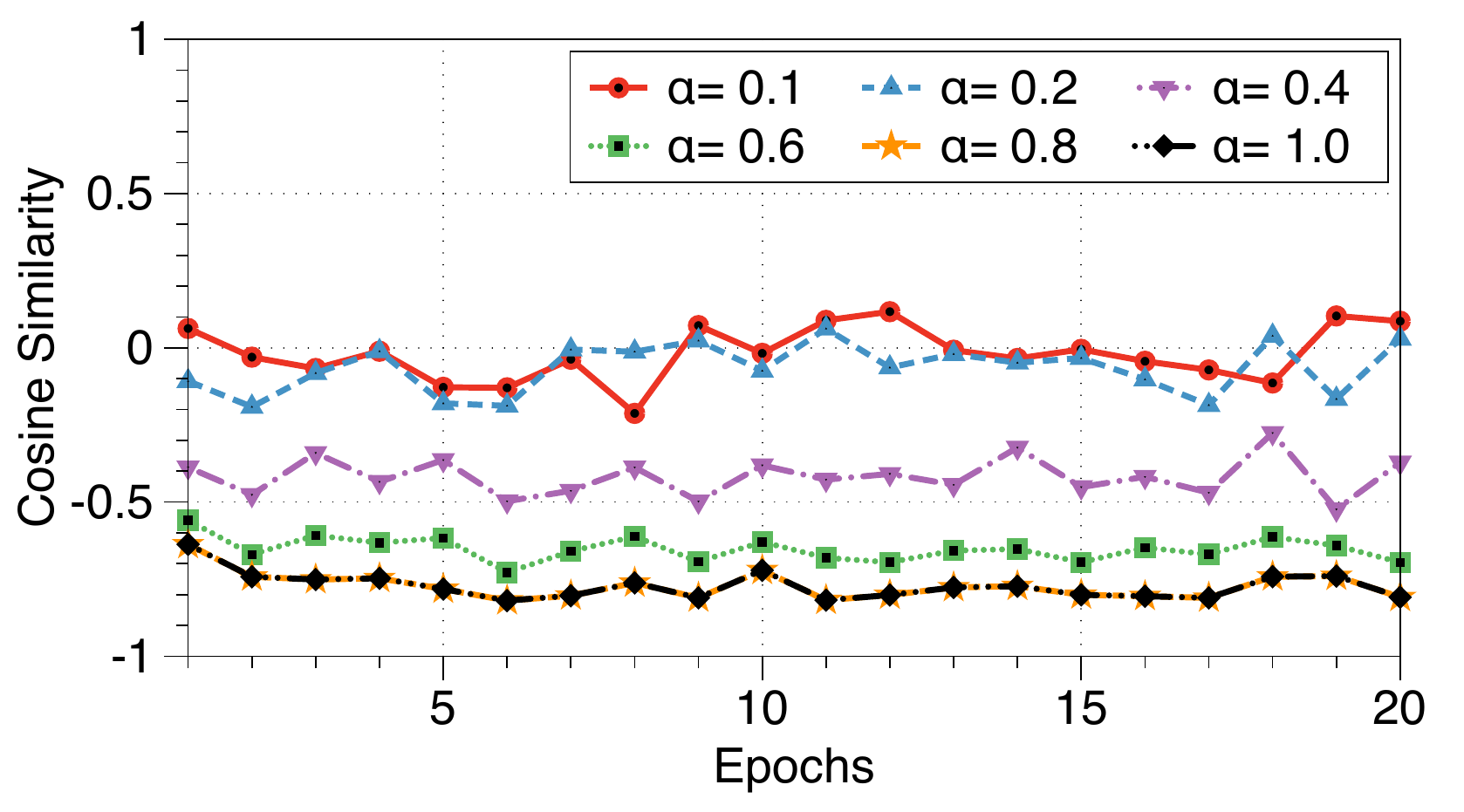}
	\end{subfigure}
	\begin{subfigure}[t]{0.32\textwidth}
		\centering 
		\includegraphics[width=\textwidth]{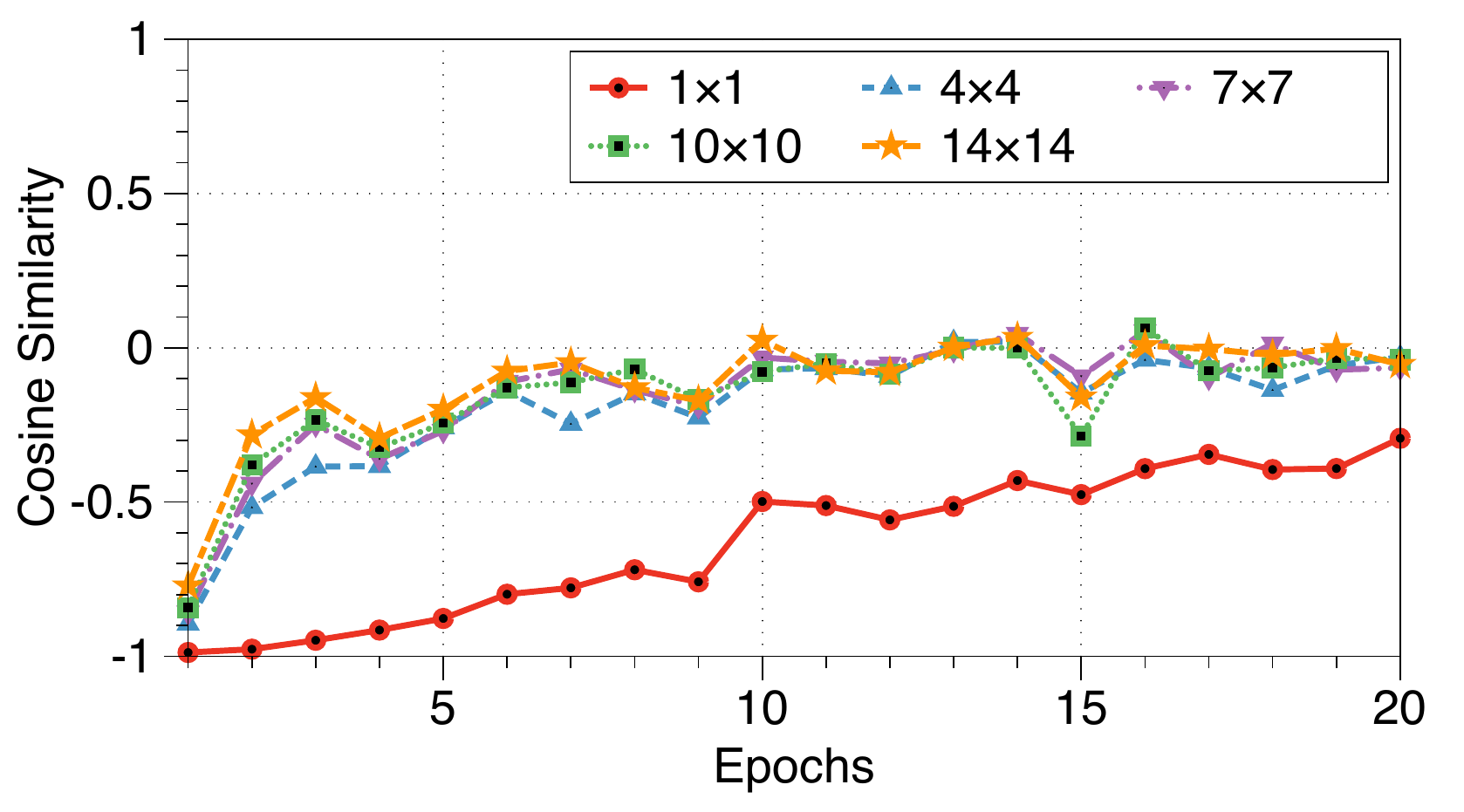}
	\end{subfigure}
	\caption{\textbf{Gradient Analysis Based on the Intensity of Poisoning Mechanisms.} We illustrate the magnitude ratios (the upper row) and the orientation differences (the lower row) during training in three scenarios: (1) an LR model is trained from scratch on multiple poisons that cause feature collisions; (2) we re-train an MLP model on multiple poisons causing feature collisions; and (3) the same MLP model is re-trained on 1\% of poisons that cause feature insertion. Each column corresponds to each scenario.}
	\label{fig:gradient-analysis}
\end{figure*}


\begin{figure}[h]
	\centering
	\begin{subfigure}[t]{0.11\linewidth}
		\centering
		\includegraphics[width=\textwidth]{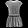}
		\caption*{\small $0.0$}
		\vspace{1.2em}
	\end{subfigure}
	\begin{subfigure}[t]{0.11\linewidth}
		\centering 
		\includegraphics[width=\textwidth]{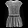}
		\caption*{\small $0.1$}
		\vspace{1.2em}
	\end{subfigure}
	\begin{subfigure}[t]{0.11\linewidth}
		\centering 
		\includegraphics[width=\textwidth]{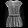}
		\caption*{\small $0.2$}
		\vspace{1.2em}
	\end{subfigure}
	\begin{subfigure}[t]{0.11\linewidth}
		\centering
		\includegraphics[width=\textwidth]{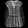}
		\caption*{\small $0.4$}
		\vspace{1.2em}
	\end{subfigure}
	\begin{subfigure}[t]{0.11\linewidth}
		\centering 
		\includegraphics[width=\textwidth]{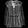}
		\caption*{\small $0.6$}
		\vspace{1.2em}
	\end{subfigure}
	\begin{subfigure}[t]{0.11\linewidth}
		\centering 
		\includegraphics[width=\textwidth]{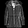}
		\caption*{\small $0.8$}
		\vspace{1.2em}
	\end{subfigure}
	\begin{subfigure}[t]{0.11\linewidth}
		\centering 
		\includegraphics[width=\textwidth]{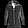}
		\caption*{\small $1.0$}
		\vspace{1.2em}
	\end{subfigure}

	\begin{subfigure}[t]{0.11\linewidth}
		\centering 
		\includegraphics[width=\textwidth]{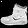}
		\caption*{\small $1\times1$}
		\vspace{0.3em}
	\end{subfigure}
	\begin{subfigure}[t]{0.11\linewidth}
		\centering 
		\includegraphics[width=\textwidth]{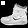}
		\caption*{\small $4\times4$}
		\vspace{0.3em}
	\end{subfigure}
	\begin{subfigure}[t]{0.11\linewidth}
		\centering
		\includegraphics[width=\textwidth]{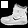}
		\caption*{\small $7\times7$}
		\vspace{0.3em}
	\end{subfigure}
	\begin{subfigure}[t]{0.11\linewidth}
		\centering 
		\includegraphics[width=\textwidth]{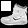}
		\caption*{\small $10\times10$}
		\vspace{0.3em}
	\end{subfigure}
	\begin{subfigure}[t]{0.11\linewidth}
		\centering 
		\includegraphics[width=\textwidth]{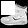}
		\caption*{\small $14\times14$}
		\vspace{0.3em}
	\end{subfigure}
	%
	\caption{\textbf{FashionMNIST Images Used in Our Gradient Analysis.} The upper row contains the interpolated samples with different interpolation ratios $\alpha$, and the bottom row shows the samples with different sizes of additional features.}
	\label{fig:mechanism-samples}
\end{figure}

In our analysis, we first craft a set of poisons that can cause the feature collision (or feature insertion) during training.
We then train a model on the poisoned training set and store the averaged gradients computed from the clean and poison samples---\ie the estimation of an individual gradient---over an epoch.
To compare the differences between the clean and poison gradients, we compute:
(1) the ratio of the magnitude (in $\ell_2$-norm) of the poison gradients to the clean gradients, and 
(2) the orientation differences by using the cosine similarity score.
All our analysis were carried out on FashionMNIST\footnote{More details on the datasets we used are described in \S\ref{subsec:evaluation-exp-setup}}.

\topic{Poisons:} We use the following techniques to craft poisons:

\subbullet{Feature Collision:}
To cause the feature collision by poisons, we use the watermarking technique.
We randomly choose the same number of samples from the two classes (\ie dress and coat) in the training set $\mathcal{D}_{tr}$ of FashionMNIST; for each pair of a dress and coat images, we overlay the coat image to the dress and label the resulting image still as a dress.
We control the intensity of feature collisions by modifying the interpolation ratio $\alpha$.
In Figure~\ref{fig:mechanism-samples}, we show the poison samples crafted with the different $\alpha$ in the upper row; the more we overlay the coat image to the dress image, the more dissimilar an interpolated image and its label (dress) become.

\subbullet{Feature Insertion:}
To exploit the feature insertion by poisons, we utilize backdooring mechanisms~\cite{BadNet19}.
We first randomly select 1\% of the training samples from any class and attach a small white-square in the top-left corner of each image.
We then assign the label coat for the patched images and add them to the training set.
Here, we control the intensity of feature insertion by increasing the patch size: $\{1\!\times\!1, 4\!\times\!4, 7\!\times\!7, 10\!\times\!10, 14\!\times\!14\}$ as shown in the bottom row of Figure~\ref{fig:mechanism-samples}.
As the patch size increases, the model is required to update its parameters more.


\begin{table*}[t]
\centering
\begin{threeparttable}
\adjustbox{width=\textwidth}{%
	\begin{tabular}{@{}c|c|cccc|cc|cc||c|cc@{}}
		\toprule
		\multirow{2}{*}{\textbf{Goal}} & \multirow{2}{*}{\textbf{Prior work}} & \multicolumn{4}{c|}{\textbf{Knowledge}} & \multicolumn{2}{c|}{\textbf{Capability}} & \multicolumn{2}{c||}{\textbf{Target}} & \multirow{2}{*}{\textbf{\begin{tabular}[c]{@{}c@{}}Poisoning\\Mechanism\end{tabular}}} & \multicolumn{2}{c}{\textbf{Gradient differences}} \\ \cmidrule(l){3-10} \cmidrule(l){12-13}
		&  & $D_{tr}$ & $\mathcal{X}$ & $\mathcal{Q}$ & $f_{\theta}$ & \textbf{\# Poisons} & \textbf{Injection} & \textbf{Task} & \textbf{Model} &  & \textbf{Magnitude} & \textbf{Orientation} \\ \midrule \midrule
		\multirow{7}{*}{\rotatebox[origin=c]{90}{\textbf{Indiscriminate}}} & Barreno~\etal~\cite{Barreno:ML10} & \tcircle{0} & \tcircle{0} & \tcircle{0} & \tcircle{0} & M. & Sc. & Bi. & Bayes & - & - & - \\
		& Biggio~\etal~\cite{Biggio:ICML12} & \tcircle{0} & \tcircle{0} & \tcircle{0} & \tcircle{0} & M. & Sc. & Bi. & SVM & Collision & High & High \\
		& Zhao~\etal~\cite{Zhao:IJCAI17} & \tcircle{0} & \tcircle{0} & \tcircle{0} & \rotatebox[origin=c]{90}{\tcircle{180}}  & M. & Sc. & Bi. / Mc. & SVM & Collision & High & High \\
		& Steinhardt~\etal~\cite{Steinhardt:NIPS17} & \tcircle{0} & \tcircle{0} & \tcircle{0} & \tcircle{0} & M. & Sc. & Bi. & SVM & Collision & High & High \\
		& Mu$\tilde{n}$oz-Gonz\'{a}lez~\etal~\cite{Gonzalez:CCS17} & \rotatebox[origin=c]{90}{\tcircle{180}} & \tcircle{0} & \tcircle{0} & \tcircle{0} & M. & Sc. & Bi. / Mc. & LR & Collision & High & High \\
		& Jagielski~\etal~\cite{Jagielski:Oakland18} & \rotatebox[origin=c]{90}{\tcircle{180}} & \tcircle{0} & \tcircle{0} & \tcircle{0} & M. & Sc. & Bi. & LR &  Collision & High & High \\
		& Ma~\etal~\cite{Ma:IJCAI19} & \tcircle{0} & \tcircle{0} & \tcircle{0} & \tcircle{0} & M. & Sc. & Bi. & Linear & Collision & High & High \\ \midrule
		\multirow{4}{*}{\rotatebox[origin=c]{90}{\textbf{Targeted}}} & Barreno~\etal~\cite{Barreno:ML10} & \tcircle{0} & \tcircle{0} & \tcircle{0} & \tcircle{0} & M. & Sc. & Bi. & Bayes & - & - & - \\
		& Suciu~\etal~\cite{Suciu:USENIX18} & \tcircle{360} & \tcircle{0} & \tcircle{0} & \tcircle{0} & M. & Re. / Sc. & Mc. & DNN & Collision \& Insertion & High & Unstable / High \\
		& Shafahi~\etal~\cite{Ali:NIPS18} & \tcircle{360} & \tcircle{0} & \tcircle{0} & \tcircle{0} & S. / M. & Re. / Sc. & Mc. & DNN & Collision \& Insertion & High & Unstable / High \\
		& Zhu~\etal~\cite{Zhu:ICML19} & \tcircle{360} & \tcircle{0} & \tcircle{360} & \tcircle{360} & S. / M. & Re. / Sc. & Mc. & DNN & Collision \& Insertion & High & Unstable / High \\ \bottomrule
	\end{tabular}
}
{
	\begin{tablenotes}[para]
		\item \small \textbf{Knowledge:} \tcircle{360} = Black-box, \tcircle{0} = White-box, \rotatebox[origin=c]{90}{\tcircle{180}} = Both 
			\hspace{1.4em} \textbf{\# Poisons:} S. = Single-Poison, M. = Multi-Poison
	\end{tablenotes}
}
{
	\begin{tablenotes}[para]
		\item \small \textbf{Injection:} Sc. = From Scratch, Re. = From Re-training
			\hspace{1.8em} \textbf{Target Task:} Bi. = Binary classification, Mc. = Multi-class classification
	\end{tablenotes}
}
\end{threeparttable}
\caption{\textbf{Unifying Data Poisoning Attacks.} We listed indiscriminate attacks at the top half while targeted attacks are at the bottom half. From the left, the columns 3-6 contains the attacker's knowledge considered by authors, and we specify the capability of an attacker in the columns 7-8. We include the target tasks and models used in the columns 9-10. In the columns 11-13, we identify the poisoning mechanism used in each work and the shared differences expected to be seen in the gradients.}
\label{tbl:taxonomy}
\end{table*}

\subsection{Gradient Analysis: Feature Collision}
\label{subsec:collision-gradients}

Here, we focus on two scenarios where (1) a model is trained from scratch on the training set containing multiple poisons, and (2) we update a trained model on the poisoned training set.
They represent common indiscriminate and targeted poisoning scenarios, respectively.
In the first scenario, we train a linear regression (LR) model on a subset of the FashionMNIST dataset.
We use the Adam~\cite{kingma2014adam} optimizer and train the model for 40 epochs with the batch size 300 and the learning rate 0.01.
The subset is a binary dataset comprising of samples from the dress and coat classes; the training set includes 10,800 samples (5,400 from each class), and the testing set contains 2000 samples.
We construct the poisoned training set by adding 100 interpolated dress samples.
In the second scenario, we first train a multi-layer perceptron (MLP) with two hidden layers on the entire FashionMNIST dataset.
We then re-train the model on the poisoned training set containing the same 100 poisons.
We use the SGD optimizer and re-train the model for 20 epochs with the batch size 100 and the learning rate 0.04.

The first and second columns in Figure~\ref{fig:gradient-analysis} illustrate the results from our feature collision analysis.
Overall, we observe that the magnitude ratio between the gradients from individual poison and clean samples is high during training in both the scenarios---\ie the magnitude of poison gradients are larger than that of clean gradients.
We also found the ratio becomes larger when we re-train a model with the same poisons; in this case, the model already has learned about the clean samples so the gradients computed from the clean data are smaller than the poison gradients.
In terms of the orientation differences, we can see the cosine similarity scores are less than zero in both cases.
This indicates the features are colliding---\ie the information from the poison gradients is in contrast to that from the clean ones.
Moreover, we identified that, as the intensity of the feature collision increases, the magnitude ratio and orientation difference become more stable and visible.
For instance, in the first scenario, the case of $\alpha\!=\!1.0$ shows fewer oscillations of the magnitude ratio and cosine similarity score than the $\alpha\!=\!0.1$ case.
In the re-training scenario, when $\alpha$ approaches one, the ratio increases, and the cosine similarity gets closer to minus one.

\subsection{Gradient Analysis: Feature Insertion}
\label{subsec:insertion-gradients}
In targeted poisoning, an attacker can also exploit the feature insertion to cause local misclassifications on a few target samples.
Since the feature insertion is useful against the re-training of a high-capacity model such as a neural network, we consider the scenario where the victim updates an MLP model on the poisoned training set.
We take the MLP model in \S\ref{subsec:collision-gradients}, trained on the entire FashionMNIST, and construct the poisoned training set by adding 1\% of patched samples to the original (clean) training set.
We then re-train the model on the poisoned training set for 20 epochs by using the SGD optimizer with the batch size 100 and the learning rate 0.01.

The last column in Figure~\ref{fig:gradient-analysis} shows the results from our feature insertion analysis.
In the earlier epochs, we found similar patterns to the results in \S\ref{subsec:collision-gradients}, there is a significant difference in the magnitude of the gradients computed from the poisons and clean samples.
However, we can also see the differences that we observed in the earlier epochs are reduced during training and, at the end of the re-training, both the magnitude ratio and cosine similarity become closer to zero.
This implies that the model learns the new features coming from our poisons during re-training with minimal collisions with the existing features; thus, the attacker can cause misclassifications of the test-time samples including the new features without hurting the model's original behaviors.



\section{Unifying Data Poisoning Attacks}
\label{sec:taxonomy}

In this section, we unify the attack surface exploited by data poisoning attacks based on our analysis of their impact on the magnitude ratio and orientation differences between the clean and poison gradients.
We start with an overview of the existing data poisoning attacks based on the taxonomy structured around the knowledge and capability of an attacker~\cite{Suciu:USENIX18}.
We then focus on poisoning mechanisms that the indiscriminate or targeted attacks use and identify shared features that can be observed in the magnitude ratio and orientation differences.
Building on this intuition, we lay out essential properties for a generic defense against poisoning attacks.


\subsection{Indiscriminate Poisoning Attacks}
\label{subsec:taxonomy-indiscriminate}

\topic{Overview of Existing Attacks:}
In Table~\ref{tbl:taxonomy}, we listed the indiscriminate attacks at the top half.
Most (not all) work~\cite{Barreno:ML10, Biggio:ICML12, Zhao:IJCAI17, Steinhardt:NIPS17, Gonzalez:CCS17, Jagielski:Oakland18, Ma:IJCAI19} has focused on attacking simple models such as linear regression (LR) models or support vector machines (SVMs) in the binary classification setting (the columns 9-10).
Mu$\tilde{n}$oz-Gonz\'{a}lez~\etal~\cite{Gonzalez:CCS17} performed attacks on neural networks; however, they turned out to be ineffective as the accuracy drop caused was less than 1\%.
In the columns 3-6, we can see the indiscriminate attacks that consider the white-box attacker who knows the training data $\mathcal{D}_{tr}$, the features $\mathcal{X}$, the model $f_{\theta}$, and the training algorithm $\mathcal{Q}$.
Using this knowledge, they craft poisons that can maximize the test-time loss (or the training loss as a proxy for the test-time loss~\cite{Steinhardt:NIPS17}).
This crafting process is commonly formulated as an optimization problem, the attacker first chooses several samples from $\mathcal{D}_{tr}$ and finds perturbations to their features $\mathcal{X}$.
The solution is obtained through gradient descent methods by exploiting the knowledge about the target model $f_{\theta}$ and training algorithm $\mathcal{Q}$.
Some scenarios consider limited knowledge of an attacker about the training data or the target model~\cite{Zhao:IJCAI17, Gonzalez:CCS17, Jagielski:Oakland18}; however, the attacker can exploit transferability to compute poisons by using auxiliary training data or surrogate models.

\topic{Estimated Impact on Gradients:}
To maximize the test-time loss during training, the attacker needs to craft poisons that cause the largest disturbance to the gradients computed from the clean data.
Hence, the indiscriminate attackers exploit the feature collision (see column 11 in Table~\ref{tbl:taxonomy}).
As the main purpose of the indiscriminate attack is to cause a significant accuracy drop, the attacker increases the intensity of feature collision by blending multiple poisons into the training set.
This scenario is similar to our analysis of feature collision in \S\ref{subsec:collision-gradients} when we train an LR model on the binary training set containing 100 poisons.
Thus, we expect to observe a similar magnitude ratio and orientation difference between the poison and clean gradients across the indiscriminate poisoning attacks---\ie cosine similarity becomes -1, and the magnitude ratio is high.
We specify this intuition in the columns 12-13.

\subsection{Targeted Poisoning Attacks}
\label{subsec:taxonomy-discriminate}

\topic{Overview of Existing Work:}
%
The bottom half of Table~\ref{tbl:taxonomy} shows the targeted attacks.
Most work~\cite{Suciu:USENIX18, Ali:NIPS18, Zhu:ICML19}, except~\cite{Barreno:ML10}, performed targeted attacks on large capacity models such as deep neural networks (DNNs). 
In the columns 3-6, we show that initial work on targeted poisoning~\cite{Barreno:ML10} considers the white-box attacker who knows $\mathcal{D}_{tr}$, $\mathcal{X}$, $f_{\theta}$, and $\mathcal{Q}$; however, the following work~\cite{Suciu:USENIX18, Ali:NIPS18} identified that an adversary can perform effective attacks without the knowledge of the training set $\mathcal{D}_{tr}$.
The most recent work~\cite{Zhu:ICML19} demonstrates the successful attacks without knowing the target model $f_{\theta}$ and its training algorithm $\mathcal{Q}$ by exploiting the transferability of poisons across different DNNs.
In targeted poisoning, the attacker crafts poisons using the test samples in the target class.
For example, if an attacker wants to misclassify a small subset of dog images (targets) into fish, the attacker first picks a few test samples in the fish class.
In the creation process, the attacker minimizes the distance between the poisons and targets in the internal representation space of a model and the perturbations in the input features $\mathcal{X}$.
This makes targeted attacks inconspicuous---\ie the poisons are perceptually indistinguishable to a human, but they affect the model's decision locally without a significant accuracy drop.

\topic{Estimated Impact on Gradients:}
To minimize the accuracy drop caused by poisoning samples, a targeted attacker exploits both the feature collision and insertion (see column 11 in Table~\ref{tbl:taxonomy}):
(1) To cause the local misclassifications, the attacker can cause collisions with the features important for the target classifications, but the model does not rely on them for the entire classification.
(2) On the other hand, the attacker inserts new features to the target model and increases their importance on the classification of targets in training.
Considering that the targeted attacks were more successful when the attacker blends poisons during the re-training of a victim model, we expect to observe the magnitude ratio and orientation difference showed in \S\ref{subsec:insertion-gradients}---\ie the ratio becomes high, and the orientation differences 
become unstable or slightly less than -1.
However, as shown in Figure~\ref{fig:gradient-analysis}, cosine similarity can decrease and become 0 when the attacker reduces the intensity---the number of poisons---of both the mechanisms.


\section{Mitigate Poisoning with Gradient Shaping}
\label{sec:defense}

The unified view of data poisoning attacks in \S\ref{sec:taxonomy} outlines a series of requirements for effective defenses.
Here, we introduce \emph{gradient shaping}, a property for anti-poison defenses, that satisfies these requirements.
Using this property, we taxonomize the existing defenses against data poisoning attacks and discuss the effectiveness and limits of those mechanisms.
We then instantiate a defense by leveraging differentially private (DP) optimizers and demonstrate its effectiveness in \S\ref{sec:evaluation}.

\subsection{Gradient Shaping}
\label{subsec:gradient-shaping}

Our unified view of data poisoning provides two key insights.
During training, (1) the $\ell_2$-norm of an individual gradient from poison is larger than the norm of a clean gradient on average, and
(2) there is an orientation difference between the poison and clean gradients.
While the extent of both the observations varies depending on the intensity of an attack, they are common across data poisoning attacks.
Given that poison gradients update the model at each batch, the total change in model parameters is exacerbated over training epochs.

\begin{figure}[t]
	\centering
	\includegraphics[width=\linewidth]{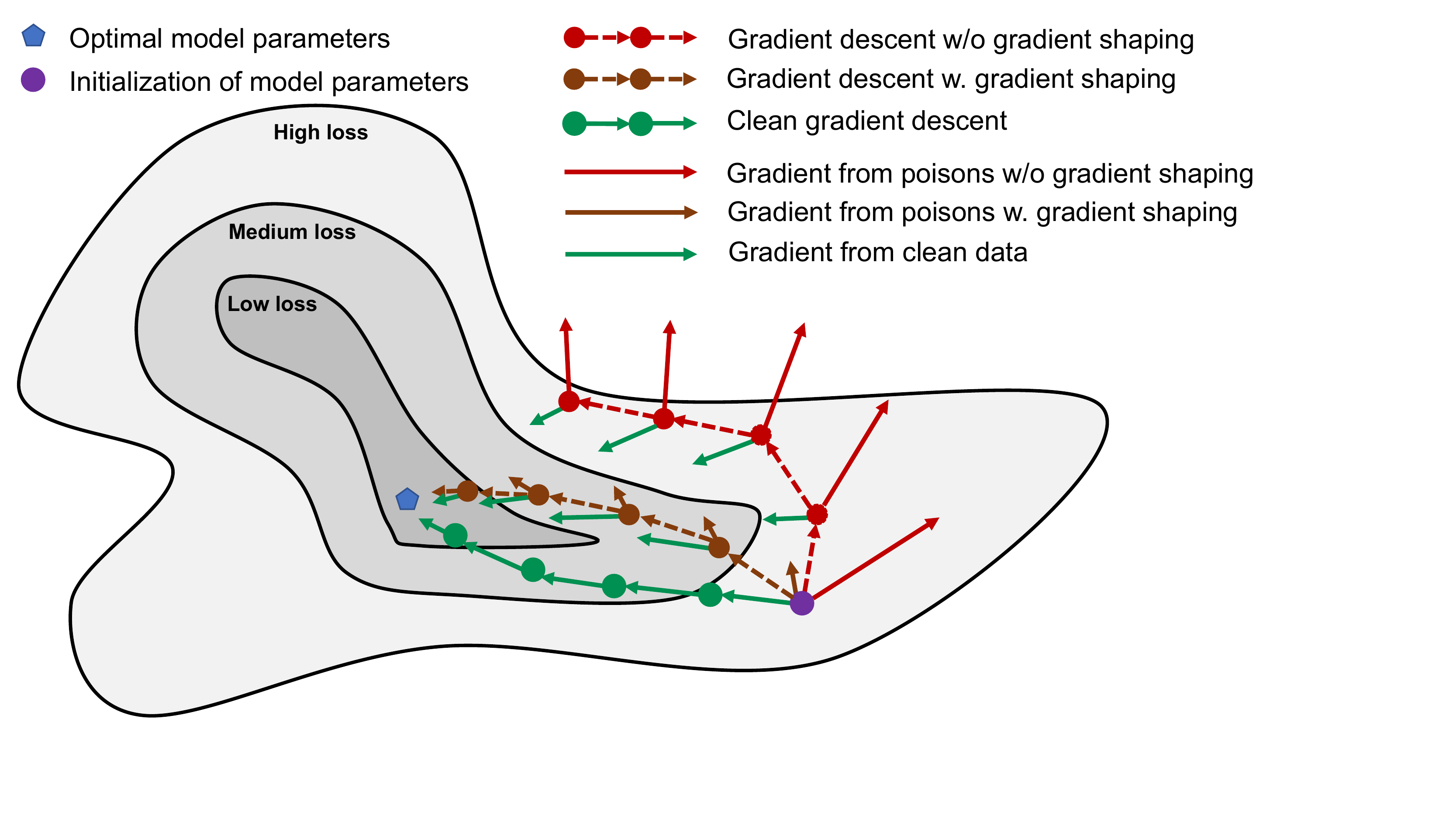}
	\caption{\textbf{Illustration of How Gradient Shaping Mitigates Poisoning.} This is a contour line visualization of the model's loss for different values of its parameters; each arrow corresponds to the gradients computed from poisons or clean samples, each dashed arrow indicates the step taken when computing a single model update, and each dot is a set of parameters obtained after one or more steps of gradient descent from the random initialization. This abstract representation helps understand how the property mitigates data poisoning.}
	\label{fig:shaping-illustration}
\end{figure}

\topic{Gradient Shaping:}
By controlling both the magnitude ratio and orientation difference of the gradients, we could intuitively safeguard ML models against data poisoning attacks.
We refer to this property as \emph{gradient shaping}.
Any defense mechanism is used to minimize the differences in gradients before the model uses them to update its parameters.
In Figure~\ref{fig:shaping-illustration}, we illustrate how this property reduces the differences in gradients during the training of a model with poisons 
in the context of indiscriminate attacks.
When we train a model on the poisoned training set, we compute clean gradients (the {\color{moss} green} lines in the figure) and the poison gradients (the {\color{red} red} lines in the figure) at each iteration.
We sum the two gradients and update the model parameters.

If the data does not include poisons, the training algorithms updates the model parameters following the trajectory shown in the {\color{moss} green} dashed-lines---\ie it updates the parameters in the direction of reducing loss.
However, when the data contains poisons, the updates at each iteration are affected by the poison gradients (see the {\color{red} red} dashed-lines). Thus, the model parameters land on the surface with high loss, which leads to the accuracy drop of the trained model.
%
If gradient shaping is used, the magnitude and orientation differences between the poison and clean gradients are reduced (see the {\color{brown} brown} lines).
In this case, the updates of the model parameters deviate less, following the trajectory illustrated in the {\color{brown} brown} dashed-lines, and 
land on the surface with low loss.


\subsection{Existing Poisoning Defenses}
\label{subsec:existing-defenses}

Table~\ref{tbl:taxonomy_defense} outlines defenses against poisoning.
Prior work~\cite{Barreno:ML10, Steinhardt:NIPS17, Jagielski:Oakland18, Diakonikolas:ICML19, Suciu:USENIX18}  focused primarily on outlier removal (also known as data sanitization). 
In outlier removal, the defender considers outliers as poisons and removes them from the training data, which meets the requirements of gradient shaping as it clears away the poison gradients by removing a set of poisons.

However, outlier removal is brittle: they identify poisons based on analysis of nearest neighbors, training loss, and dimensionality reduction techniques---all of which are dependent on the training data $\mathcal{D}_{tr}$ and/or the model $f_{\theta}$ and its parameters $\theta$.
Worse so, when the attack uses inconspicuous poisons, \eg targeted poisoning, it is difficult to detect them by the sanitization techniques~\cite{athalye2018obfuscated}; as witnessed in the abundant work in evasion attacks~\cite{goodfellow2014explaining}, malicious samples with similar input representations often produce distinctly different gradients.
An extended version of the RONI defense called tRONI~\cite{Suciu:USENIX18} examines the misclassification of targets during training; however, tRONI assumes a defender knows the targets that an attacker wishes to misclassify.
Moreover, outlier removal scales poorly as it increases the computational overheads during training.
They require an iterative analysis of the training samples~\cite{Barreno:ML10, Steinhardt:NIPS17, Jagielski:Oakland18, Suciu:USENIX18} or rely on robust optimizations~\cite{Diakonikolas:ICML19} using higher-order derivatives. Approaches from robust optimization could indeed apply here, but they were found by Jagielski \etal to perform poorly in the presence of adversary-induced poisoning~\cite{Jagielski:Oakland18}.


\begin{table}[t]
	\centering
	\begin{threeparttable}
		\adjustbox{width=0.80\linewidth}{%
			\begin{tabular}{@{}l|c|c|ccc|c@{}}
				\multirow{1}{*}[-0.45in]{\textbf{Name}} & 
					\multirow{1}{*}[0.5em]{\rotate{Mechanism}} & \rotate{Considered Attack} & \multirow{1}{*}[-0.2em]{\rotate{Know $\mathcal{D}_{tr}$}} & \multirow{1}{*}[0.6em]{\rotate{Know $\mathcal{Q}, f_{\theta}$}} & \multirow{1}{*}[1.3em]{\rotate{Know Targets}} & \multirow{1}{*}[2.6em]{\rotate{Comp. Overhead}} \\ \midrule \midrule
				RONI~\cite{Barreno:ML10} & OR & I & \cmark & \xmark & \xmark & \cmark \\
				Certified Defense~\cite{Steinhardt:NIPS17} & OR & I & \cmark & \cmark & \xmark & \cmark \\
				TRIM~\cite{Jagielski:Oakland18} & OR & I & \cmark & \cmark & \xmark & \cmark \\
				SEVER~\cite{Diakonikolas:ICML19} & OR & I & \cmark & \cmark & \xmark & \cmark \\
				tRONI~\cite{Suciu:USENIX18} & OR & T & \cmark & \xmark & \cmark & \cmark \\ \bottomrule
				\textbf{Ours} & \textbf{DP} & \textbf{I \& T} & \xmark & \xmark & \xmark & \xmark \\ \bottomrule
			\end{tabular}
		}
		{
			\begin{tablenotes}[para]
				\item \small \hspace{1.6em} \textbf{Attacks:} I = Indiscriminate, T = Targeted
			\end{tablenotes}
		}
		{
			\begin{tablenotes}[para]
				\item \small \hspace{1.6em} \textbf{Mechanism:} OR = Outlier Removal, DP = DP Optimizers
			\end{tablenotes}
		}
	\end{threeparttable}
	\caption{\textbf{An Overview of Existing Poisoning Defenses\protect\footnotemark.}}
	\label{tbl:taxonomy_defense}
\end{table}

\footnotetext{Ma \etal~\cite{Ma:IJCAI19} considers differential privacy~\cite{Dwork:Book11} to provide theoretical bounds on the number of poisons that a model can resist during training. However, we exclude this work from our taxonomy as the bound was evaluated on synthetic attacks, not the poisoning attacks in recent literature.}

\subsection{Generic Defense: DP Optimizers}
\label{subsec:defense-proposal}

Recall that the requirements for an effective anti-poison defense include (1) controlling the norm of the poison gradient, and (2) restricting differences in the orientation between the poison and clean gradients.
Similar requirements are needed to ensure that learning algorithms train a model privately. 
We adopt the differential privacy framework, which can be thought of as requiring that the model updates are not influenced overly by any of the individual examples contained in the training data.
Abadi \etal~\cite{Abadi:CCS16} propose a differentially private mechanism for off-the-shelf optimizers, \eg SGD (henceforth referred to as \emph{DP optimizers}).
By choosing a pre-defined clipping norm, they bound the influence of an individual gradient to a model.
Also, they proceed to make the gradients indistinguishable by adding Gaussian noise (see Algorithm~\ref{algo:sgd-and-dp} in Appendix~\ref{appendix:dp-sgd}). 
%
Thus, we utilizes DP optimizers to meet the requirements for an effective anti-poison defense.
%
The key advantage of using DP optimizers is that they are generic mechanisms agnostic to the dataset and model used by a defender, and the techniques used to craft poisons (see Table~\ref{tbl:taxonomy_defense}).
In subsequent sections, we validate these points.

\topic{Out-of-Scope:}
%
DP optimizers are designed to control the privacy leakage ($\varepsilon$), often at a significant cost in model utility~\cite{USENIX19:EvaluateDP}.
%
%
Instead, we focus on how certain parameter configurations of DP optimizers can defend against poisoning with minimal utility loss, regardless of the privacy provided.


\section{Evaluation}
\label{sec:evaluation}

In this section, we evaluate the effectiveness of training a model with DP optimizers as a defense against data poisoning attacks.
We start with an overview of our experimental setup (\S\ref{subsec:evaluation-exp-setup}). 
We then quantify the resilience of training a model with DP optimizers against indiscriminate (\S\ref{subsec:resilience-indiscriminate}) and targeted poisoning attacks (\S\ref{subsec:resilience-targeted} and \S\ref{subsec:targeted-resilience-multi}).
In experiments, we individually vary the two parameters
---the clipping norm $C$ and the noise multiplier $\sigma$---for each attack, to analyze the impact of parameter configurations on the resilience.
To understand the impact of a specific parameter choice, we compare the magnitude ratio and orientation difference between the poison and clean gradients observed in the training without DP optimizers and those seen when we use them.
%
Moreover, we turn our attention to distinct defense scenarios where the resilience of a model itself is necessary (\S\ref{subsec:resilience-dp-trained-model}).

\subsection{Experimental Setup}
\label{subsec:evaluation-exp-setup}

Our analysis framework quantifies the effectiveness of using DP optimizers against data poisoning attacks.
Given an attack, the framework crafts poisons, trains multiple models using the specified training algorithms (SGD/Adam or DP-SGD/DP-Adam) with the poisoned training set, and reports the metrics that we define.
%
We build this framework using Python 3.73 and TensorFlow 1.14.0.
To train a model with DP optimizers, we use the open-source library, TensorFlow-Privacy~\cite{TF-Privacy}. 
%


\begin{figure*}[t]
	\centering
	\begin{subfigure}[t]{0.32\textwidth}
		\centering
		\includegraphics[width=\textwidth]{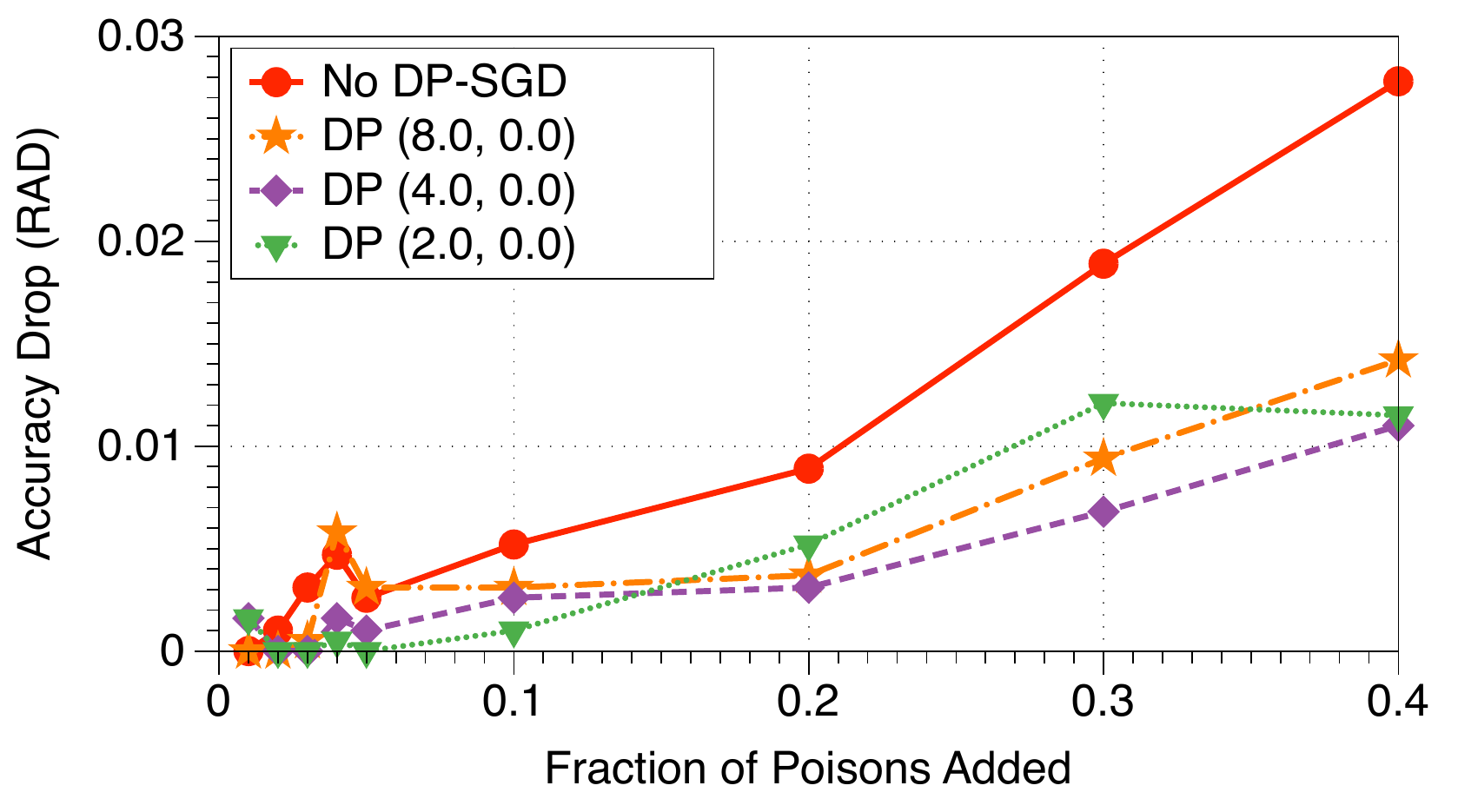}
		\caption{\textbf{LF: Impact of the Clipping Norm}}
		\vspace{1.0em}
		\label{fig:indiscriminate-resilience-a}
	\end{subfigure}
	\begin{subfigure}[t]{0.32\textwidth}   
		\centering 
		\includegraphics[width=\textwidth]{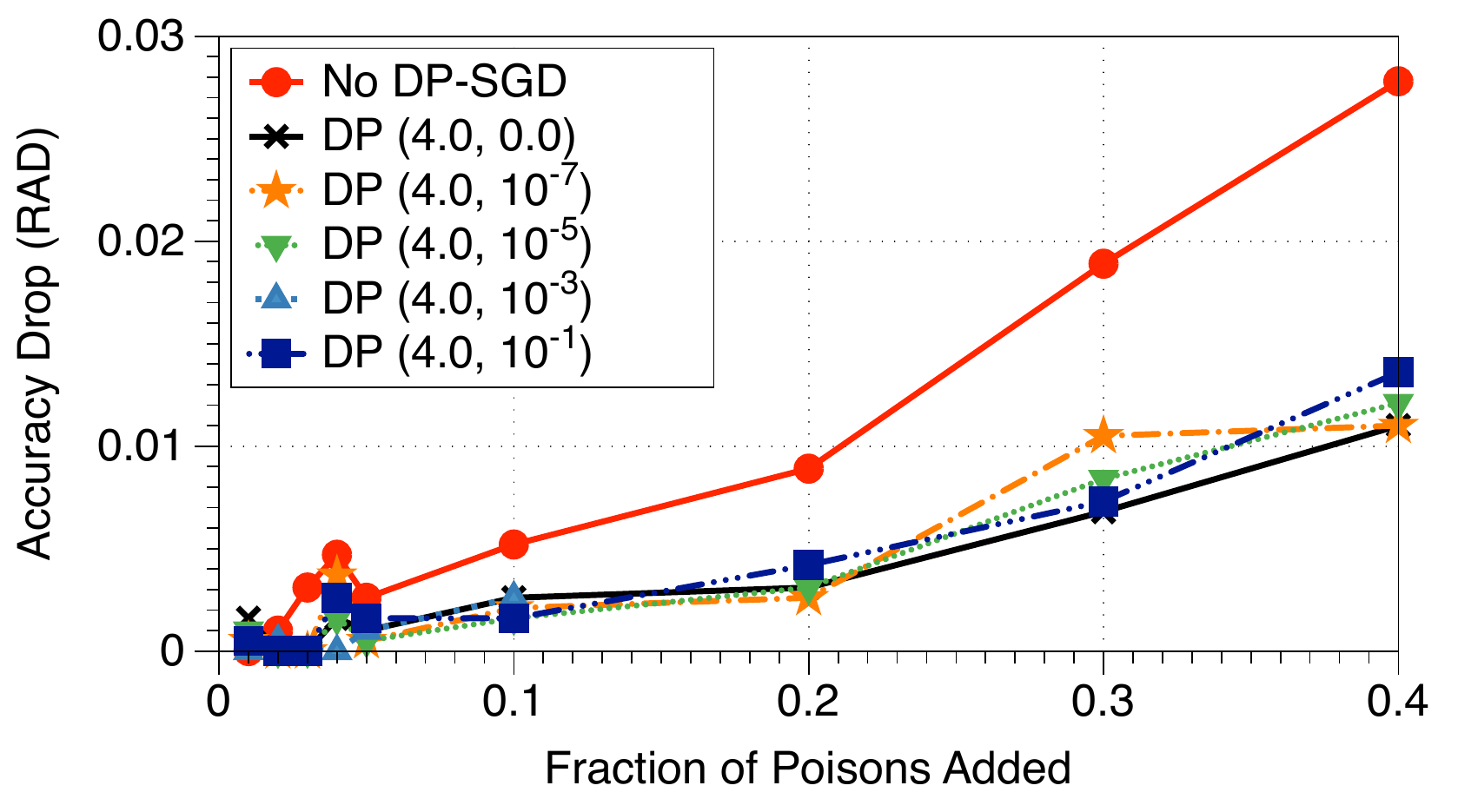}
		\caption{\textbf{LF: Impact of the Noise Multiplier}}
		\vspace{1.0em}
		\label{fig:indiscriminate-resilience-b}
	\end{subfigure}
	\begin{subfigure}[t]{0.33\textwidth}
		\centering 
		\includegraphics[width=\textwidth]{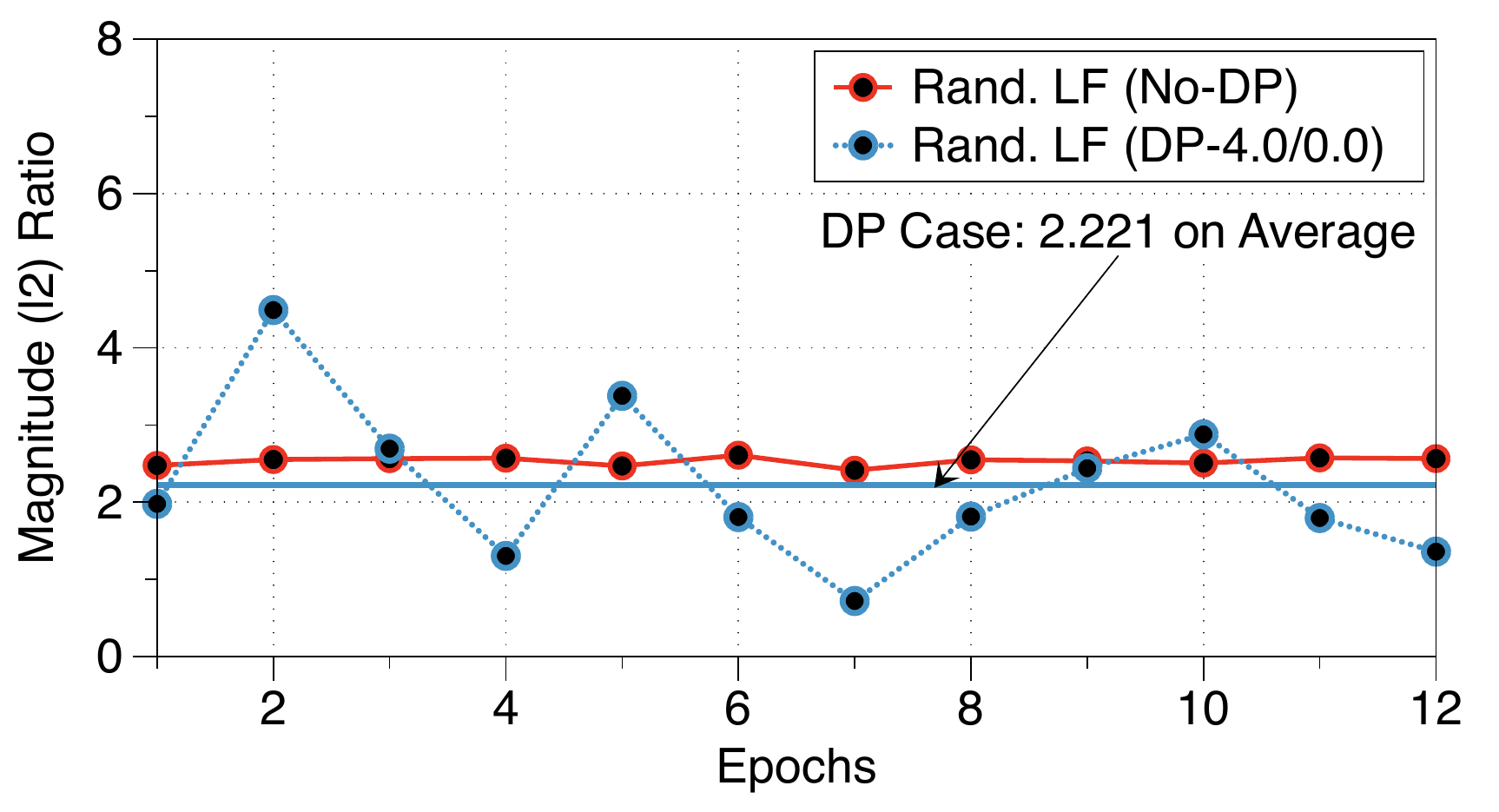}
		\caption{\textbf{LF: Magnitude Difference}}
		\vspace{1.0em}
		\label{fig:indiscriminate-resilience-c}
	\end{subfigure}
	\begin{subfigure}[t]{0.32\textwidth}
		\centering
		\includegraphics[width=\textwidth]{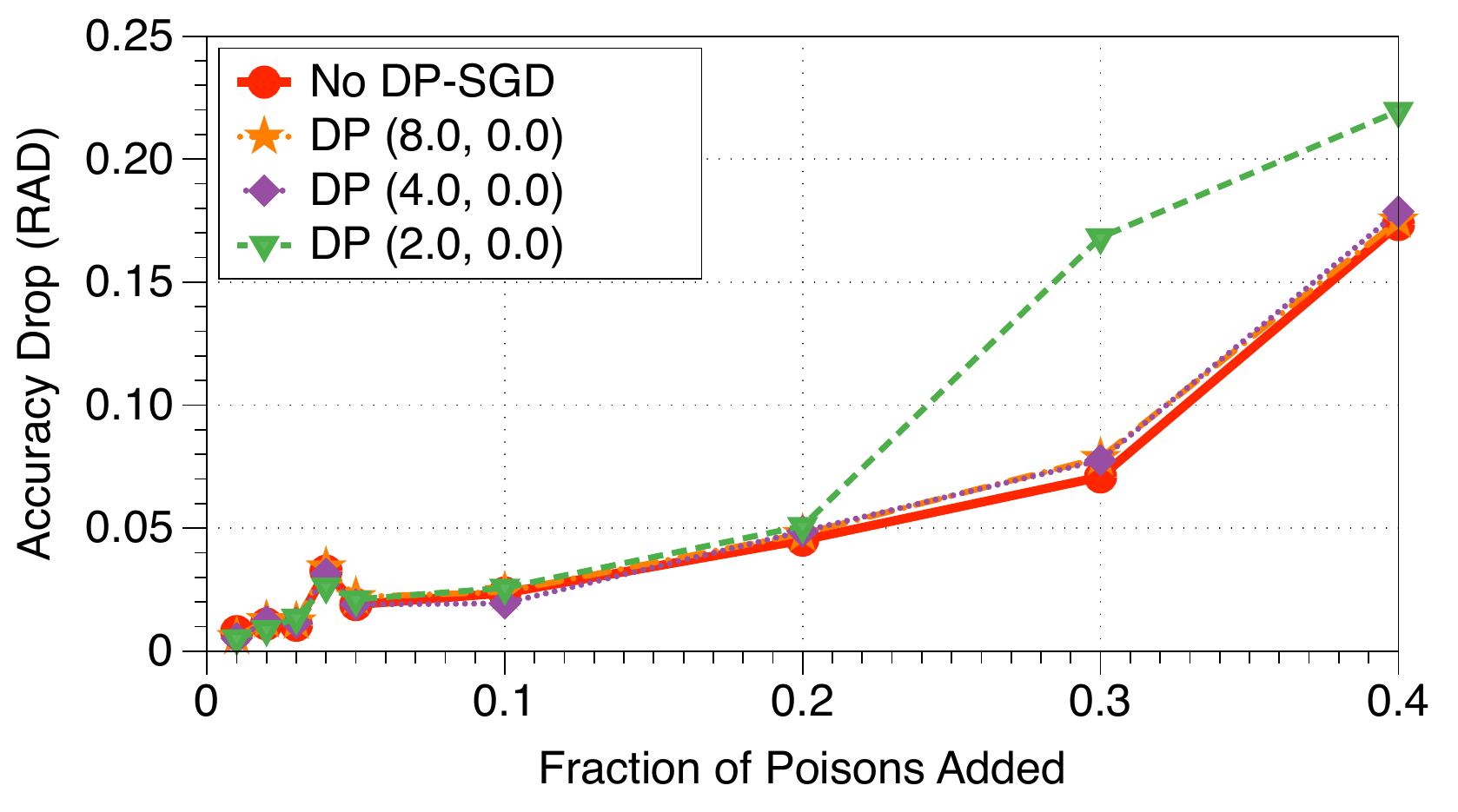}
		\caption{\textbf{SOTA: Impact of the Clipping Norm}}
		\vspace{0.6em}
		\label{fig:indiscriminate-resilience-d}
	\end{subfigure}
	\begin{subfigure}[t]{0.32\textwidth}   
		\centering 
		\includegraphics[width=\textwidth]{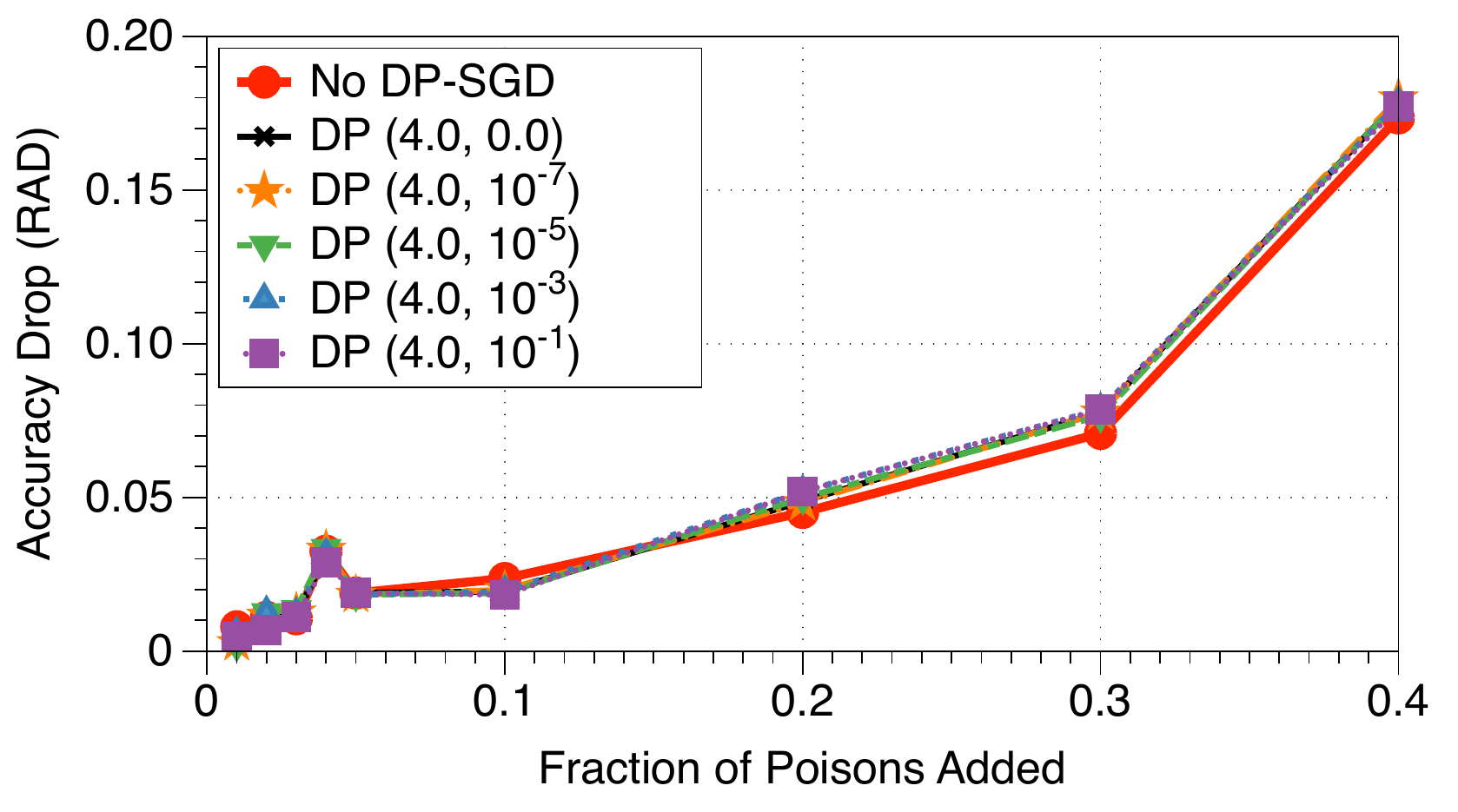}
		\caption{\textbf{SOTA: Impact of the Noise Multiplier}}
		\vspace{0.6em}
		\label{fig:indiscriminate-resilience-e}
	\end{subfigure}
	\begin{subfigure}[t]{0.33\textwidth}
		\centering 
		\includegraphics[width=\textwidth]{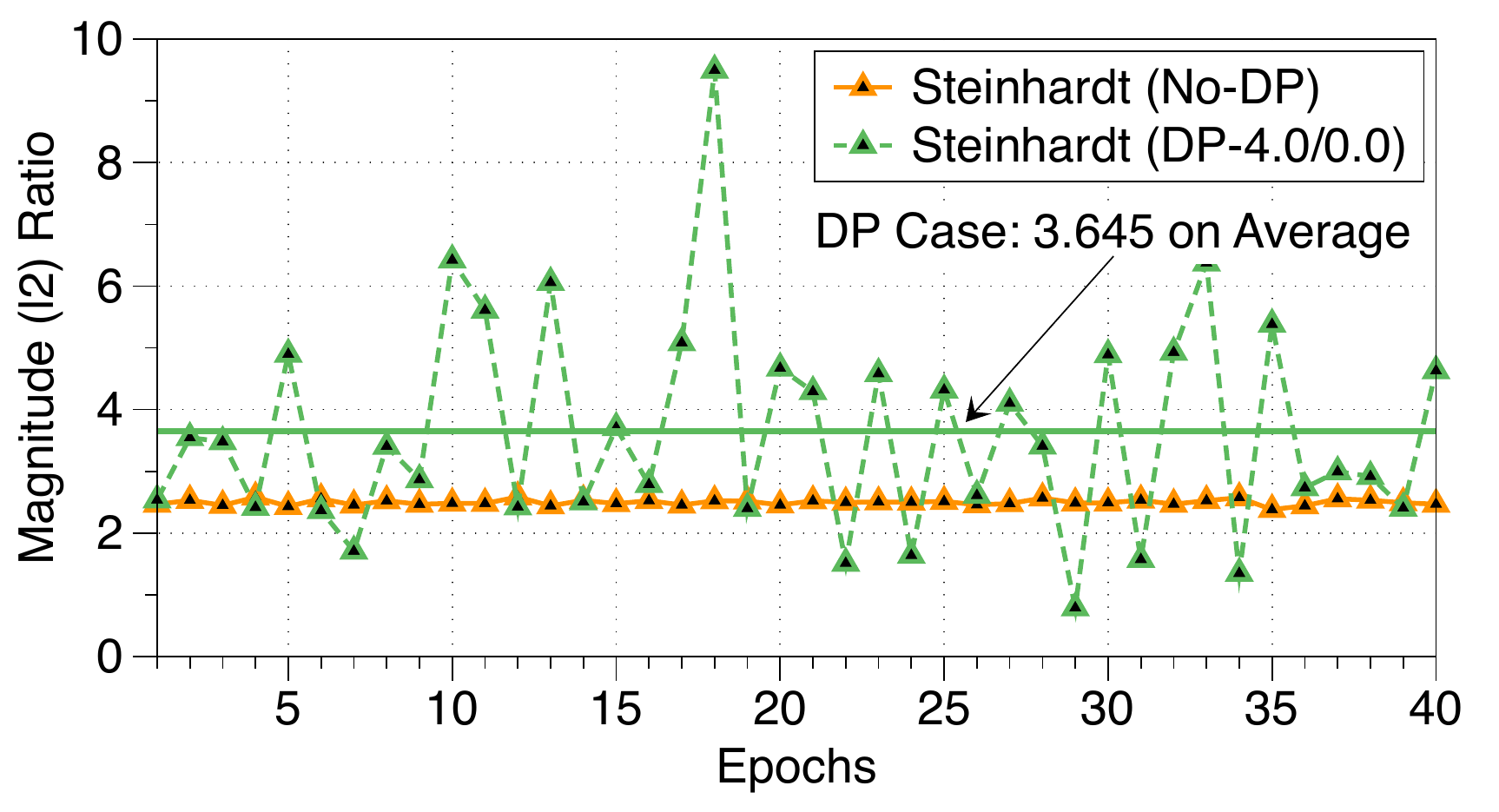}
		\caption{\textbf{SOTA: Magnitude Difference}}
		\vspace{0.6em}
		\label{fig:indiscriminate-resilience-f}
	\end{subfigure}
	\caption{\textbf{Effectiveness of Training Models with DP Optimizers against Indiscriminate Poisoning Attacks.} We illustrate the RADs of the vanilla- and DP-models in the random LF attacks (upper) and the SOTA attacks (lower). We show that training with DP-Adam is effective in mitigating the random LF attacks whereas DP-Adam cannot defeat the SOTA attacks. In the right column, we also show the magnitude difference as a ratio between poison and clean gradients in both the attacks.}
	\label{fig:indiscriminate-resilience}
\end{figure*}

\topic{Datasets:}
We conduct our analysis with three datasets: Purchase-100~\cite{Purchases}, 
FashionMNIST~\cite{FashionMNIST:Xiao17}, and CIFAR-10~\cite{CIFAR10:krizhevsky09}.
Purchase-100 consists of 200k customer purchase records of size 100 each (corresponding to the 100 frequently purchased items), and the records are categorized into 100 classes based on the customers' purchase style.
Here, we use 10k randomly-chosen records for training and 10k randomly-selected non-training samples for the test set~\cite{USENIX19:EvaluateDP}.
FashionMNIST is composed of 28x28 grayscale images of 70k fashion products from 10 categories, with 7k images per class, which contains 60k training and 10k testing samples.
CIFAR-10 includes 32x32 pixels, colored natural images of 10 classes, containing 50k training and 10k testing samples.

\topic{Models:} We consider a logistic regression (LR), a multi-layer perceptron (MLP), and a convolutional neural network (CNN).
%
%
We include the network configurations that we used in Appendix~\ref{appendix:network-arch}.
For Purchase-100, we use LR and MLP models; however, for the FashionMNIST and CIFAR10, we use 
MLP and CNN models because the LR models have poor accuracy ($\lt$ 50\%) on the test set.
In all figures and discussion of results, we add the prefix vanilla- and DP- to denote models trained with SGD and DP-SGD respectively.

\topic{Metrics:} Since the indiscriminate attacker aims to cause significant accuracy drop of a model over the test set, we utilize the relative accuracy drop (RAD) 
to measure the attacker's success.
RAD is the accuracy drop of a model caused by poisoning over the accuracy of the clean model---the larger the RAD, more effective the attack. 
For targeted poisoning, we consider an attack to be successful when the target becomes misclassified at any epoch during re-training without causing significant accuracy degradation.
Specifically, successful attacks are those where RAD $\lt$ 0.05, the same threshold used by Suciu \etal~\cite{Suciu:USENIX18}.
Moreover, we measure the attack intensity as the number of poisons added to the clean training set.
For indiscriminate attacks, we denote the intensity as a ratio of the number of poisons to the number of clean samples. 
In targeted attacks, the intensity is the 
number of poisons.


\subsection{Mitigating Indiscriminate Poisoning}
\label{subsec:resilience-indiscriminate}

\topic{Experimental Methodology:}
Indiscriminate poisoning is known to be effective against binary classification tasks that utilize linear models~\cite{Nelson:LEET08, Barreno:ML10, Biggio:ICML12, Steinhardt:NIPS17, Jagielski:Oakland18}; thus, we focus our analysis on the LR models trained on the subset of FashionMNIST used in \S\ref{subsec:collision-gradients}.
Our analysis considers two attacks: (1) the random label-flipping~\cite{Barreno:ML10} (LF) that manipulates the labels of clean samples, and (2) the state-of-the-art (SOTA) attack formulated by Steinhardt~\etal~\cite{Steinhardt:NIPS17}.
For each attack, we first construct the poisoned training sets that include varying number of poisons synthesized using one of the attacks specified above.
On each poisoned training set, we train models with DP-Adam using different 
clipping norms and noise multipliers and compare their RAD with that of the vanilla-model.

Figure~\ref{fig:indiscriminate-resilience} illustrates the RAD of the LR models constructed using the above methodology.
We display (1) the results from the random LF attacks in the upper row, and (2) the results from the SOTA attacks in the bottom row.
%
%
For our analysis with DP-Adam, we choose the clipping norm from $\{8.0, 4.0, 2.0, 1.0, 0.1\}$ and the noise multiplier from $\{10^{-1}, 10^{-2}, 10^{-3}, 10^{-5}, 10^{-6}, 10^{-7}\}$\footnote{We use these clipping norms since the median value of parameter updates observed during the training of a model lies in that range (as recommended by Abadi~\etal~\cite{Abadi:CCS16}). Once we identify which clipping norm provides the best resilience, we examine noise multipliers that do not cause an accuracy drop more than 10\% of the model trained with the clipping norm.}.
The batch size and the learning rate are fixed to $300$ and $0.01$ respectively, and we train a model over $40$ epochs.
For each model, we increase the intensity of our attacks by blending 0, 1, 2, 3, 4, 5, 10, 20, 30, and 40\% of poisons like~\cite{Steinhardt:NIPS17}.
For the vanilla-models, we observe that the RAD caused by the SOTA attack ($\sim$0.19) is significantly higher than the random LF attack ($\sim$0.03).
We also observe that when the attackers blend more poisons, 
the trained model generally suffers from a larger RAD.

\topic{Impact of the Clipping Norm:}
%
We first examined whether using only the clipping norm can reduce the success of an indiscriminate attack, as using the noise multiplier causes the utility loss of a model.
We found that \emph{setting the clipping norm to a particular value in [2.0, 8.0] can reduce RAD caused by random LF attacks by $2\times$}.
Figure~\ref{fig:indiscriminate-resilience-a} shows the RAD of DP-models trained on the data containing different numbers of poisons from the random LF attacks.
The DP-models have smaller RAD than vanilla-models.
In particular, we achieve the lowest RAD when the clipping norm is 4.0---the DP-model trained with 40\% of poisons has 0.011 (RAD) whereas the vanilla-model shows 0.028 (RAD).
We also examine the clipping norms in $\{1.0, 0.1\}$; however, they could not achieve a RAD smaller than 0.011.

In contrast, we observed that \emph{using the clipping norm cannot reduce the RAD of a model caused by the SOTA attacks}.
Figure~\ref{fig:indiscriminate-resilience-d} shows the DP-models trained with the clipping norm in [4.0, 8.0] could not lead to a smaller RAD in all the attacks.
Also, when we use the smaller clipping norm 2.0, the RAD of the DP-models becomes worse than that of the vanilla-models.
Training with 40\% of poisons, the RAD of the DP-model is 0.217 whereas that of the vanilla-model is 0.178.
To understand why DP optimizers are ineffective in the SOTA attacks, we conduct an extensive analysis in \S\ref{sec:failure-case-analyses}.


\begin{figure*}[t]
\centering
	\begin{subfigure}[t]{0.32\textwidth}
	    \centering
	    \caption*{\textbf{Purchase-100 (LRs)}}
	    \vspace{-0.6em}
	    \includegraphics[width=\textwidth]{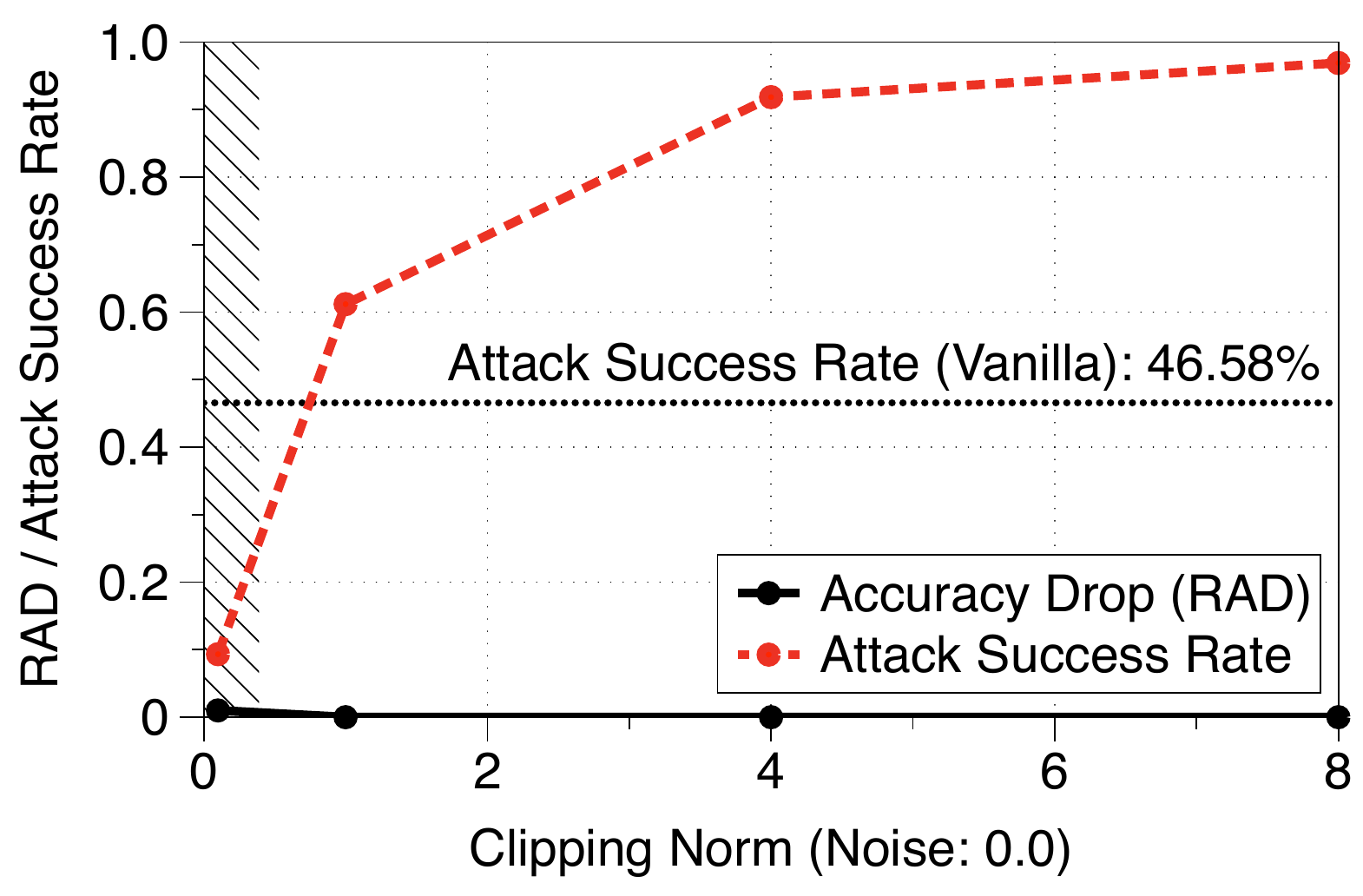}
	\end{subfigure}
	\sfvrule
	\begin{subfigure}[t]{0.32\textwidth}   
	    \centering 
	    \caption*{\textbf{FashionMNIST (MLPs)}}
	    \vspace{-0.6em}
	    \includegraphics[width=\textwidth]{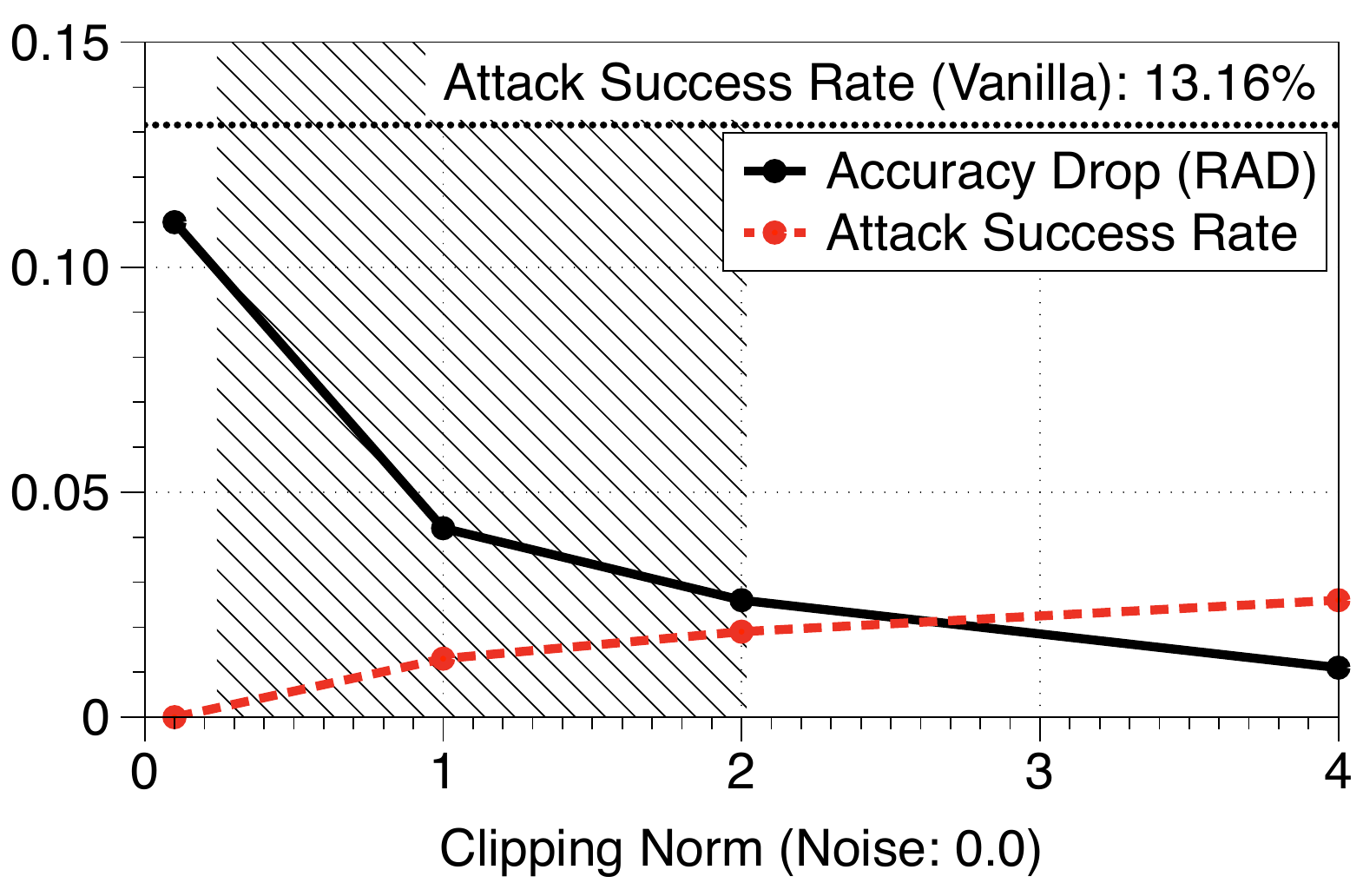}
	\end{subfigure}
	\sfvrule
	\begin{subfigure}[t]{0.32\textwidth}
	    \centering
	    \caption*{\textbf{CIFAR-10 (CNNs)}}
	    \vspace{-0.6em}
	    \includegraphics[width=\textwidth]{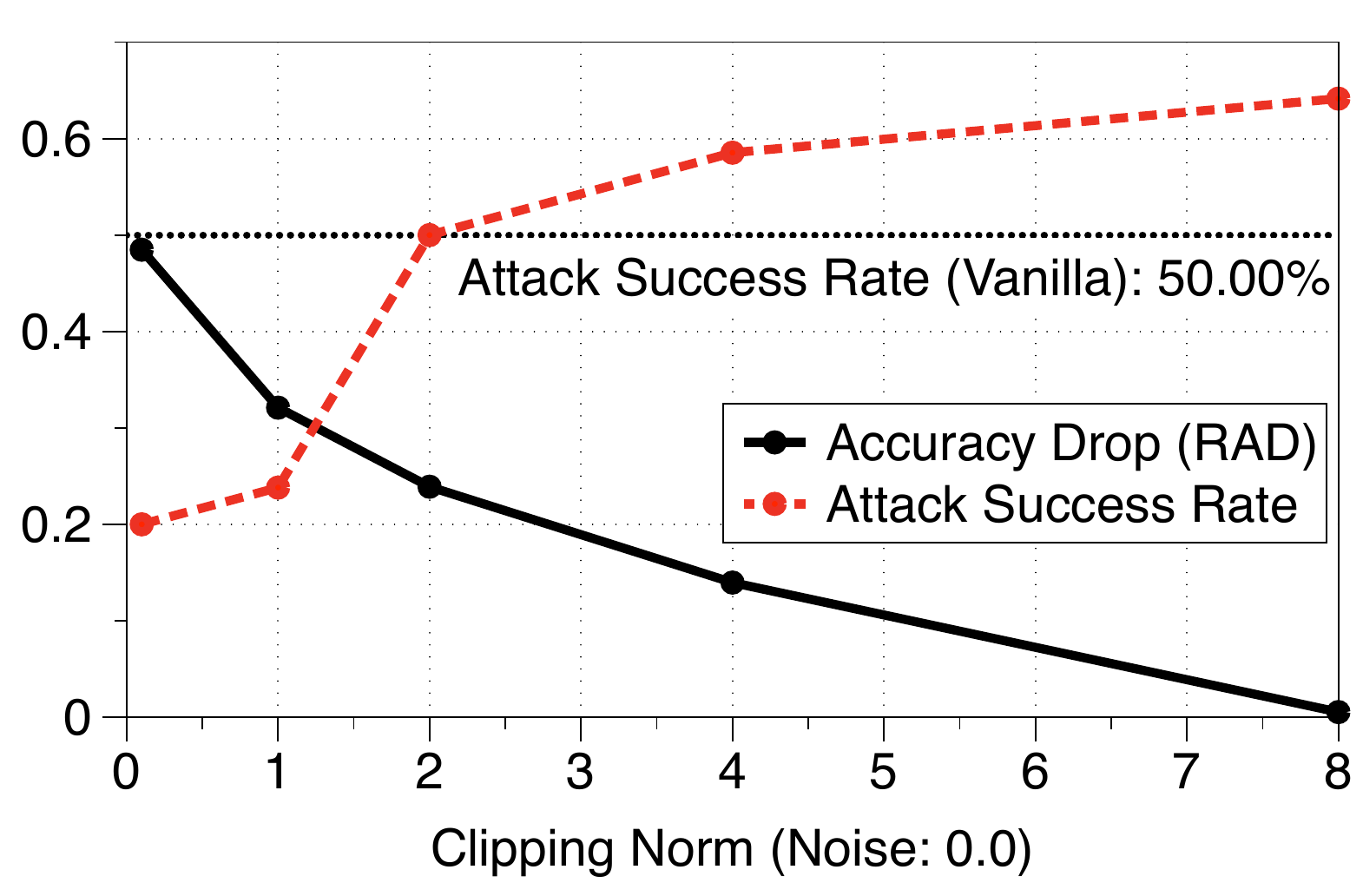}
	\end{subfigure}
	%
	%
	\begin{subfigure}[t]{0.32\textwidth}  
	    \centering 
	    \includegraphics[width=\textwidth]{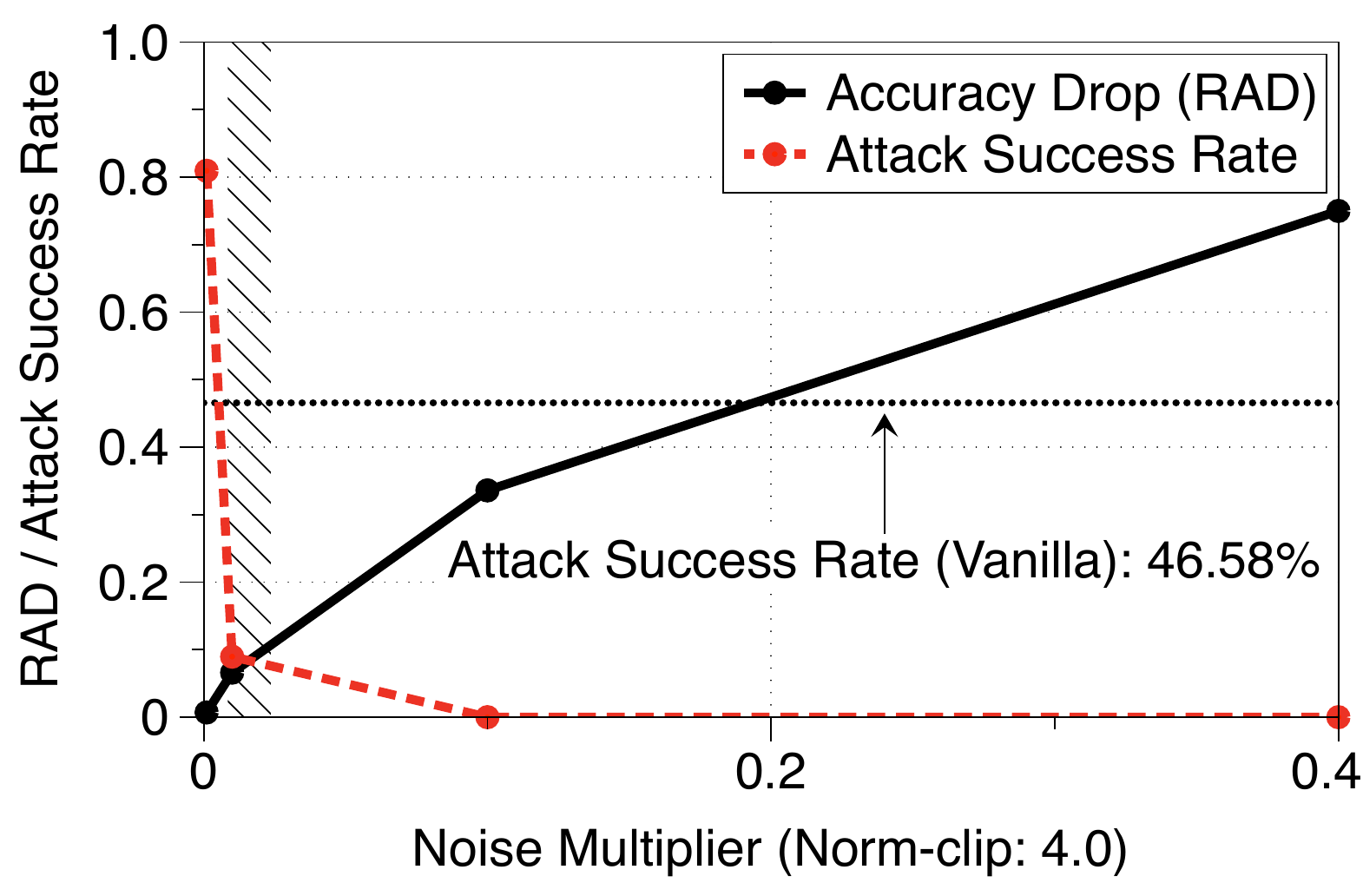}
	\end{subfigure}
	\sfvrule
	\begin{subfigure}[t]{0.32\textwidth}   
	    \centering 
	    \includegraphics[width=\textwidth]{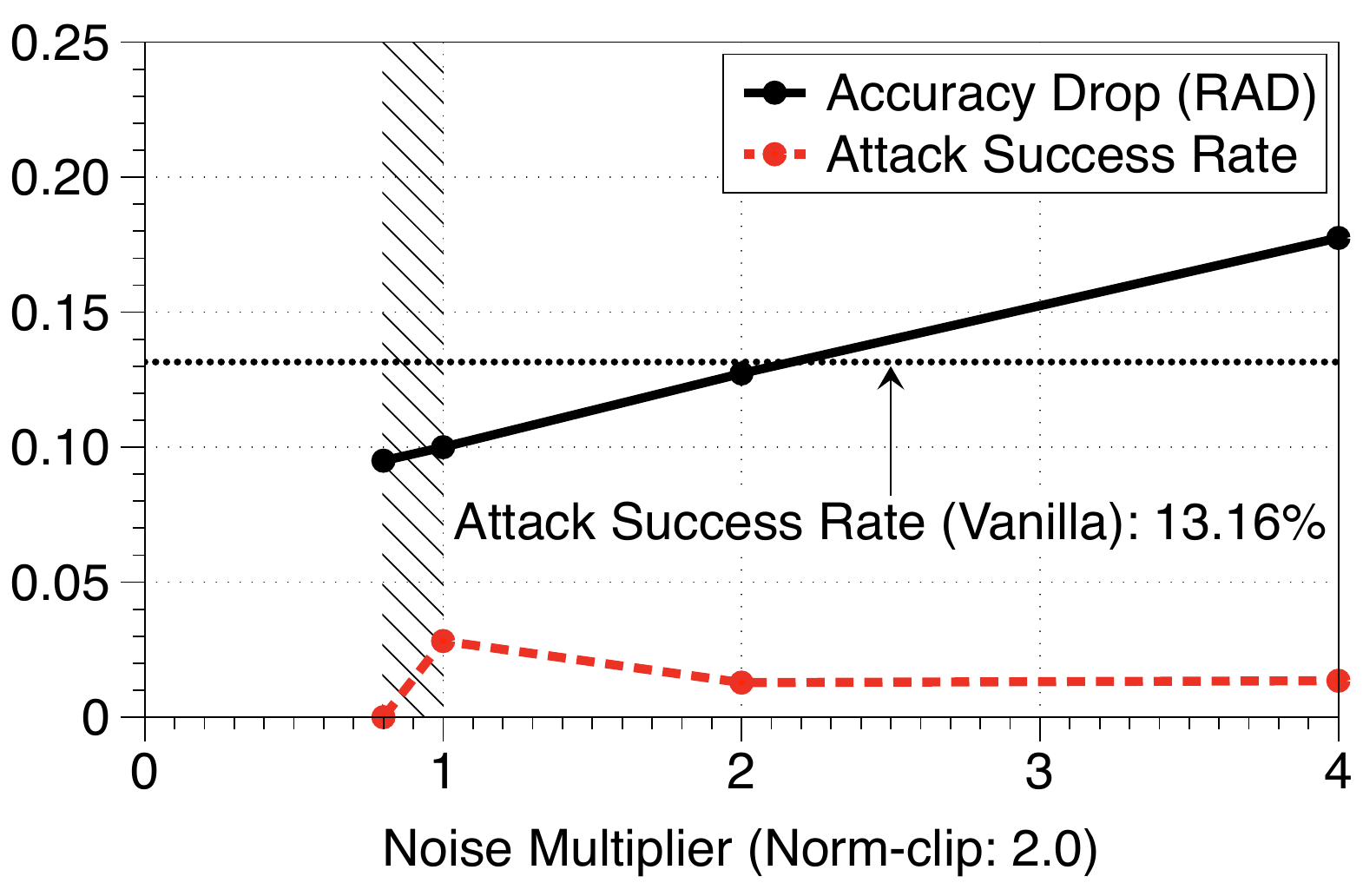}
	\end{subfigure}
	\sfvrule
	\begin{subfigure}[t]{0.32\textwidth}   
	    \centering 
	    \includegraphics[width=\textwidth]{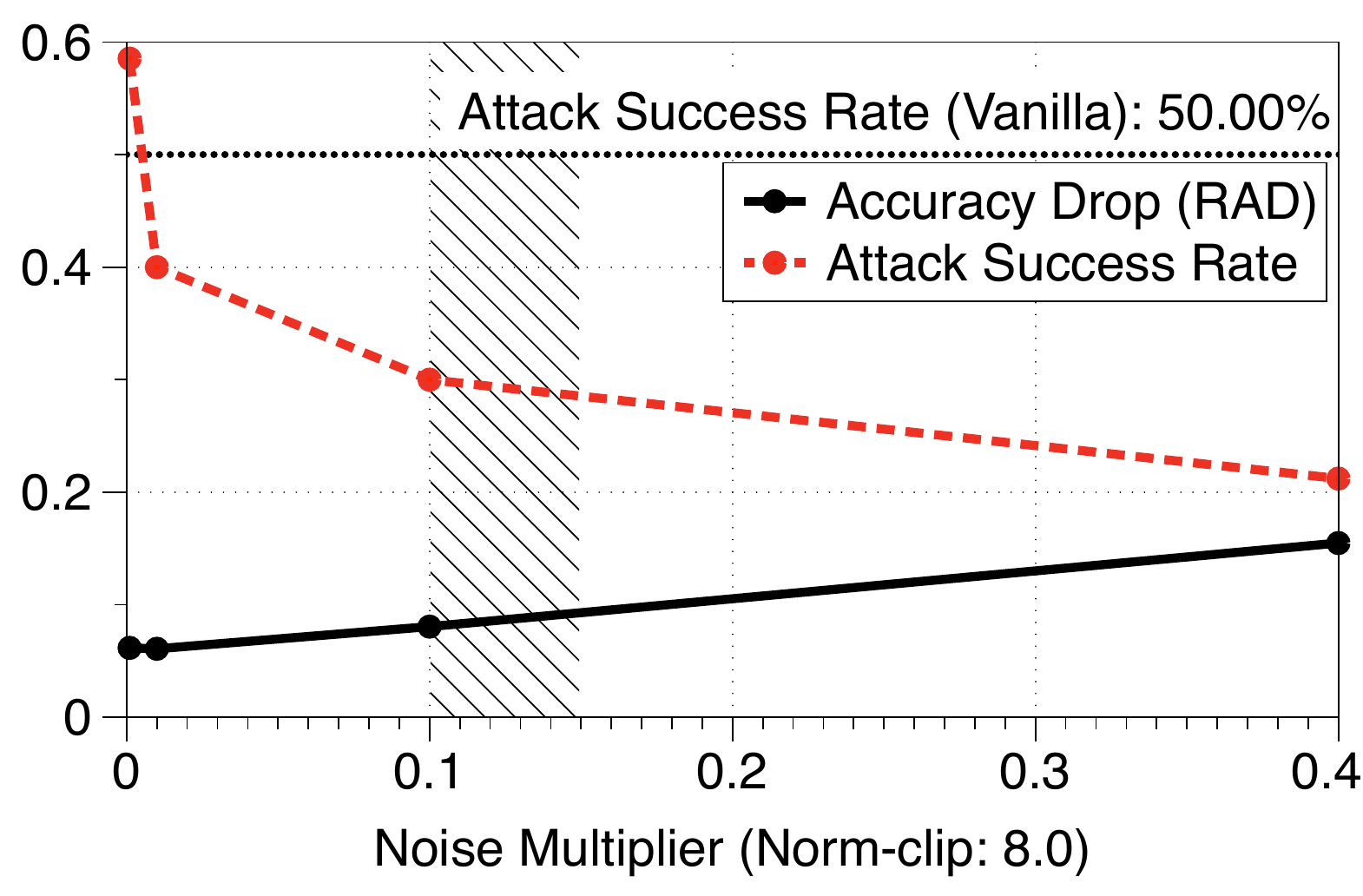}
	\end{subfigure}
	%
\caption{\textbf{Effectiveness of Training with DP Optimizers against the One-shot (Targeted) Poisoning.} We illustrate the RADs (solid lines) and the attack success rate (dashed lines) of the DP-models trained with the different choice of DP-SGD parameters. In the upper row, we only use the clipping norm whereas we vary the noise in the bottom row when the clipping norm is fixed. In the shaded area, we find DP-SGD configurations that reduces the attack success rate by half with a RAD of 0.1.}
\label{fig:targeted-resilience}
\end{figure*}

\topic{Impact of the Noise Multiplier:}
%
%
Here, we evaluate whether combining the noise multipliers with a particular choice of clipping norm can reduce the RAD caused by the attacks further than individually using the parameters.
We set the clipping norm to 4.0---the best setting found from our analysis---and vary the noise between $\{10^{-1}, 10^{-3}, 10^{-5}, 10^{-7}\}$.
Figure~\ref{fig:indiscriminate-resilience-b} and~\ref{fig:indiscriminate-resilience-e} show that \emph{a defender could not benefit from combining the noise multiplier with a specific clipping norm}.
In the random LF attacks, the DP-models shows more RAD when we combine the noise multipliers
and the clipping norm 4.0 than the models trained without the noise.
In the SOTA attacks, using the noise multiplier could not provide any benefit for the defender as the RAD of the DP-models is similar to that of the vanilla-models.
We revisit this in \S\ref{sec:failure-case-analyses}.

\topic{Impact on the Gradients:}
In \S\ref{subsec:taxonomy-indiscriminate}, we identified that indiscriminate poisoning attacks induce contrasting magnitude and orientations between the gradients from poisons and clean samples during training.
Hence, for the attack cases where using the clipping norm is an effective defense, the magnitude and orientation differences in training has to be reduced.
We found that is the case.
In Figure~\ref{fig:indiscriminate-resilience-c} and~\ref{fig:indiscriminate-resilience-f}, we compare the magnitude differences observed in the training of the vanilla- and DP-model with 40\% of poisons from both the random LF and SOTA attacks.
We set the clipping norm to 4.0.
%
%
For the random LF attack, we observe the magnitude ratio decreases when the clipping norm is used; on average, the ratio is 2.527 in the vanilla-model and 2.221 in the DP-model.
However, in the SOTA attack, the ratio becomes higher in the DP-model (3.645) than what we see in the vanilla-model (2.497).
%
%
%
This implies that the magnitude of poison gradients is smaller in the DP-model than that in the vanilla-model; thus, \emph{during training of a model with DP optimizers, the influence of poisons on the model is less than that in the vanilla training}.



\subsection{Mitigating Targeted Poisoning}
\label{subsec:resilience-targeted}

\topic{Experimental Methodology:} 
We evaluate the effectiveness of our defense against the realistic, worst-case targeted attacker formulated by~\cite{Ali:NIPS18}.
%
%
This attack considers the white-box adversary who has the full knowledge of the target model and its parameters.
By exploiting this internal information, the attacker becomes inconspicuous, but effective---\ie the adversary crafts poisons perceptually indistinguishable by a human but can cause misclassification on targets with small number of poisons, \eg a single poison.
Moreover, the attack does not modify the original label of poisons (clean-label).
%
This is currently considered a worst-case attack because such inconspicuous poisons are difficult to be filtered out by using the existing outlier-based defenses in \S\ref{subsec:existing-defenses}.
To maximize the influence of poisons on a model, the attacker blends them into the training set used for re-training of the model.
We denote the case where the attacker uses a single poison as \emph{one-shot} and \emph{multi-poison} when they use multiple.

To evaluate our defense comprehensively, we attack 100 test-time samples in a specific class, and for each target, we choose 100 other testing samples in another class close to the target in the penultimate layer's representation space as candidates for crafting poisons.
Our attacker has computational constraints\footnote{We choose the same constraints as what is used in Shafahi~\etal~\cite{Ali:NIPS18}.}: (1) as the poison crafting is an optimization procedure, we bound the attacker within 10k iterations, and (2) during this optimization, we only use the poisons in a specific proximity of the target in the representation space---\ie an $\ell_2$ distance of 3.5.
%
%
During the attacks, we consider the one-shot attack to be successful when any one of the poisons lead to the misclassification on a target with RAD $\le$ 0.05, and the multi-poison attack is successful when a set of poisons cause misclassification.
We define the \emph{attack success rate} as a ratio of the number of misclassified targets over the total 100.

Figure~\ref{fig:targeted-resilience} shows the success rate of the one-shot poisoning attacks and the RAD caused by training with DP-SGD in three different models (LR, MLP and CNN) trained using Purchase-100, FashionMNIST, and CIFAR-10 respectively.
%
%
During our training of a model with DP-SGD, we use the clipping norm in $\{8.0, 4.0, 2.0, 1.0, 0.1\}$ and the noise multiplier in $\{0.001, 0.01, 0.1, 0.4, 0.8, 1.0, 2.0, 4.0\}$.
The batch size is fixed to 100, and we use the learning rate 0.08 for the LR and MLP models and 0.02 for CNN models.
We first train a model from scratch on the clean training set for 100 epochs and then re-train the same model for 50 epoch with the same training set containing poisons.
Since using the noise multiplier 
decreases a model's utility, we first examine the impact of the clipping norm by setting the noise multiplier to zero.
Then, we fix the clipping norm to a specific value and repeat the same set of analyses while varying the noise multiplier.
%

\topic{Impact of the Clipping Norm:}
We observe that \emph{setting the clipping norm to a small value can decrease the success rate of the one-shot poisoning attack significantly}.
This result is consistent with our intuition in \S\ref{subsec:collision-gradients}; since the attacker exploits feature collision locally, the magnitude difference between the poison and clean gradients becomes high, and the orientation difference oscillates during re-training.
Thus, suppressing the parameter updates from the poison by setting the clipping norm can be an effective defense.
In the LR models trained on Purchase-100, using the clipping norm 0.1 reduces the attack success rate from 46.58\% to 9.33\% with RAD $\lt$ 0.1.
For the MLP models trained on FashionMNIST, we also observe that the attacker's success rate decreases to 1.33\% with a RAD of 0.04 when we use the clipping norm of 1.0.
Moreover, in the CNN models trained on CIFAR-10, setting the clipping norm to 0.1 also reduces the success rate by more than $2\times$---from 50.00\% to 21.00\%.
However, we sacrifice the model's utility 0.48 in RAD to achieve the resilience.


\begin{figure*}[t]
	\centering
	\begin{subfigure}[t]{0.322\linewidth}
		\centering
		\includegraphics[width=\textwidth]{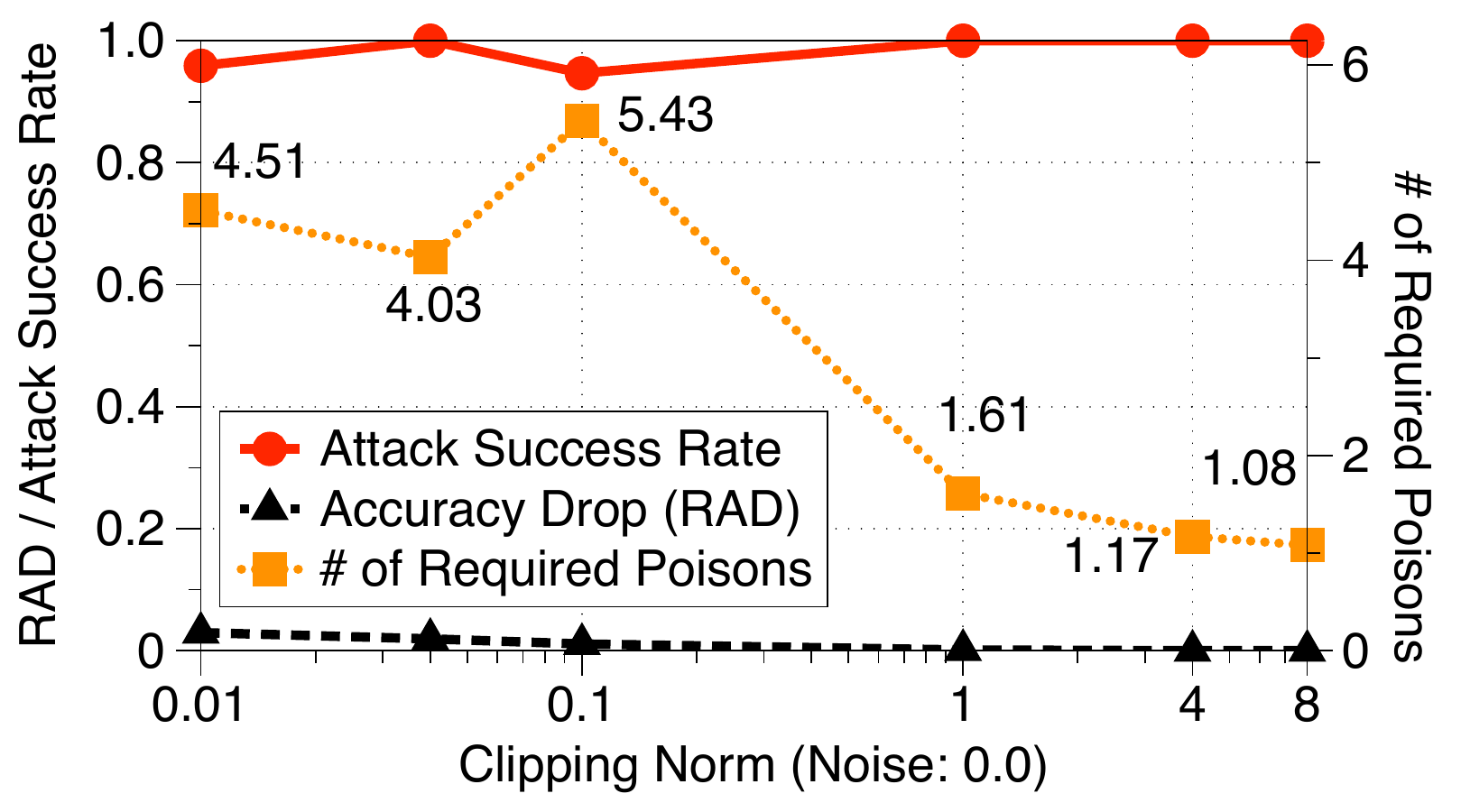}
		\caption{\textbf{LR: Impact of the Clipping Norm}}
		\vspace{0.6em}
		\label{fig:targeted-resilience-multi-a}
	\end{subfigure}
	\begin{subfigure}[t]{0.322\linewidth}  
		\centering 
		\includegraphics[width=\textwidth]{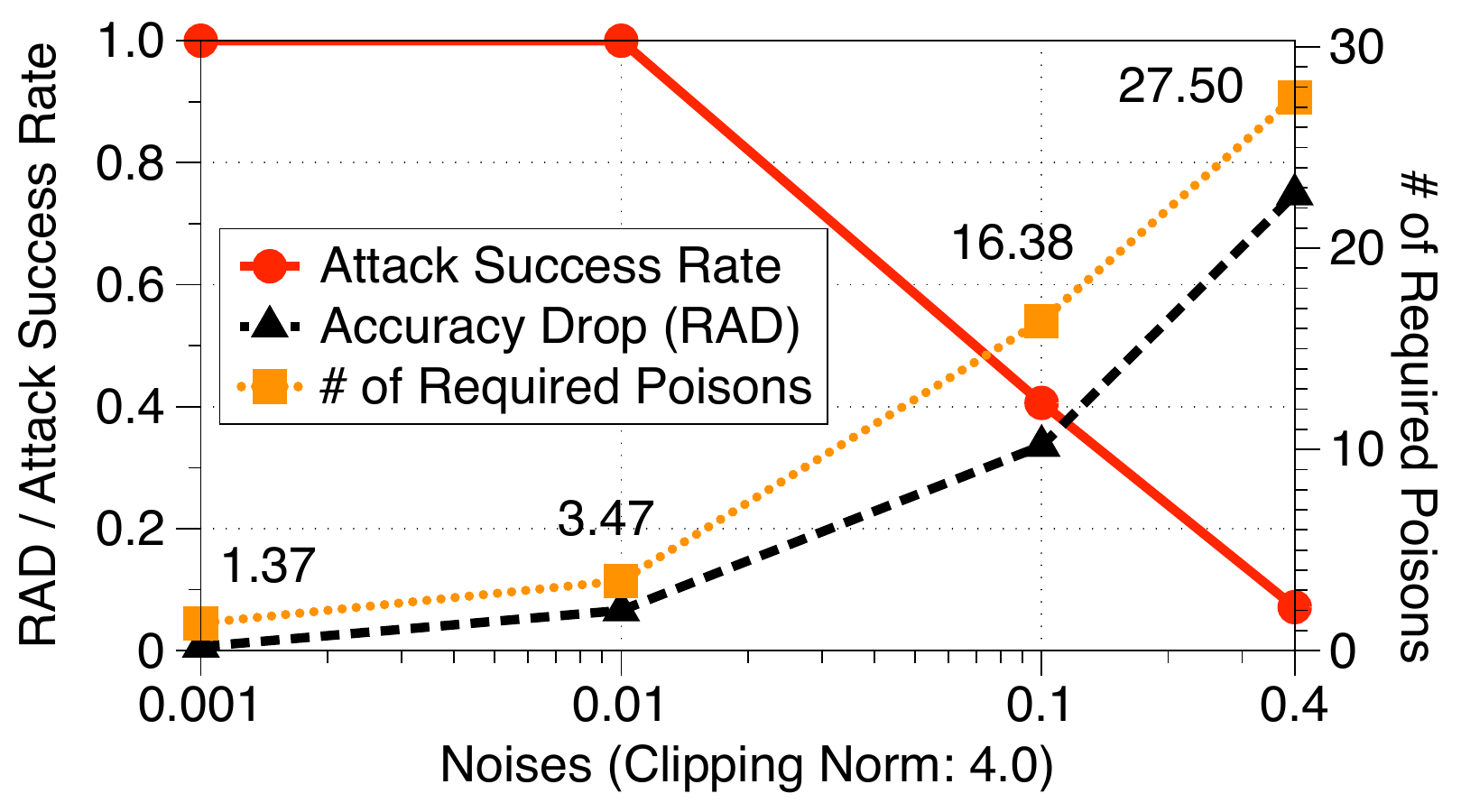}
		\caption{\textbf{LR: Impact of the Noise Multiplier}}
		\vspace{0.6em}
		\label{fig:targeted-resilience-multi-b}
	\end{subfigure}
	\begin{subfigure}[t]{0.322\linewidth}  
		\centering 
		\includegraphics[width=\textwidth]{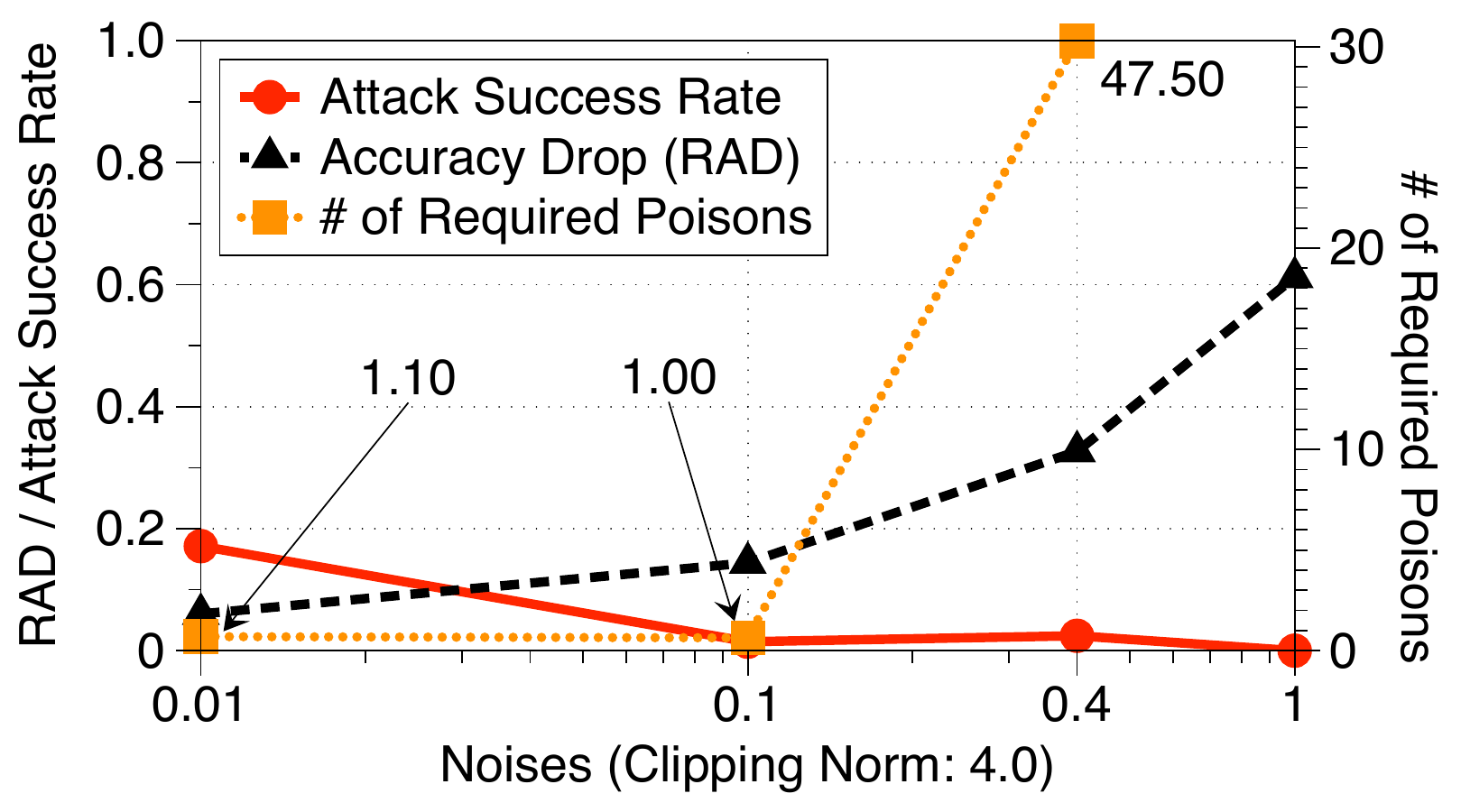}
		\caption{\textbf{MLP: Impact of the Noise Multiplier}}
		\vspace{0.6em}
		\label{fig:targeted-resilience-multi-c}
	\end{subfigure}
	\caption{\textbf{Effectiveness of Training with DP Optimizers against the Multi-Poison (Targeted) Attacks.} We illustrate the success rate of the attacker (solid lines), the averaged number of required poisons for a successful attack (dotted lines), and the RADs (dashed lines) of the DP-models trained with the difference choices of DP-SGD/Adam parameters on Purchase-100.}
	\label{fig:targeted-resilience-multi}
\end{figure*}

\topic{Impact of the Noise Multiplier:}
Our previous analysis raises a question: can we achieve better resilience with the same RAD by combining the noise multiplier?
To answer this question, we choose the clipping norm from our analysis results and examine different noise multipliers.
The results are shown in the second row of Figure~\ref{fig:targeted-resilience}.
We found that \emph{combining the noise multiplier with a specific value of the clipping norm reduces the attacker's success rate further with RAD $\lt$ 0.1}.
In Purchase-100, we use the noise multiplier 0.01 with the clipping norm 4.0, and we decrease the success rate of the attacker to 8.97\%.
We achieve 0\% attack success rate with the clipping norm 2.0 and the noise multiplier 0.8 in FashionMNIST.
In CIFAR-10, using only the clipping norm 0.1, we can reduce the attacker's success rate by $2\times$, but we lose the model's utility by 0.48 in RAD.
However, when we use the noise multiplier 0.4, the attacker's success rate drops to 21.21\%, similar to the results by using only the clipping norm, but the utility loss is much smaller ($0.15$ in RAD).
%

\begin{figure}[h]
	\centering
	\includegraphics[width=0.72\linewidth]{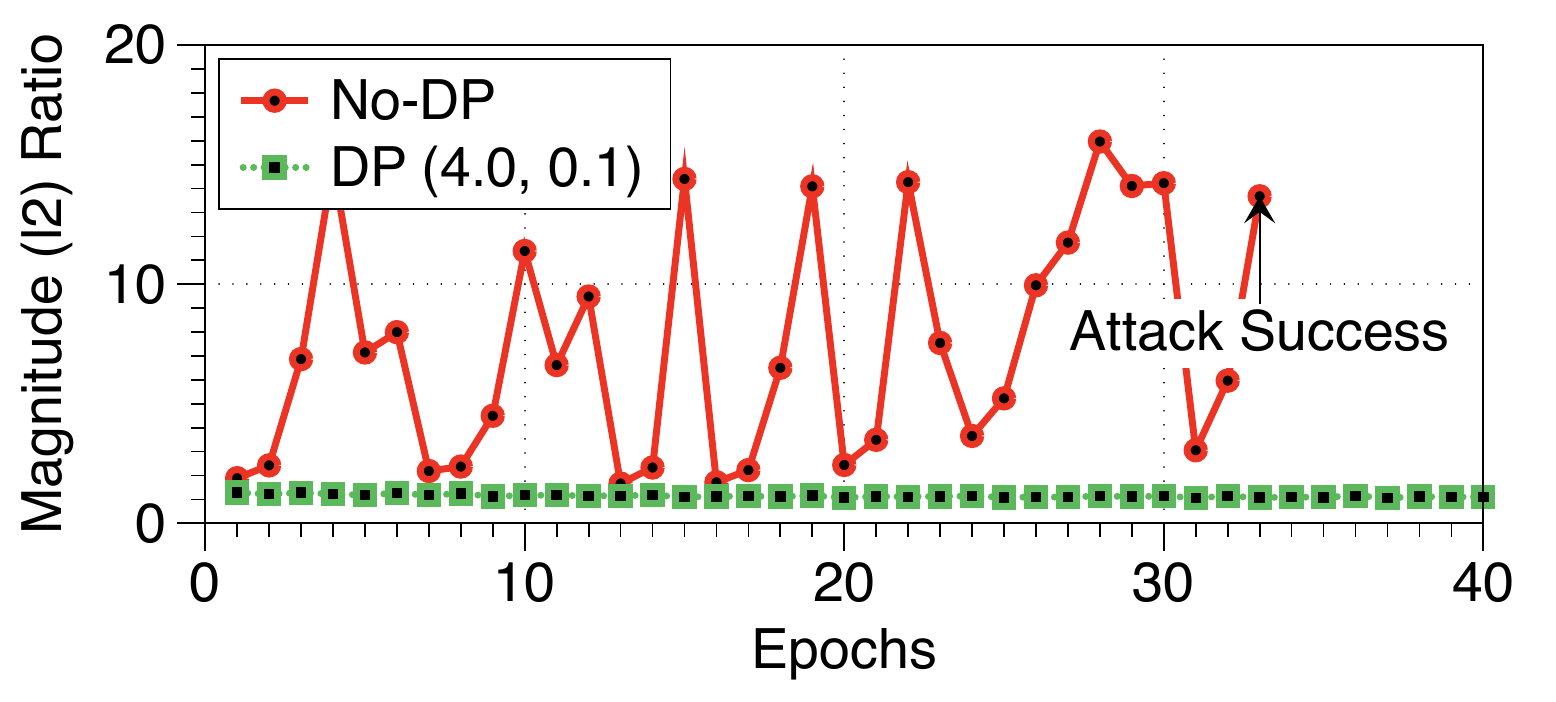}
	\includegraphics[width=0.72\linewidth]{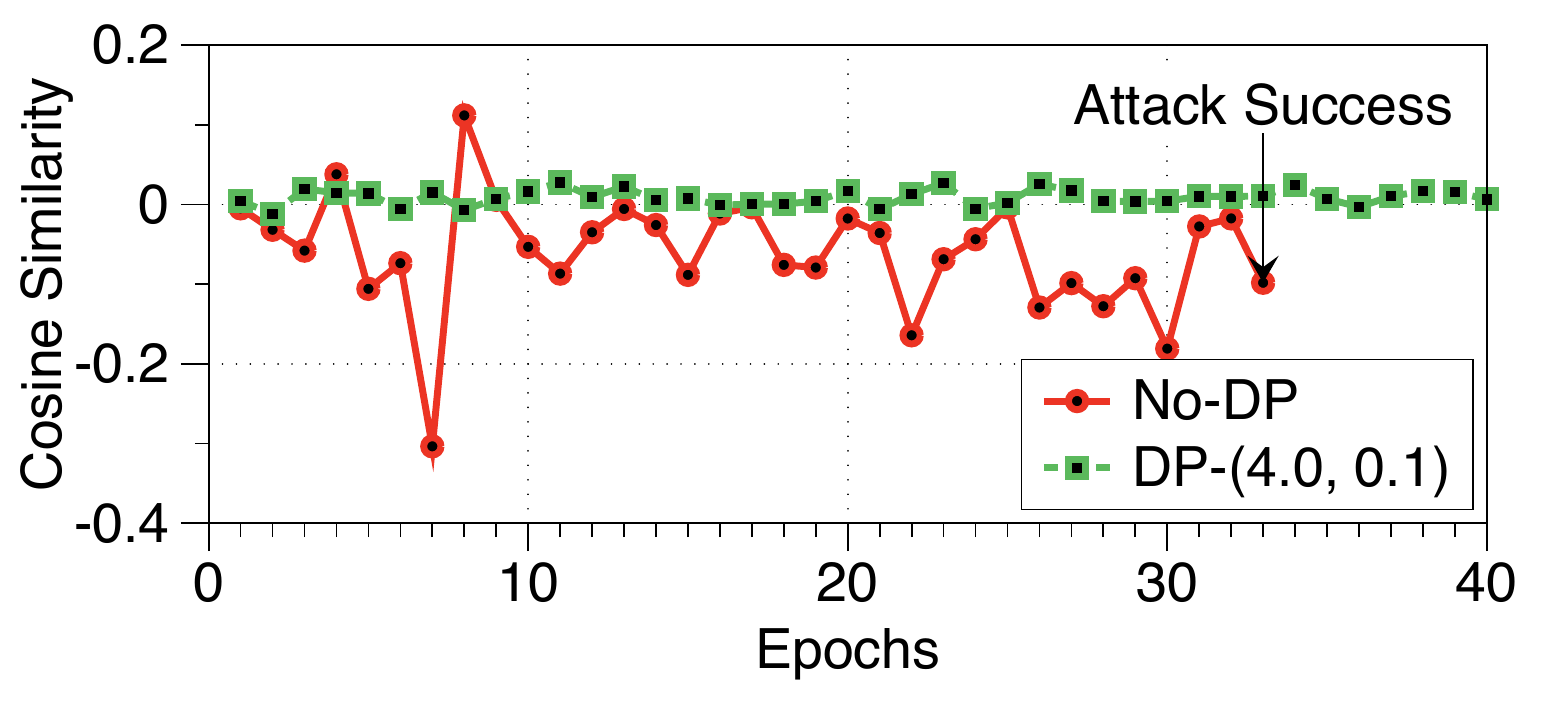}
	\caption{\textbf{Gradient-level Differences in One-shot Attacks.} We compare the magnitude ratio and orientation difference seen in re-training (Purchase-100). For the same target, the attack succeeds with SGD and fails when we use DP-SGD.}
	\label{fig:targeted-magnitudes}
\end{figure}

\topic{Impact on the Gradients in Training:}
Here, we conduct an analysis of the impact of re-training with DP-SGD on the magnitude ratio and orientation difference.
If the DP-SGD is an effective anti-poison defense against the one-shot attacks, the magnitude ratio becomes smaller and the orientation difference is stabilized when we re-train a model with the optimizer.
Figure~\ref{fig:targeted-magnitudes} illustrates our analysis results.
We compare the magnitude ratios between the re-training process with SGD and that with DP-SGD in the upper plot; the orientation differences are shown in the lower plot.
For DP-SGD, we use the clipping norm 4.0 and the noise multiplier 0.1.

We found that \emph{re-training of a model with DP-SGD reduces the magnitude ratios between the poison and clean gradients and stabilizes the orientation differences}.
The magnitude ratios decrease from 7.46 to 1.14 on average over 50 epochs.
Also, the standard deviation of the orientation differences seen in the successful attack is 0.073, whereas the value becomes 0.01 when we use DP-SGD during re-training.



\subsection{Mitigating Multi-Poison Attacks}
\label{subsec:targeted-resilience-multi}
In this subsection, we extend our previous analysis by conducting the multi-poison attacks on the LR and MLP models trained on Purchase-100.
Figure~\ref{fig:targeted-resilience-multi} illustrates our results: we show (1) the attacker's success rate, (2) the number of poisons required for a successful attack on average, and (3) the RAD of a model caused by DP-SGD.
We use the same set of clipping norms and noise multipliers as in \S\ref{subsec:resilience-targeted}.

\topic{Impact of the Clipping Norm:}
We found that \emph{setting the clipping norm to a small value increases the number of required poisons for an successful attack with RAD $\lt$ 0.05, but this could not reduce the success rate of the attack}.
First, in Figure~\ref{fig:targeted-resilience-multi-a}, we can see the attacker consistently achieves a success rate over 94.67\% in all the clipping norms.
However, we found that the number of required poisons on average increases from 1.08 up to 5.43 as we decrease the clipping norm from 8.0 to 0.1, with the small amount of utility loss (0.011 in RAD).
We also examine the clipping norms smaller than 0.1---\ie 0.4 and 0.01; nevertheless, they do not provide more benefits to a defender.
The number of required poisons for an attack saturates, and the utility loss starts to increase.

\topic{Impact of the Noise Multiplier:}
%
Can we reduce the success rate of the attacker by combining the noise multiplier with a specific value of the clipping norm?
To answer this question, we repeat the same experiments with the fixed clipping norm (4.0) and vary the noise multipliers.
Figure~\ref{fig:targeted-resilience-multi-b} and~\ref{fig:targeted-resilience-multi-c} illustrates our results in LR and MLP models trained on Purchase-100.
We found that \emph{using the noise multiplier with the clipping norm is helpful to reduce the attack success rate and to increase the number of required poisons for an successful attack, but this comes with the significant utility loss}.
In the LR models with the noise multiplier 0.4, the attacker's success rate becomes 7.14\%, and the number of required poisons is 27.50.
In the MLP models, when we use the noise multiplier 1.0, we make the success rate of the attacker 0\%.
However, in both cases, the utility loss of the LR and MLP models are 0.748 and 0.612 in RAD respectively.

\topic{Impact on the Gradients in Training:}
DP-SGD works as a defense by clipping the norm of an individual gradient and adding Gaussian noise to it (see Algorithm~\ref{algo:sgd-and-dp}).
Hence, \emph{the multi-poison attacker can neutralize the impact of the clipping norm by using multiple poisons}.
Contrary to the one-shot attack where each batch includes at-most one poison, the multi-poison attack enforces each batch to contain more than one poison.
%
Even if each poison gradient is bounded by a small clipping norm, the influence of total poison gradients to the model parameter updates during training can be sufficient to cause a successful attack.

To decrease the success rate of the multi-poison attacker, we need to set the noise multiplier.
The noise added to each gradient prevents the gradients computed from multiple poisons orienting towards a similar direction.
Thus, their sum in re-training is insufficient to cause the misclassification.
This is also true that \emph{the attacker can neutralize the noise by blending multiple poisons---\ie the expected sum of the noise added to each gradient is zero}; however, due to the randomness, the number of required poisons to remove the noise is a lot more than what the attacker needs to evade the clipping norm.



\subsection{Distinct Defense Scenarios}
\label{subsec:resilience-dp-trained-model}

In this section, we consider the distinct defense scenarios such as transfer learning cases where a defender can only use DP-SGD in the specific stage of the training process.


\begin{table}[b]
\centering
\adjustbox{width=\linewidth}{%
\begin{tabular}{@{}ccccc@{}}
	\toprule
	\textbf{Attack} & \textbf{Base Model} & \textbf{Re-train} & \textbf{Success Rate} & \textbf{\# Poisons} \\ \midrule
	\multirow{3}{*}{\textbf{One-Shot}} & Vanilla & w/o DP & 46.58\% (34/73) & 1 \\
	& Vanilla & w. DP & 17.81\% (13/73) & 1 \\
	& DP-Model & w/o DP & 16.67\% (13/78) & 1 \\ \midrule
	\multirow{3}{*}{\textbf{Multi-poison}} & Vanilla & w/o DP & 100.0\% (73/73) & 1.79 \\
	& Vanilla & w. DP & 61.64\% (45/73) & 2.47 \\
	& DP-Model & w/o DP & 100.0\% (78/78) & 3.47 \\ \bottomrule
\end{tabular}
}
\caption{\textbf{Effectiveness of Gradient Shaping in Distinct Defense Scenarios.} For each attack, we show the success rate and the number of required poisons for a successful attack.}
\label{tbl:distinct-retrain}
\end{table}

\topic{Re-training a Vanilla-Model with DP-SGD:}
This happens when a defender cannot train a model from scratch.
For example, it is difficult to train a large model~\cite{GoogleNMT:ArXiv16} from scratch with DP-SGD as the training takes a week on a super-computer cluster.
%
%
Considering such a scenario, we evaluate whether re-training of a vanilla-model with DP-SGD on the poisoned training set can be resilient against the targeted attacks.

We take the vanilla LR model trained on Purchase-100 and re-train the model using two methods: (1) we continue to train with Adam or (2) use DP-Adam with the clipping norm 4.0 and the noise multiplier 0.01.
During re-training, we perform both the one-shot and multi-poison attacks.
Table~\ref{tbl:distinct-retrain} shows the attacker's success rate and the number of required poisons for a successful attack.
We found that \emph{a defender can make the re-training of a vanilla model resilient against targeted poisoning by using DP optimizers}.
In the one-shot attack, the success rate decreased from 46.58\% to 17.81\%.
In the multi-poison attack, the attacker's success rate decreased from 100.00\% to 61.64\%, and the number of required poisons increased from 1.79 to 2.47, which shows better resilience than the cases in \S\ref{subsec:targeted-resilience-multi}.

%

\topic{Re-training a DP-Model with SGD:}
We now 
direct our attention to the scenario where we re-train a DP-model with DP-Adam on the poisoned training data.
Here, we take the DP-LR model trained on Purchase-100 using the same clipping norm and noise multiplier and re-train the model using Adam.
During re-training, we perform the one-shot and multi-poison attacks.
Our results is in Table~\ref{tbl:distinct-retrain}.
We observe that \emph{when a model trained with DP is re-trained with vanilla optimizer, the model becomes resilient against targeted poisoning even if we do not use our defense during re-training}.
The one-shot attacker's success rate on the model re-trained with Adam is 16.67\% whereas the same attacks are 46.58\% successful when we re-train a vanilla-model with Adam.
In the multi-poison cases, the success rate of the attack does not decrease, but the number of required poisons are 3.47 on average, which is higher than the vanilla-model case (1.79).
This result implies that when a defender distributes a teacher model for transfer learning, it is safer to train the model with a DP optimizer.
To understand this resilience, we analyze the decision boundary of a DP-model and include the discussion in Appendix~\ref{appendix:boundary-analysis}.



\section{Discussion}
\label{sec:failure-case-analyses}

In this section, we discuss the scenarios where DP optimizers become ineffective.
Our previous analysis identified that:
(1) training an LR model with DP-Adam cannot mitigate the attacks formulated by Steinhardt~\etal~\cite{Steinhardt:NIPS17} (\S\ref{subsec:resilience-indiscriminate}), and
(2) setting the noise multipliers to a high value accompanies a significant utility loss of a trained model (\S\ref{subsec:targeted-resilience-multi}).
We 
conduct analysis of those cases to understand the limits of DP optimizers when we use them as a mechanism for realizing gradient shaping and discuss potential improvements to address them.


\begin{figure}[t]
	\centering
	\begin{subfigure}{0.74\linewidth}
		\centering
		\includegraphics[width=\linewidth]{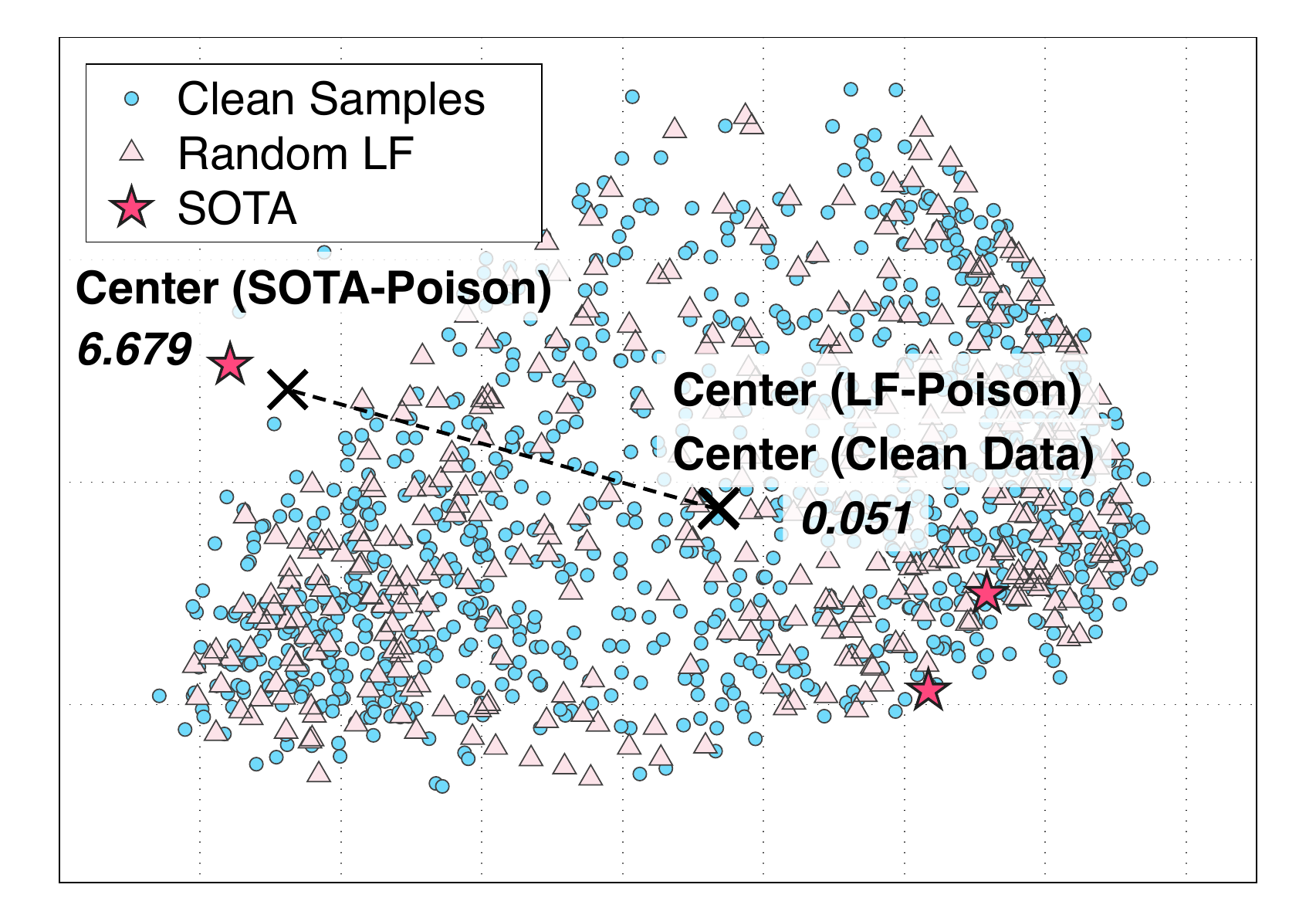}
	\end{subfigure}
	\hfill
	\parbox{.25\linewidth}{
		\centering
		\begin{subfigure}{\linewidth}
			\centering
			\includegraphics[width=0.6\linewidth]{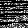}
			\caption*{\textbf{Poison (Dress)}}
			\vspace{1.6em}
		\end{subfigure}
		\vfill
		\begin{subfigure}{\linewidth}
			\centering
			\includegraphics[width=0.6\linewidth]{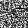}
			\caption*{\textbf{Poison (Coat)}}
		\end{subfigure}
	}
	\caption{\textbf{Distribution of Clean Data and Poisons of the Indiscriminate Attacks.} We visualize the clean data and poisons from the random LF and SOTA~\cite{Steinhardt:NIPS17} attacks in \S\ref{subsec:resilience-indiscriminate}. On the left, we show two poisons used in the SOTA attack.}
	\label{fig:steinhardt-case}
\end{figure}

\subsection{Case Study: Steinhardt's Attack}
\label{subsec:steinhardt-attack}

We compare the distribution of poison samples to understand why we cannot defeat the SOTA attack.
We found that the SOTA attack uses \emph{unrealistic poisons} that can exploit the weakness of linear models.
Figure~\ref{fig:steinhardt-case} shows the distribution differences between the clean and poison samples from the random LF and SOTA attacks.
With the 10,800 clean samples and 4,320 (40\%) poisons from each attack, we perform principal component analysis (PCA)~\cite{PCA:Pearson} to reduce the dimension and then use KMeans~\cite{KMeans}
on the 2-dimensional data for clustering.
In the figure, we observe that the poisons from the SOTA attack are unrealistic; they consists of the same 3,970 coat samples (a single point in the left) and 350 dress samples in two types (two points in the right).
When we train an LR model on this data, the model first fits its decision boundary that splits the poisons well and then adjusts the boundary to classify clean samples (see Appendix~\ref{appendix:steinhardt-training} for details).
Contrarily, the 4,320 poisons from the random LF attack have the similar distribution to the clean samples; thus, the model trained on the data can learn its boundary from the majority of clean samples and become resilient to the poisons.

\subsection{Case Study: Multi-Poison Attack}
\label{subsec:multi-poison-attack}

Our analysis of the multi-poison attacks \S\ref{subsec:targeted-resilience-multi} shows that a defender can reduce the success rate of the attacker and increase the number of required poisons by setting the noise multiplier to a high value.
However, this results in a significant utility loss of a trained model.
This is because \emph{DP-SGD consistently adds the same amount of noise to the gradients computed from clean samples and poisons during training whereas gradient's magnitudes become smaller as a model learns about the training data}.
To solve this problem, the amount of noise added to each gradient should be proportional to its magnitude.
One simple solution is to monitor the average $\ell_2$-norm of an individual gradient in the previous epoch and adjust the amount of noise.
Since we take the averaged magnitude of an individual gradient, the attacker, who attempts to adapt to our solution, would have to manipulate a large fraction of the data to influence the noise multiplier.
However, this solution incurs computational overheads, and the mechanism does not provide the privacy guarantee anymore.
We leave this research direction---to come up with an ideal mechanism that realizes gradient shaping---as future work.


\section{Related Work}
\label{sec:related}

\topic{Gradient Regularization as a Defense}
Prior work, especially in the context of neural networks, has proposed imposing regularization penalties on a model's gradients to improve the accuracy~\cite{drucker1991double}, interpretability~\cite{ross2018improving} or robustness against adverarial examples~\cite{lyu2015unified}.
In contrast to our threat model, by explicitly penalizing the input gradients, these mechanisms aim to regulate the test-time predictions, while assuming a clean training set.  
We propose gradient shaping against training-time attacks to suppress the adverse gradient signatures poisons produce during training.
Further, gradient penalties rely on computationally restrictive ``double backpropagation'', whereas we implement gradient shaping with a more efficient DP-SGD mechanism.

\topic{Model Poisoning Attacks against Machine Learning:}
In the context of distributed learning scenarios, such as federated learning, recent work has proposed model poisoning attacks that directly manipulate the parameter updates (gradients) end-hosts send to the shared model~\cite{BackdoorFedLearn:arXiv18, AttackFedLearn:ICML19,Krum:NIPS17, Coordinate:ICML18}.
These attacks have been considered effective when the attacker and the victim interactively train a model.
Model poisoning is out-of-scope for our work because in the poisoning attacks we consider, the adversary manipulates the training set, not the gradients.

\topic{Privacy Attacks in Machine Learning:}
Due to the high capacity of machine learning models, especially neural networks, models trained on private data, such as health-care or face datasets, may potentially leak sensitive information about their training sets.
Prior work~\cite{Carlini:USENIX19, Shokri:IEEE17, MLLeak:NDSS19, CCS15:MInversion} has demonstrated attackers who aim to extract sensitive information from trained models. 
DP-SGD~\cite{Abadi:CCS16} is developed as a tool to protect the model from these attacks by clipping and adding noise to the gradients during training. 
We use DP-SGD as an exemplar tool to demonstrate the feasibility of gradient shaping as a defense against data poisoning attacks.

\section{Conclusions}
\label{sec:conclusion}

This work tackles data poisoning in machine learning with a unifying view of the threat landscape.
We focus on a common element of all poisoning attacks: they manipulate gradients computed during training to update models.
We identified two main artifacts shared by various forms of poisoning---(1) gradients computed on poisoned data have significantly higher magnitudes than their counterparts on clean data, and (2) their orientations also differ.
%
%
Building on this analysis, we next introduced \emph{gradient shaping}---the prerequisite for an attack-agnostic defense to poisoning---that bounds gradient magnitudes and minimizes angular differences.
Gradient shaping allows us to move towards a generic defense, in contrast to prior defenses that exploit attack-specific properties or rely on the identification of points that were poisoned.
To study the feasibility of gradient shaping, we consider DP-SGD---a natural candidate algorithm for training with gradient shaping because it clips and perturbs the gradients to provide privacy guarantees.
Our experiments with DP-SGD show that it reduces the model's accuracy drop in the presence of indiscriminate attacks, mitigates one-shot targeted attacks, and increases the adversary's cost in multi-poison targeted attacks.
We also observed that DP-SGD becomes ineffective against a strong, yet unrealistic, indiscriminate attack.
This highlights designing an effective gradient shaping mechanism is a promising direction towards an ideal poisoning defense.

\if\usenixcamready1

\section*{Availability}
\label{sec:availability}

Our code is available under an open-source license from: {\small \url{https://github.com/Sanghyun-Hong/Gradient-Shaping}}.

\section*{Acknowledgments}
\label{sec:ack}

We thank Dana Dachman-Soled, Furong Huang, Matthew Jagielski, Yuzhe Ma, and Michael Davinroy for their constructive feedback.
We acknowledge Tom Goldstein, Ronny W. Huang, and the University of Maryland super-computing resources\footnote{\url{http://hpcc.umd.edu}} (DeepThought2) made available for conducting the experiments reported in our paper.
This research was partially supported by the Department of Defense and Canadian Institute for Advanced Research (CIFAR).
\else\fi

{
    \footnotesize
    \bibliographystyle{plain}
    \bibliography{bibliography/security,bibliography/miscs}
    
    \vspace{2.0em}
}

\appendix

\begin{center}
	{\large \textbf{Appendix}}
	\vspace{-0.6em}
\end{center}

\section{Intuition Behind Our Gradient-Analysis}
\label{appendix:discuss-loss}

\begin{figure}[h]
	\centering
	\includegraphics[width=\columnwidth]{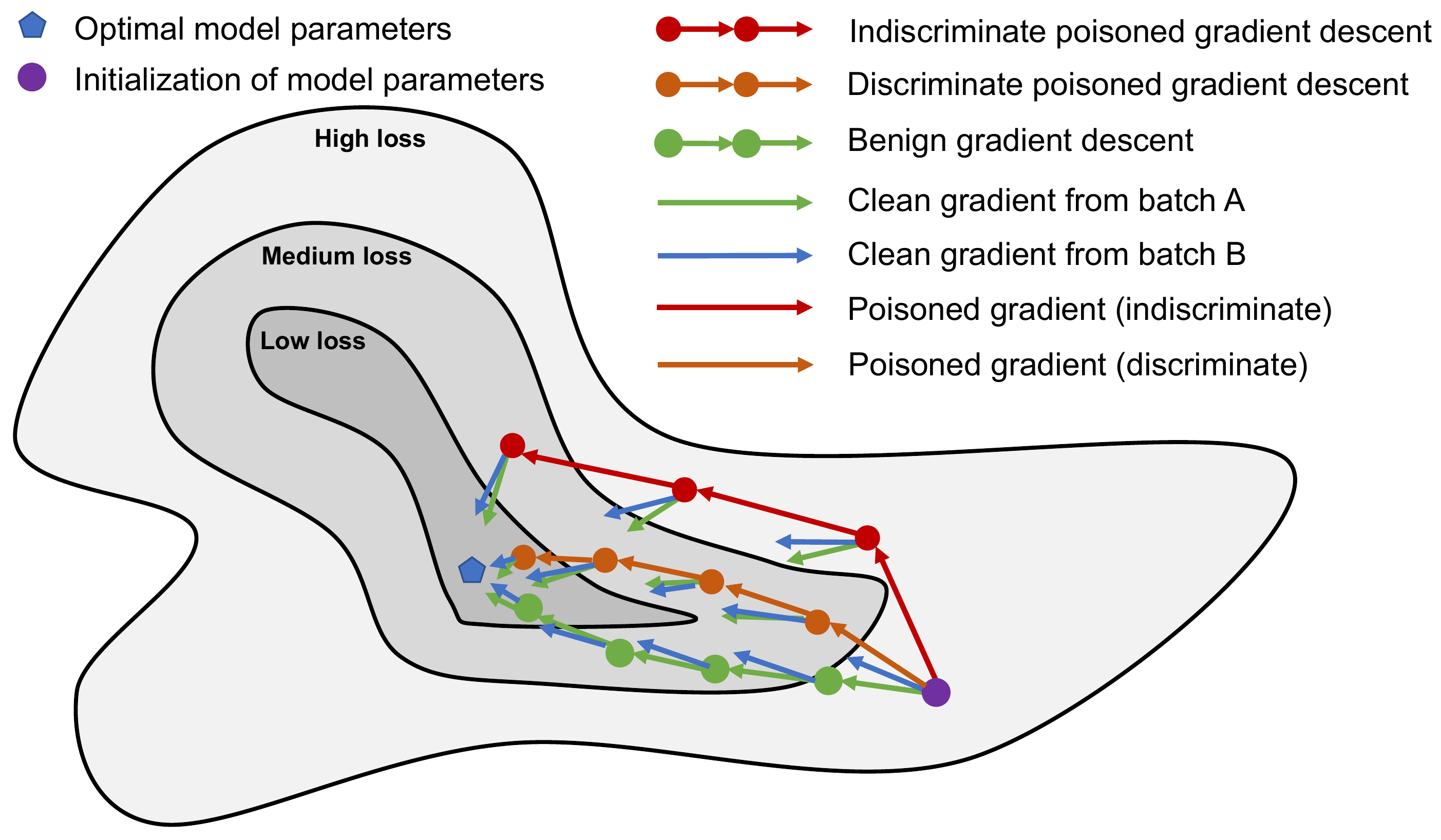}
	\caption{\textbf{Intuition Behind Gradient-Based Analysis of Poisoning.} This is a contour line visualization of the model's loss for different values of its parameters; each arrow corresponds to the step taken when computing a single model update, and each dot is a set of parameters values obtained after one or more steps of gradient descent from the random initialization. This abstract representation helps understand how poisoning impacts the loss.}
	\label{fig:motivating-example}
\end{figure}

\topic{Discussion Related to \S\ref{subsec:existing-defenses}:}
There is a limitation in characterizing the goodness of a set of parameters computed by the gradient descent sorely based on the value of the training loss.
Because the training objective for a neural network is non-convex, there exists multiple local minima with associated losses taking values comparable to the global minimum, denoted as the blue pentagon in Figure~\ref{fig:motivating-example}.
In the benign setting, all of these local minima correspond to models with comparable performance when it comes to predicting on test data.
The ML community identified that finding one of the local minima is sufficient even if it is not the global minimum~\cite{saxe2013exact, dauphin2014identifying, goodfellow2014qualitatively}.
One of these minima is the green point obtained by gradient descent on legitimate data---the green trajectory in Figure~\ref{fig:motivating-example}.
When poison is inserted in the training data, the adversary forces training to follow an alternative descent that may achieve either (1) a loss indicated by the orange dot that is \textit{similar} to the green dot in the case of \textit{targeted} poisoning attacks, or (2) a higher loss value for the red dot when the attack is \textit{indiscriminate}.
\emph{This distinction makes it difficult to characterize poisoning solely based on the loss achieved upon completion of training, in particular when the attack is targeted.
Instead, one should capture the trajectory taken by gradient descent in the presence of poisoned data}.
Thus, we focus on the norm and orientation of gradients.

\section{Differentially Private (DP) SGD}
\label{appendix:dp-sgd}


\begin{algorithm}[b]
\SetAlgoLined
\LinesNumbered

\SetKwInOut{Input}{Input}
\SetKwInOut{Output}{Output}
\Input{%
$\mathcal{D}_{tr}$: training set, $\mathcal{L}$: loss function, $\eta_t$: learning rate, \textcolor{blue}{$C$: clipping norm, $\sigma$: noise multiplier}}
\Output{%
$\theta$: model parameters, \textcolor{black}{$\varepsilon$: overall privacy cost}}

\textbf{initialize} $\theta_0$ randomly\;
\For{t $\in [T]$}{
    $\mathbf{x}_t \gets$ a mini-batch of random samples\;
    $\mathbf{g}_t \gets \nabla_{\theta_t}\mathcal{L}(f_{\theta}(\mathbf{x}_t), \mathbf{y}_t)$ \tcp*{compute gradient}
    \textcolor{blue}{$\bar{{\mathbf{g}_t}} \gets \mathbf{g}_t/\max(1, \frac{||\mathbf{g}_t(\mathbf{x}_t)||_2}{C})$} \tcp*{clip gradient\quad\quad}
	\textcolor{blue}{$\bar{{\mathbf{g}_t}} \gets \bar{{\mathbf{g}_t}} + \mathcal{N}(0, {\sigma^2}{\mathbb{\mathbf{I}})}$} \tcp*{add noise\quad\hspace{1.2em}\quad}
    $\theta_{t+1} \gets \theta_t - {\eta_t}{\bar{{\mathbf{g}_t}}}$ \tcp*{descent: update $\theta$}
}

\Return $\theta_{T}$, \textcolor{black}{$\varepsilon$}\;

\caption{Differentially Private (DP) SGD.}
\label{algo:sgd-and-dp}
\end{algorithm}

\topic{Details of DP-SGD Discussed in \S\ref{subsec:gradient-shaping}:}
To train a model with a provable privacy guarantee, we commonly use DP-SGD~\cite{Abadi:CCS16}---\ie a simple modification to the popular training mechanism, mini-batch SGD.
In Algorithm~\ref{algo:sgd-and-dp}, we highlighted the modifications in {\color{blue}blue}.
For each sample in a mini-batch, DP-SGD first bounds the gradient computed from a sample on the predefined value $C$ (line 5) and adds Gaussian noise to the gradient proportionally (line 6).
DP-SGD also provides an accounting mechanism that measures the total privacy budget spent up to a certain iteration, which enables to estimate the worst-case privacy leakage $\varepsilon$ of a model.
In practice, an ML expert controls the \emph{clipping norm} $C$ and the variance $\sigma$ (\emph{noise multiplier}) of the distribution where the noise is drawn to train a model that achieves a reasonable accuracy and privacy guarantee.
The training procedure stops when the total privacy expenditure exceeds a privacy leakage of $\varepsilon$.
In our work, we do not utilize DP-SGD's privacy accounting mechanism; we use the algorithm as a tool to realize gradient shaping.

\section{Neural Network Architectures}
\label{appendix:network-arch}


\begin{table}[h]
\centering
\adjustbox{max width=0.86\linewidth}{
    \begin{tabular}{@{}cc|cc@{}}
    \toprule
    \multicolumn{2}{c|}{\textbf{MLP}} & \multicolumn{2}{c}{\textbf{CNN}} \\ \midrule
    \textbf{Layer Type} & \textbf{Layer Size} & \textbf{Layer Type} & \textbf{Layer Size} \\ \midrule
    FC (R) & 256 & Conv (R) & 3x3x16 \\
    FC (-) & \# classes & Conv (R) & 3x3x32 \\
    - & - & MaxPool & 2x2 (p: 0.5) \\
    - & - & FC (R) & 64 \\
    - & - & FC (-) & \# classes \\ \midrule
    \end{tabular}
}
\caption{\textbf{Neural Network Architectures.} The architectures of neural networks (MLP and CNN) used in our evaluation.}
\label{tbl:network-architectures}
\end{table}

\topic{Information Relevant to \S\ref{subsec:evaluation-exp-setup}:}
We show our two baseline networks (an MLP and CNN) in Table~\ref{tbl:network-architectures}.
The description in a parenthesis indicates the activation functions used (R: ReLU, and '-': None).
The output of each network is equal to the number of classes: 10 for FashionMNIST and 100 for CIFAR10 and Purchase-100.

\section{Why is the DP-Model Resilient?}
\label{appendix:boundary-analysis}

\begin{figure}[h]
	\centering
	\includegraphics[width=\linewidth]{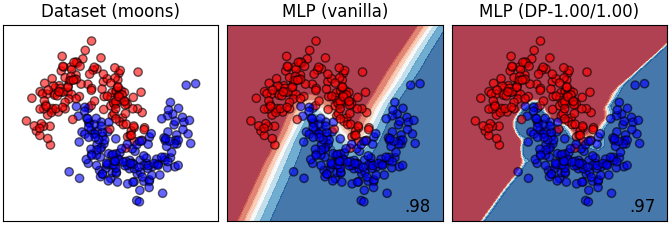}
	\caption{\textbf{Decision Boundaries of Vanilla and DP Models.} The boundary learned by the model trained with DP-SGD (the right) are more complex than the vanilla model (in the middle) and has higher confidence in decisions on the samples near it.}
	\label{fig:regularization-boundaries}
\end{figure}

\topic{Discussion Related to \S\ref{subsec:resilience-dp-trained-model}:}
To understand why the DP-model can reduce the success rate of the targeted attacks even if the model is trained with SGD, we conduct the decision boundary analysis of two models: one trained without DP-SGD and the 
one with.
For this experiment, we utilize the 2-dimensional, two-moons dataset~\cite{Moons}
that consists of 700 training and 300 testing samples.
We trained two MLP models. 
We set both the clipping norm and noise multiplier to 1.0; the models achieve the accuracy of 0.98 and 0.97 respectively.

\begin{figure}[t]
	\centering
	\includegraphics[width=0.8\linewidth]{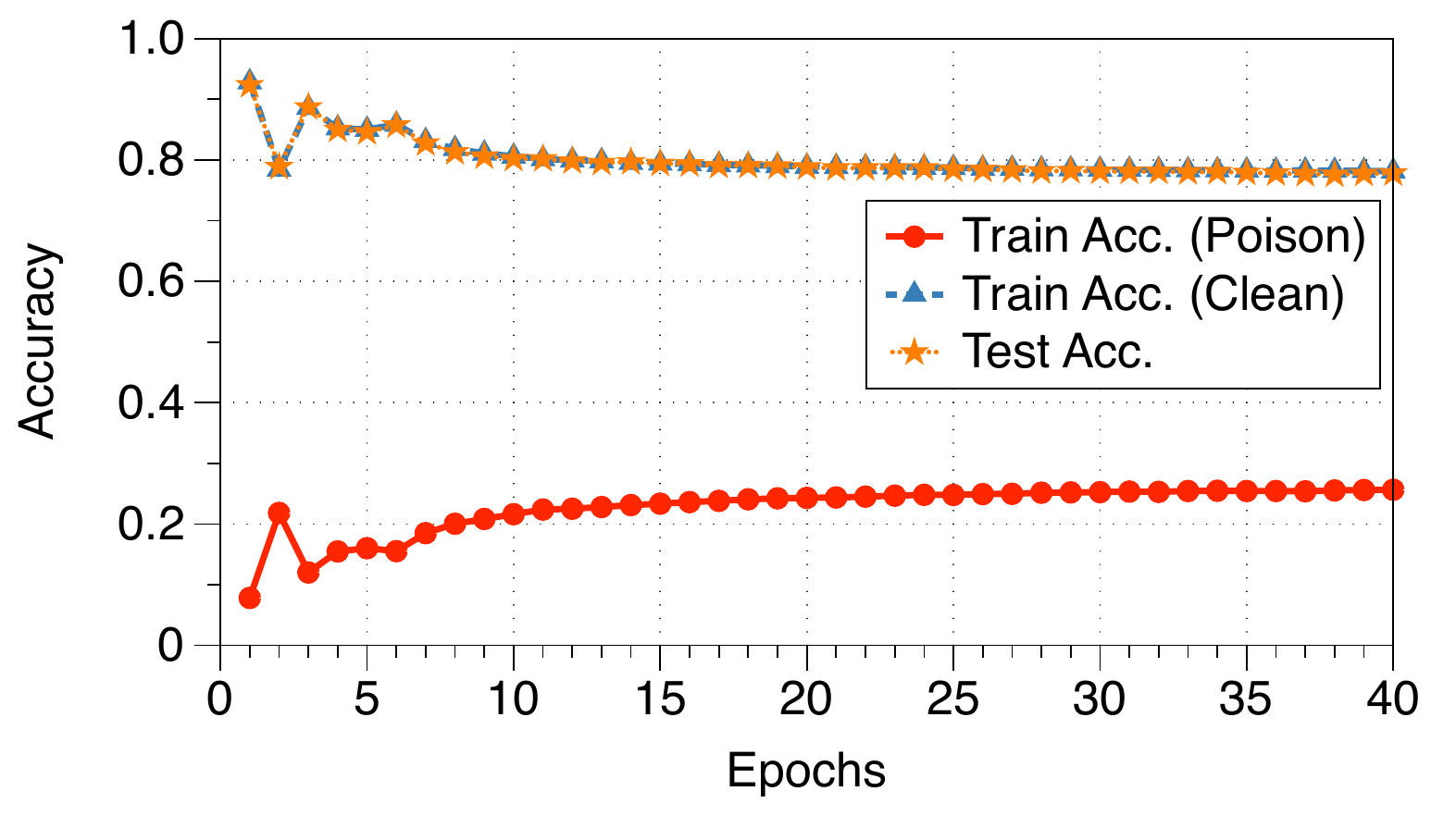}
	\includegraphics[width=0.8\linewidth]{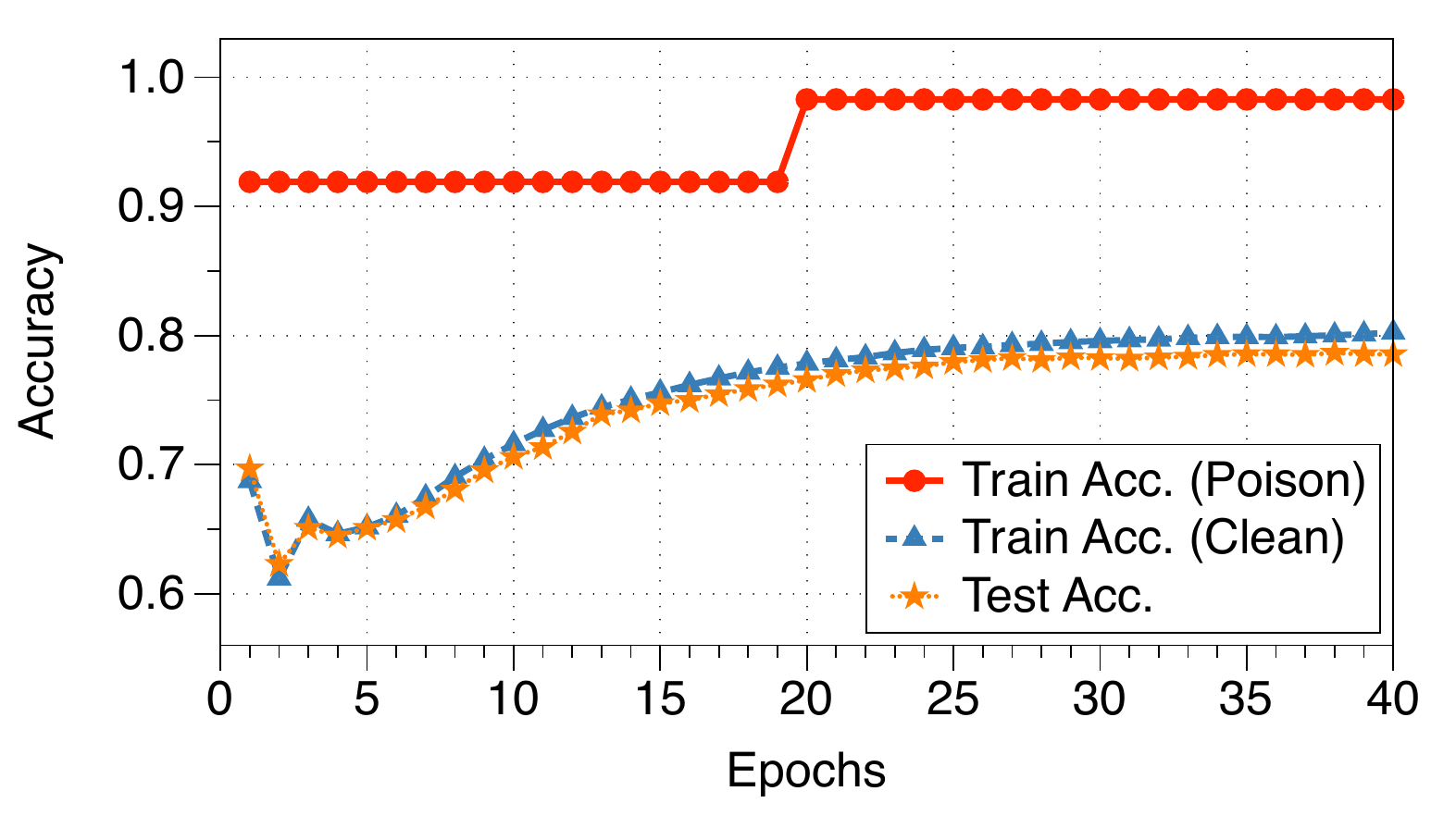}
	\caption{\textbf{The Accuracy of a Model over Clean and Poison Samples During Training.} We plot the case of the random LF attacks in the upper figure and the SOTA attack case in the lower plot. In both cases, we use 40\% of poisons.}
	\label{fig:training-accuracy-analysis}
\end{figure}

Figure~\ref{fig:regularization-boundaries} illustrates the decision boundary of the vanilla- and DP-model.
The leftmost figure shows the distribution of the two-moons dataset with the color {\color{red}red} and {\color{blue}blue} corresponding to each class, and in the other two figures, we display the boundary in the figure's background.
The contour colors indicates the decision confidence: if the color is darker, the higher the confidence is.
The white area in between is where the decision boundary lies.

Here, we found that, when a model is trained with DP-SGD in a way that the model achieves the best possible accuracy, the model learns complex decision boundary---\ie it overfits to the training data.
We also observe that, in the DP-model, the white-area where a model is uncertain about its decisions becomes narrower.
This means the overall confidence of the model's decision will increase; thus, the amount of parameter updates---the sum of the gradients required for misclassifications of targets---will increase.
In consequence, in the one-shot attack, the success rate of the attacker decreases and the multi-poison attack has to use more poisons.

\section{Analysis of Training-time Accuracy in the Indiscriminate Poisoning Attacks}
\label{appendix:steinhardt-training}

\topic{Discussion Related to \S\ref{subsec:steinhardt-attack}:}
The training-time accuracy observed during training is an indicator that shows whether the model learns its decision boundary based on a specific set of samples.
Hence, we monitor the accuracy of a model over the clean data and poisons during training in the random LF and the SOTA attacks formulated by~\cite{Steinhardt:NIPS17}.
Figure~\ref{fig:training-accuracy-analysis} illustrates the training-time accuracy monitored in both the attacks.
In the random LF attack (upper), the accuracy of a model over clean samples is more than 80\% over 40 epochs whereas the accuracy over the poisons is below 30\%.
We can see, in this case, the model achieves 80\% accuracy---the same as the accuracy over clean data---over the testing set.
On the other hand, the training-time accuracy of a model over the poisons formulated by the SOTA attack is over 90\% whereas the accuracy over the clean samples is below 80\% over 40 epochs.
This means the LR model trained in the SOTA attack formulates the decision boundary based on the poisons and cannot be modified easily during training.
From our analysis in \S\ref{subsec:steinhardt-attack}, we know the poisons consist of a single image of the class coat and two images of the class dress; thus, the model uses these poisons as pivots for the linear decision boundary and, during training, it is changed by the clean samples marginally.

\section{Trade-offs Between Model's Utility and Privacy Leakage}
\label{appendix:acc-privacy-trade-offs}


\begin{figure*}[h]
\centering
\begin{subfigure}[t]{0.48\textwidth}
    \centering
    \includegraphics[width=\textwidth]{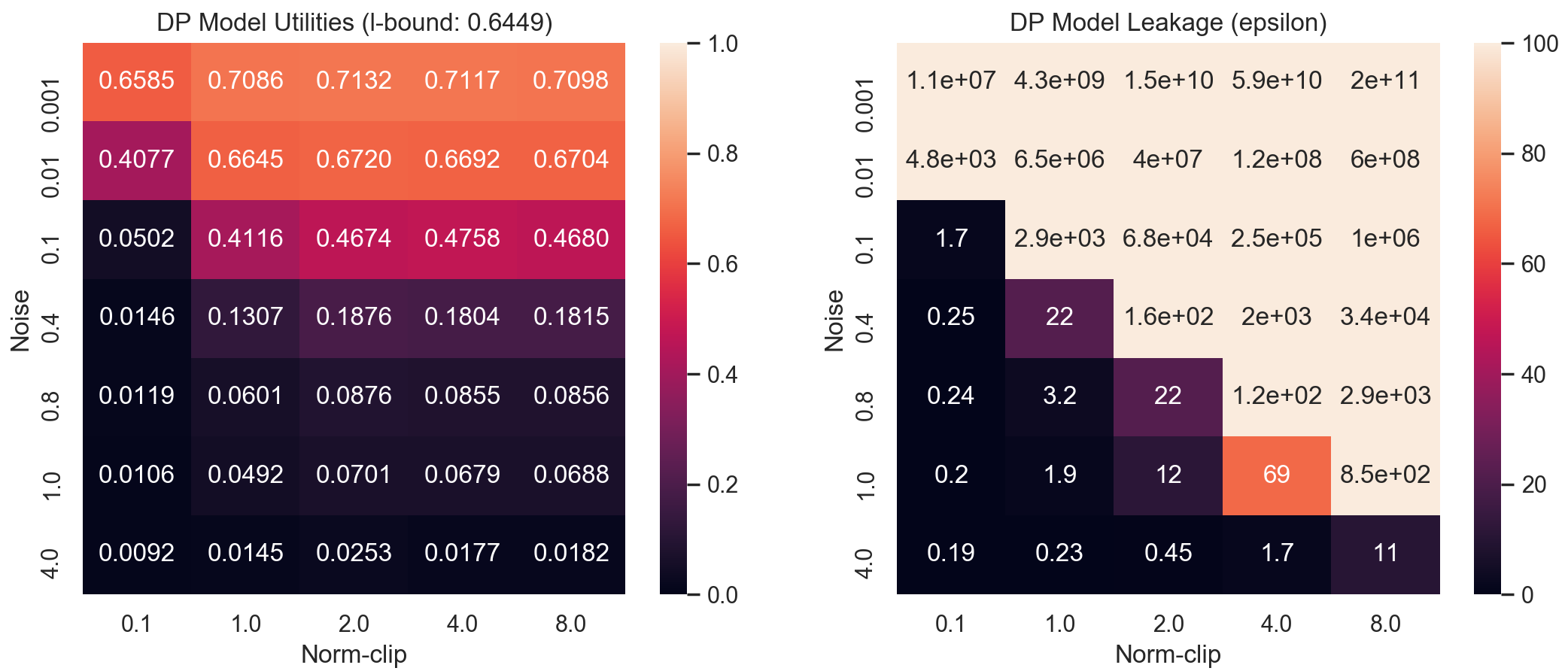}
    \caption*{\textbf{Logistic Regression (Purchases-100)}}
\end{subfigure}
\begin{subfigure}[t]{0.48\textwidth}  
    \centering 
    \includegraphics[width=\textwidth]{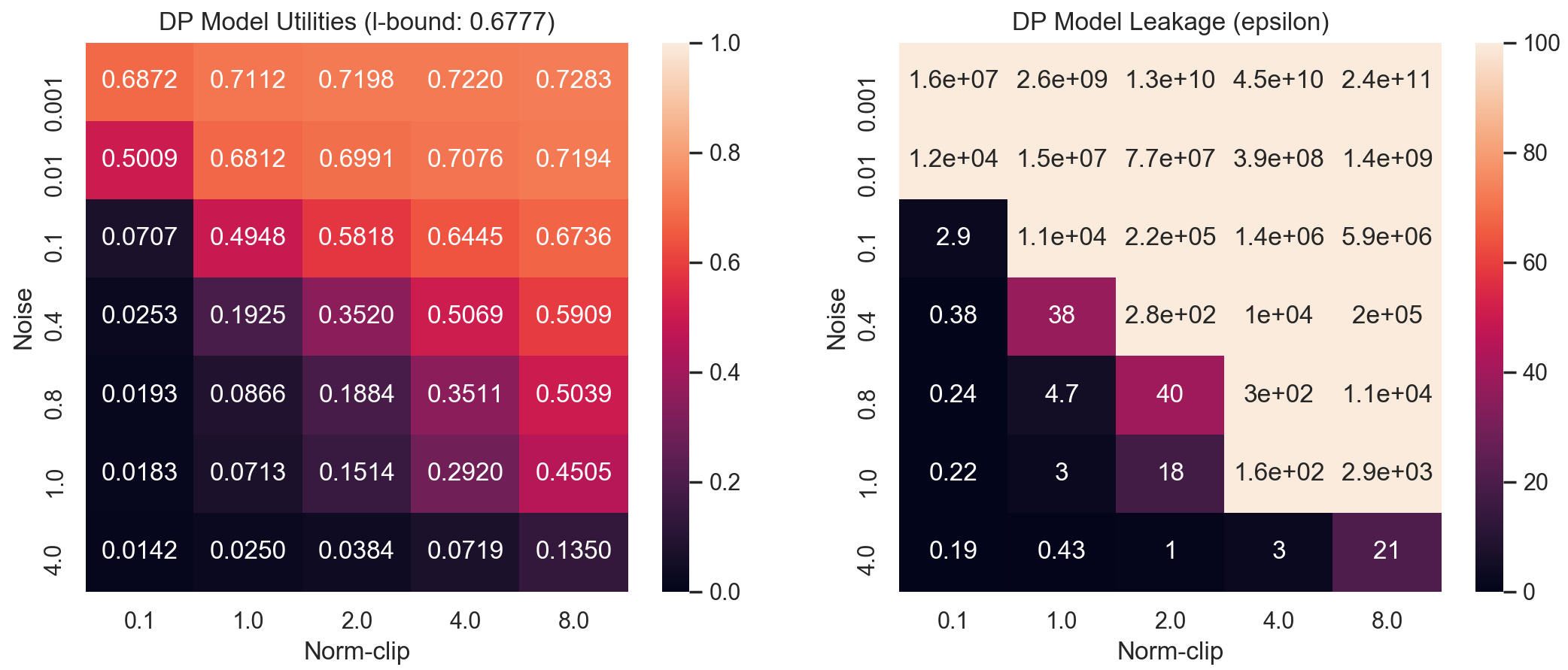}
    \caption*{\textbf{Neural Network (MLP) (Purchases-100)}}
\end{subfigure}
\vskip\baselineskip
\begin{subfigure}[t]{0.48\textwidth}   
    \centering 
    \includegraphics[width=\textwidth]{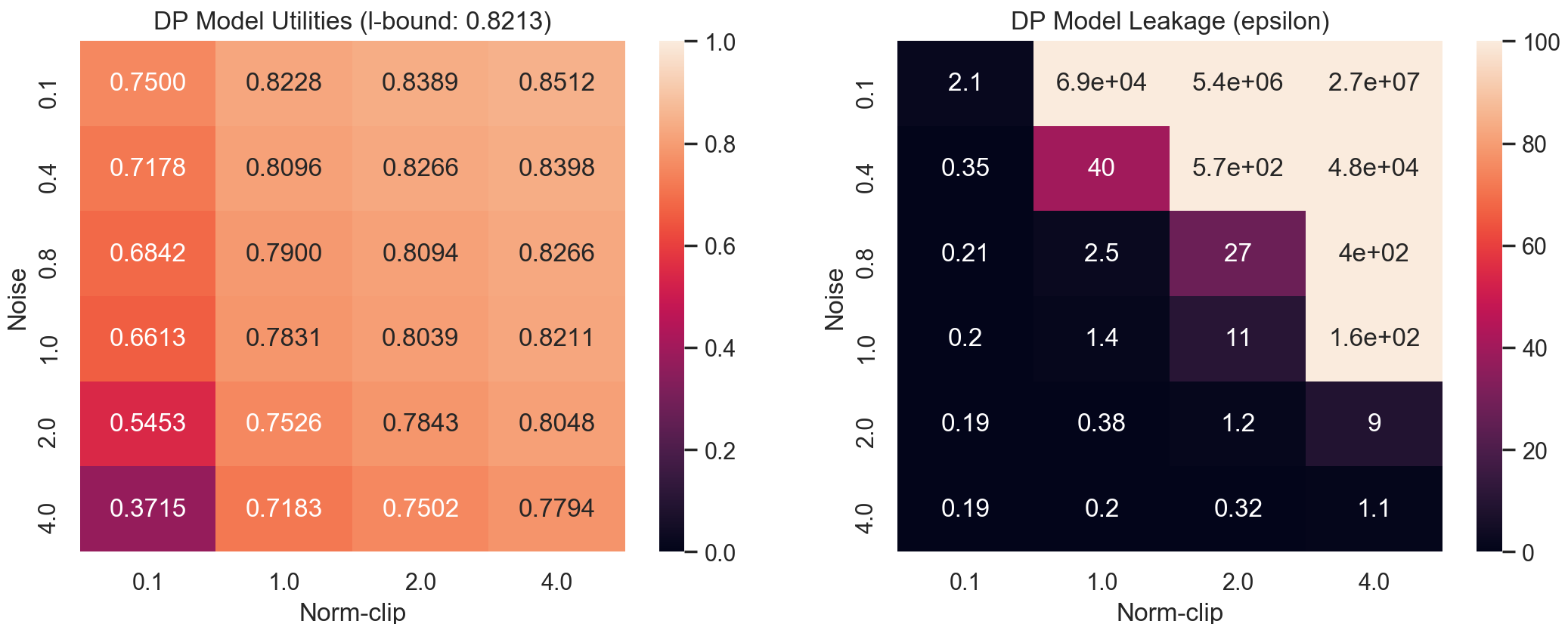}
    \caption*{\textbf{Neural Network (MLP) (FashionMNIST)}}
    \vspace{0.8em}
\end{subfigure}
\begin{subfigure}[t]{0.492\textwidth}   
    \centering 
    \includegraphics[width=\textwidth]{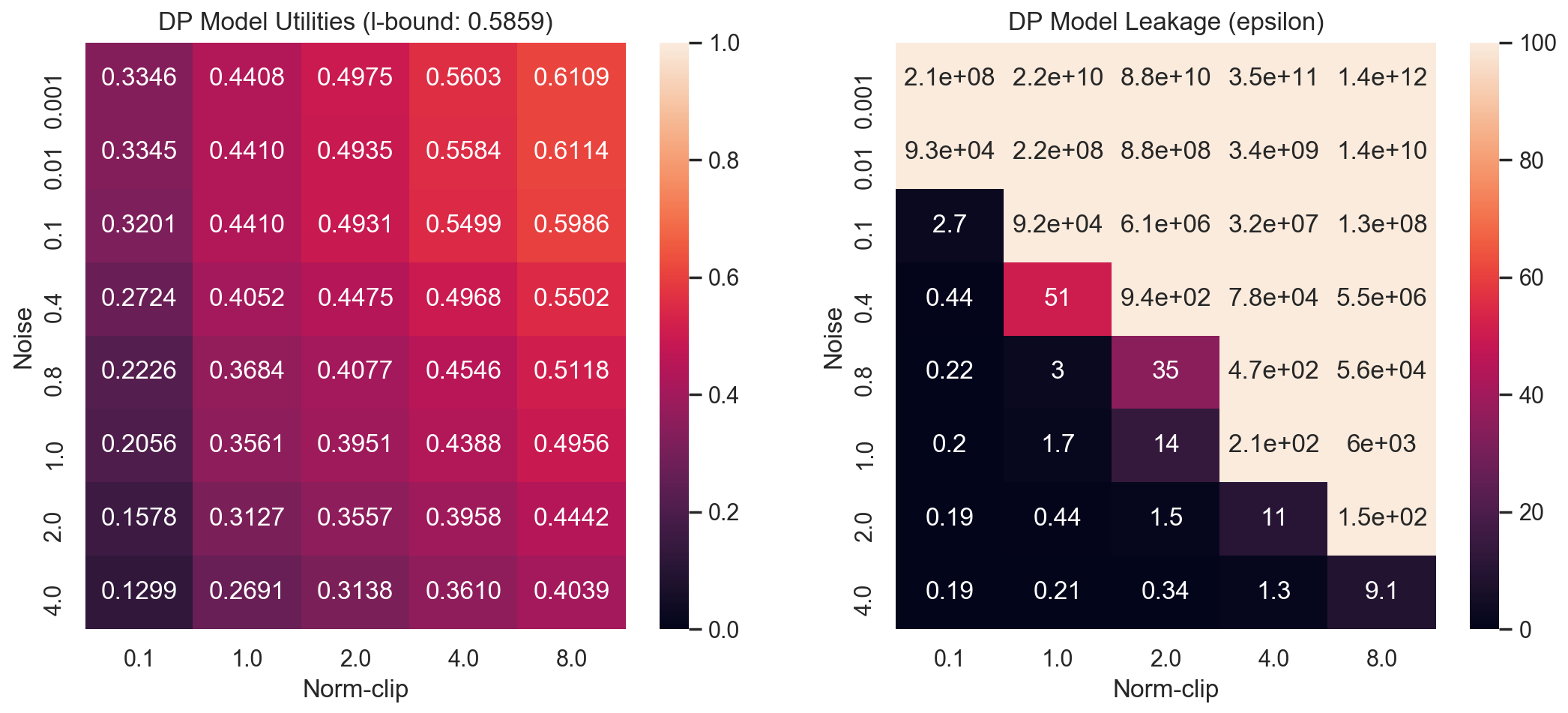}
    \caption*{\textbf{Neural Network (CNN) (CIFAR-10)}}
    \vspace{0.8em}
\end{subfigure}
\caption{\textbf{Utility of Models Trained with Different DP-SGD Parameters.} We illustrate the utility of models trained with various DP-SGD parameters. We see that the model's utility decreases when we use a stronger privacy guarantee---a smaller clipping norm and higher noise multiplier. We examine three models (LR, MLP, and CNN) trained on Purchases-100, FashionMNIST, and CIFAR-10, respectively.}
\label{fig:model-tradeoffs}
\end{figure*}

\topic{Analysis of the Utility Loss When We Use DP-SGD (\S\ref{sec:evaluation}):}
Here, we discuss the trade-offs between the model's utility and privacy leakage ($\varepsilon$) in Figure~\ref{fig:model-tradeoffs}.
DP-SGD/-Adam is designed to control the leakage of a model by adding noise to gradients, which inherently causes the performance degradation of a resulting model.
Observe that one can minimize the noise added to each gradient (ergo improve the utility of the model learned) by minimizing the clipping norm.
However, setting a very small value for the clipping norm destroys important information carried in the gradients.
Similarly, one can choose to retain this information by choosing a large value for the clipping norm.
This, in turn, translates to a large value of noise required to the gradients, degrading utility.


\end{document}